\ifdefined\pdfoutput\pdfoutput=1\fi %
\documentclass[a4paper,11pt]{article}
\usepackage[T1]{fontenc}
\usepackage{lmodern}
\usepackage[a4paper,top=3cm,bottom=2.6cm,left=2.6cm,right=2.6cm]{geometry}
\usepackage{amsmath,amssymb,amsfonts,mathtools}
\usepackage{graphicx,booktabs,array,tabularx,enumitem}
\usepackage[dvipsnames]{xcolor}
\usepackage{tikz}
\usetikzlibrary{arrows.meta,decorations.pathreplacing,calc}
\usepackage[font=small,labelfont=bf]{caption}
\usepackage{microtype}
\usepackage[section]{placeins}
\usepackage[numbers,sort&compress]{natbib}
\usepackage[colorlinks=true,linkcolor=blue!50!black,citecolor=green!40!black,urlcolor=red!50!black]{hyperref}
\hypersetup{pdftitle={Level repulsion and the quantum viscosity of near-extremal black branes},pdfauthor={Panos Betzios, Olga Papadoulaki, Maria Myrto Pegioudi}}
\IfFileExists{orcidlink.sty}{\usepackage{orcidlink}}{}
\ifdefined\orcidlink\else
  \DeclareRobustCommand{\orcidlink}[1]{\,\href{https://orcid.org/#1}{\textsuperscript{\textsf{\tiny iD}}}}
\fi
\numberwithin{equation}{section}
\setlist{itemsep=2pt,topsep=4pt}
\setlist[itemize]{leftmargin=1.5em}
\setlist[enumerate]{leftmargin=1.9em}

\allowdisplaybreaks
\graphicspath{{./}{figures/}{figs/}}

\makeatletter
\renewcommand\section{\@startsection{section}{1}{\z@}{-3.5ex \@plus -1ex \@minus -.2ex}{2.3ex \@plus.2ex}{\normalfont\Large\bfseries}}
\renewcommand\subsection{\@startsection{subsection}{2}{\z@}{-3.25ex\@plus -1ex \@minus -.2ex}{1.5ex \@plus .2ex}{\normalfont\large\bfseries}}
\makeatother

\makeatletter
\newcommand{\arxiv}[1]{\in@{/}{#1}%
  \ifin@[\href{https://arxiv.org/abs/#1}{\texttt{#1}}]%
  \else[\href{https://arxiv.org/abs/#1}{\texttt{arXiv:#1}}]\fi}
\makeatother

\newcommand{\etasc}{\eta^{\mathrm{s.c.}}}
\newcommand{\sinc}{\operatorname{sinc}}
\newcommand{\erf}{\operatorname{erf}}
\newcommand{\erfc}{\operatorname{erfc}}
\newcommand{\Ai}{\operatorname{Ai}}
\newcommand{\cJ}{\mathcal{J}}
\newcommand{\cI}{\mathcal{I}}
\newcommand{\cG}{\mathcal{G}}
\newcommand{\cP}{\mathcal{P}}
\newcommand{\cE}{\mathcal{E}}
\newcommand{\cO}{\mathcal{O}}
\newcommand{\cH}{\mathcal{H}}
\newcommand{\cR}{\mathcal{R}}
\providecommand{\bra}[1]{\langle #1|}
\providecommand{\ket}[1]{|#1\rangle}
\newcommand{\vev}[1]{\left\langle #1\right\rangle}
\newcommand{\dd}{\mathrm{d}}
\newcommand{\blockhead}[1]{\multicolumn{4}{@{}>{\raggedright\arraybackslash}p{\textwidth}@{}}{#1}}

\begin{document}

\thispagestyle{empty}
\begin{flushleft}
\vspace*{0.4cm}
{\LARGE\sffamily\bfseries Level repulsion and the quantum viscosity of near-extremal black branes\par}
\vspace{0.9cm}
{\large\sffamily\bfseries
Panos Betzios\,\orcidlink{0000-0002-5350-9404},$^{a}$ Olga Papadoulaki\,\orcidlink{0000-0001-5302-2930}$^{b}$ and Maria Myrto Pegioudi$^{b}$\par}
\vspace{0.5cm}
{\small\itshape
$^{a}$Department of Physics and Astronomy, Ghent University,\\ Krijgslaan 281-S9, 9000 Gent, Belgium\\[2pt]
$^{b}$CPHT, CNRS, \'Ecole polytechnique, Institut Polytechnique de Paris,\\ 91120 Palaiseau, France\par}
\vspace{0.4cm}
{\small\textit{E-mail:} \href{mailto:panos.betzios@ugent.be}{\texttt{panos.betzios@ugent.be}}, \href{mailto:olga.papadoulaki@polytechnique.edu}{\texttt{olga.papadoulaki@polytechnique.edu}},\\ \href{mailto:maria-myrto.pegioudi@polytechnique.edu}{\texttt{maria-myrto.pegioudi@polytechnique.edu}}\par}
\end{flushleft}

\vspace{0.7cm}
\noindent\textsc{Abstract:} Below the gap scale of the AdS$_2$ throat, quantum fluctuations make the shear viscosity of near-extremal black branes grow as $(CT)^{-1/2}$ in the disk approximation of JT gravity, and the one-loop $\eta/s$ diverges. To cure this divergence, we include the non-perturbative completion of JT gravity as a matrix integral. The Kubo formula then probes universal features of level correlations. At finite extremal entropy $S_0$, level repulsion makes the DC viscosity vanish. A finite, universal viscosity survives in a double-scaling limit in which the probe frequency $\omega$ is held in a fixed ratio to the level-spacing scale $\omega_{\rm sp}\equiv e^{-S_0}/C$. At fixed $\omega$, with $\omega_{\rm sp}\ll\omega\ll e^{S_0/3}\omega_{\rm sp}$ the Schwarzian growth then saturates at $\eta_{\max}\simeq0.174\,(\omega/\omega_{\rm sp})\,\etasc$, where $\etasc$ is the semiclassical viscosity, and the viscosity decreases at lower temperatures. At the extremely low temperature scale $CT\sim e^{-2S_0/3}$ the genus expansion and the one-loop entropy computations completely break down. There we start probing the edge of the spectrum and the problem reduces exactly to the Airy kernel. We find that the ratio of the quenched averages of $\eta$ and $s$ is finite and positive at every non-zero temperature. We compare our results with explicit completions defined by the string equation, in which the spectrum is bounded from below. The bulk results are the same in these completions and the quenched edge results change only by a few per cent, while the annealed quantities depend on the non-perturbative completion at order one.

\vspace{0.45cm}

\newpage
\hrule
\vspace{-0.3cm}
\tableofcontents
\vspace{0.4cm}
\hrule
\vspace{0.6cm}

\section{Introduction}
\label{sec:intro}

One of the sharpest predictions of holography concerns the shear viscosity $\eta$ of strongly
coupled plasmas. For any theory with a two-derivative gravitational dual, the ratio of the
shear viscosity to the entropy density takes the universal value
\cite{PolicastroSonStarinets2001,KovtunSonStarinets2005,BuchelLiu2004,SonStarinets2007}
\begin{equation}
\frac{\eta}{s}=\frac{1}{4\pi}.
\label{eq:KSS}
\end{equation}
The origin of this universality is geometric. A shear flow of the boundary fluid is dual to a
transverse graviton $h_{xy}$ in the bulk. At zero spatial momentum this graviton obeys the
equation of a minimally coupled massless scalar, and its low-frequency absorption cross
section by the black brane equals the area of the horizon
\cite{DasGibbonsMathur1997,KovtunSonStarinets2005,IqbalLiu2009}. The entropy is also
given by the area of the horizon, and the two areas cancel in the ratio. The argument holds
at any temperature at which the classical geometry is reliable. This includes the
near-extremal regime, where the viscosity of the Reissner--Nordstr\"om brane is controlled by
the AdS$_2$ throat that develops near the horizon
\cite{EdalatiJottarLeigh2010,FaulknerLiuMcGreevyVegh2011}.

The classical geometry, however, is not reliable arbitrarily close to extremality. Consider a
charged black brane at chemical potential $\mu$ and temperature $T\ll\mu$. Its near-horizon
region is a long throat. Exactly at extremality the throat is infinitely long and its geometry
is AdS$_2\times X^{d-1}$, where $X^{d-1}$ is the transverse space. At small non-zero temperature
this product form holds only approximately: the throat has a finite length, its two-dimensional
part is a black hole in nearly-AdS$_2$, and the size of $X^{d-1}$ varies slowly along it. The
dynamics of this region reduces to
Jackiw--Teitelboim (JT) gravity coupled to matter
\cite{AlmheiriPolchinski2015,MaldacenaStanfordYang2016,Jensen2016,EngelsoyMertensVerlinde2016,NayakShuklaSoniTrivediVishal2018,MoitraTrivediVishal2019,MertensTuriaci2022review}.
The only dynamical gravitational degree of freedom of the throat is a boundary graviton, the
Schwarzian mode. Its action is proportional to a coupling $C$, which, like the extremal
entropy $S_0$, is proportional to the transverse area of the brane. The Schwarzian mode is
weakly coupled for $T\gg E_{\rm gap}\equiv1/C$ and strongly coupled below this scale
\cite{IliesiuTuriaci2020,StanfordWitten2017,KitaevSuh2018}. In the strongly coupled regime
the thermodynamics receives the well-known $\frac32\log T$ correction
\cite{IliesiuTuriaci2020,IliesiuMurthyTuriaci2022}, and the arguments for transport based on
the horizon must be revisited, because the fluctuations of the horizon itself are no longer
small. The same Schwarzian sector governs the low-energy physics of the SYK model and of its
transport \cite{Sachdev2015,MaldacenaStanford2016,DavisonEtAl2017}, and it appears
universally in near-extremal limits \cite{GhoshMaxfieldTuriaci2019}.

The quantum shear viscosity in this regime was computed recently
\cite{KanargiasKiritsisMurthyPapadoulakiPorfyriadis2025,CremoniniLiLiuNian2025}. Schwarzian
corrections to the shear correlators of the near-extremal Reissner--Nordstr\"om brane were also
analysed in \cite{GouterauxRamirezSupiot2025}, the one-loop correction to $\eta/s$ in the weakly
coupled Schwarzian regime was obtained in \cite{PandoZayasZhang2025}, quantum corrections to
the fluid/gravity correspondence at low temperature were studied in
\cite{NianPandoZayasYue2025}, Schwarzian corrections to the electrical conductivity, and to
transport in the presence of momentum relaxation, were computed in
\cite{LiuNianPandoZayas2024,MaulikPandoZayasZhang2026,DasMaulikPandoZayasZhang2026}, and the viscosity of
rotating near-extremal AdS$_4$ black holes was studied in \cite{NianPandoZayas2021}. All of these
works treat the spectrum of the throat as continuous. (For a related observable, the
evaporation rate of near-extremal charged black holes, non-perturbative corrections were
studied in \cite{MohanThorlacius2024}.) The Kubo
formula expresses $\eta$ through the retarded two-point function of the stress tensor
$T_{xy}$. At low frequencies this correlator is determined by the thermal two-point function
of the operator of dimension $\Delta_\cO=1$ that is dual to $h_{xy}$ in the throat, and this
two-point function is known exactly in the Schwarzian theory
\cite{MertensTuriaciVerlinde2017,BagretsAltlandKamenev2016,Yang2018}. The result of
\cite{KanargiasKiritsisMurthyPapadoulakiPorfyriadis2025} is
\begin{equation}
\frac{\eta}{\etasc}=\erf\Bigl(\sqrt{2\pi^2CT}\Bigr)\Bigl(1+\frac{1}{4\pi^2CT}\Bigr)
+\frac{e^{-2\pi^2CT}}{\sqrt{2\pi^3CT}} ,
\label{eq:intro-disk}
\end{equation}
where $\etasc=s^{\rm s.c.}/4\pi$ is the semiclassical value, built from the
Bekenstein--Hawking entropy density $s^{\rm s.c.}$ at temperature $T$. Equation
\eqref{eq:intro-disk} interpolates between two regimes,
\begin{equation}
\frac{\eta}{\etasc}=
\begin{cases}
1+\dfrac{1}{4\pi^2CT}+\cO\bigl(e^{-2\pi^2CT}\bigr), & CT\gg1,\\[10pt]
\Bigl(\dfrac{2}{\pi^3CT}\Bigr)^{1/2}\Bigl(1+\dfrac{2\pi^2}{3}\,CT-\dfrac{2\pi^4}{15}\,(CT)^2+\dots\Bigr), & CT\ll1 .
\end{cases}
\label{eq:intro-disk-limits}
\end{equation}
For $CT\gg1$, that is for $E_{\rm gap}\ll T\ll\mu$, the Schwarzian mode is weakly coupled.
The thermally occupied states then lie far above the gap scale, the throat is well described by
its classical geometry, and the viscosity takes its semiclassical value, up to a small
one-loop correction. For $CT\ll1$, that is for $T\ll E_{\rm gap}$, the Schwarzian mode is
strongly coupled. The density of states near the bottom of the spectrum is then that of the
Schwarzian theory, $\rho(E)\propto\sqrt E$, and the viscosity \emph{grows} as $(CT)^{-1/2}$ as
the temperature is lowered: quantum fluctuations of the horizon enhance dissipation. Combined
with the one-loop entropy, $\eta/s$ first dips below $1/4\pi$ and then rises
(Figure~\ref{fig:disk}), but within the disk approximation nothing stops this rise: the one-loop ratio diverges at an exponentially small but finite temperature, which signals the breakdown of that approximation.

Equation \eqref{eq:intro-disk} is a disk-level result, in which the spectrum of the throat is
treated as continuous. JT gravity, however, is equivalent, to all orders of the genus
expansion, to a double-scaled Hermitian matrix integral
\cite{SaadShenkerStanford2019,StanfordWitten2019}, and the matrix integral provides a
non-perturbative completion of the theory.\footnote{The completion is not unique. We return to this
point in Section~\ref{sec:string}.}
In this description the boundary Hamiltonian is a large random matrix. Its spectrum is
discrete, with mean level spacing
\begin{equation}
\Delta(E)=\frac{1}{\rho(E)}\propto e^{-S_0},
\end{equation}
and the levels obey random-matrix (Wigner--Dyson) statistics
\cite{Dyson1962,Mehta2004,CotlerEtAl2017}. Once matter is included, the matrix elements of a
probe operator between energy eigenstates are random variables, whose variance is set by the
Schwarzian vertex\footnote{The energy-dependent matrix element that appears in the exact disk
two-point function, see eq.~\eqref{eq:vertex} below.}, in the spirit of the eigenstate
thermalization hypothesis
\cite{Deutsch1991,Srednicki1994,Saad2019,BlommaertMertensVerschelde2019,JafferisKolchmeyerMukhametzhanovSonner2022}.
Late-time two-point functions in this framework have been studied in detail
\cite{Saad2019,StanfordYangYao2021,IliesiuMezeiSarosi2021}. In this language the Kubo formula
has a simple meaning. The imaginary part of the retarded correlator counts the transitions
of energy $\omega$ between thermally occupied levels, weighted by their matrix elements. At
frequencies comparable to the level spacing it therefore depends on how likely it is to find
two levels at distance $\omega$, that is, on the \emph{two-level correlation function}. For a
random-matrix spectrum this probability vanishes as $\omega\to0$, a phenomenon known as level
repulsion. The
question we address in this paper is what level repulsion does to the quantum viscosity
\eqref{eq:intro-disk}, and to $\eta/s$.

\paragraph{Three issues have to be settled.} The first is the fate of the DC viscosity once the two-level
correlations are included. The second concerns the normalization. A thermal correlator is
divided by a partition function, and the disk result uses the disk partition function. A
consistent calculation should instead use the partition function of the same
non-perturbative ensemble, which includes all effects of the same origin as those in the
numerator. One must ask whether these corrections matter, and whether they change the answer
qualitatively. The third issue is the regime of very low temperature. The one-loop entropy
$S_0+\frac32\log CT$ vanishes at $CT\sim e^{-2S_0/3}$, where the disk approximation ceases to be valid and the one-loop $\eta/s$ diverges
(Figure~\ref{fig:disk}(b)), and one has to find what replaces this naive divergence.

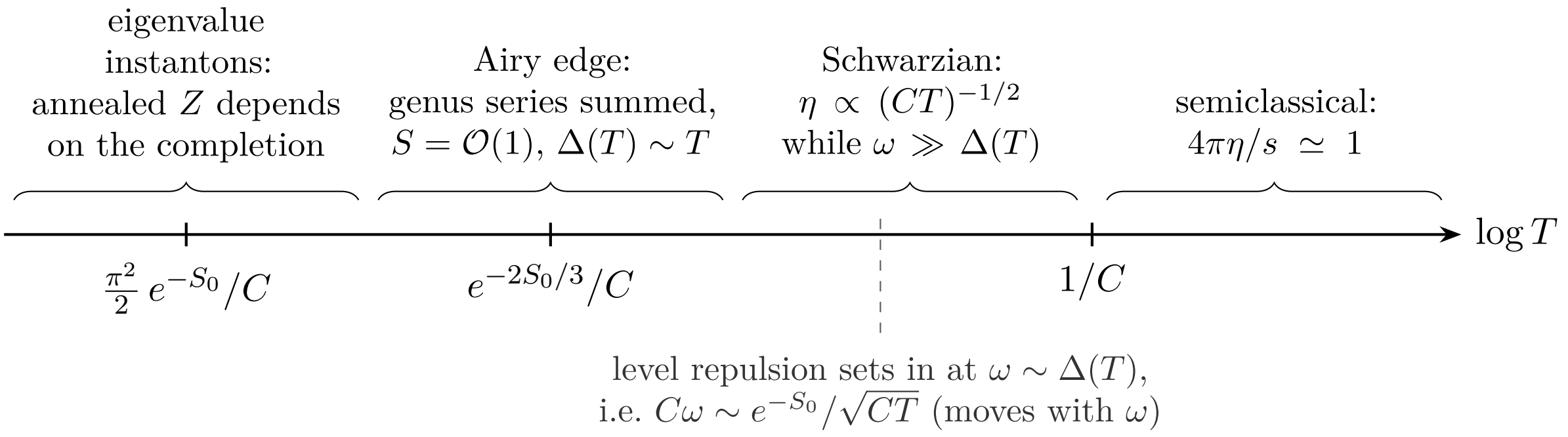
\begin{figure}[!ht]
\centering
\resizebox{\textwidth}{!}{%
\begin{tikzpicture}[x=1cm,y=1cm,>={Stealth[length=2.2mm]},
  lab/.style={align=center,font=\small,text width=3.3cm}]
  \draw[->,thick] (0,0) -- (14.8,0) node[right] {$\log T$};
  \foreach \x/\lab in {1.85/{$\tfrac{\pi^2}{2}\,e^{-S_0}/C$},5.55/{$e^{-2S_0/3}/C$},11.05/{$1/C$}}
    {\draw[thick] (\x,0.12) -- (\x,-0.12); \node[below] at (\x,-0.15) {\lab};}
  \draw[decorate,decoration={brace,amplitude=5pt}] (0.1,0.35) -- (3.6,0.35)
     node[midway,above=7pt,lab] {eigenvalue instantons:\\ annealed $Z$ depends\\ on the completion};
  \draw[decorate,decoration={brace,amplitude=5pt}] (3.8,0.35) -- (7.3,0.35)
     node[midway,above=7pt,lab] {Airy edge:\\ genus series summed,\\ $S=\cO(1)$, $\Delta(T)\sim T$};
  \draw[decorate,decoration={brace,amplitude=5pt}] (7.5,0.35) -- (10.9,0.35)
     node[midway,above=7pt,lab] {Schwarzian:\\ $\eta\propto(CT)^{-1/2}$\\ while $\omega\gg\Delta(T)$};
  \draw[decorate,decoration={brace,amplitude=5pt}] (11.2,0.35) -- (14.6,0.35)
     node[midway,above=7pt,lab] {semiclassical:\\ $4\pi\eta/s\simeq1$};
  \draw[dashed,gray!70!black] (8.9,-1.0) -- (8.9,0.25);
  \node[align=center,font=\small,gray!40!black] at (8.9,-1.6)
     {level repulsion sets in at $\omega\sim\Delta(T)$,\\ i.e.\ $C\omega\sim e^{-S_0}/\sqrt{CT}$ (moves with $\omega$)};
\end{tikzpicture}}
\caption{Schematic hierarchy of temperature scales for the viscosity of a near-extremal brane with extremal entropy
$S_0$ and Schwarzian coupling $C$. The horizontal axis is $\log T$: the three scales
shown are separated by factors that are exponentially large in $S_0$. The one-loop entropy vanishes, and the genus expansion of
the partition function breaks down, at $CT\sim e^{-2S_0/3}$. At finite $S_0$ the DC limit is
never smooth. At frequency $\omega$, level repulsion suppresses the response once the thermal
level spacing $\Delta(T)$ exceeds $\omega$.}
\label{fig:scales}
\end{figure}

The organizing principle that allows us to answer all three questions is that the
eigenvalues of a double-scaled Hermitian matrix integral form a \emph{determinantal point
process} \cite{Mehta2004,Forrester2010}. All their correlation functions are determinants
built from a single kernel $K(E_1,E_2)$. In particular, the one- and two-level densities are
\begin{equation}
\begin{gathered}
\vev{\rho(E)}=K(E,E),\\
\vev{\rho(E_1)\rho(E_2)}=K(E_1,E_1)\,K(E_2,E_2)-K(E_1,E_2)^2\qquad(E_1\neq E_2).
\end{gathered}
\end{equation}
The Kubo numerator is built from the two-level density. The partition function is the Laplace
transform of the one-level density, with the \emph{same} kernel,
\begin{equation}
Z(\beta)=\int\!\dd E\;e^{-\beta E}K(E,E).
\end{equation}
This is the precise meaning of ``normalizing by the partition function that includes the same
effects''. More precisely, it defines the annealed prescription, in which numerator and
denominator are averaged separately. The quenched prescription, which averages their ratio
over the same point process, can also be computed exactly, in terms of Fredholm determinants
(Section~\ref{sec:edge-exact}). The two prescriptions answer different questions. The annealed
one is the ratio of two ensemble averages, each of which is computed by a gravitational path
integral. The quenched one is the ensemble average of the response of a single member of the
ensemble, that is, of a quantum system with a definite Hamiltonian and a discrete spectrum.
The two agree as long as the partition function is self-averaging, which is the case for
$CT\gg e^{-2S_0/3}$ (Section~\ref{sec:DS-entropy}). They differ at order one at lower
temperatures (Section~\ref{sec:aq}), where only the quenched prescription gives a well-defined, non-negative entropy.
\paragraph{Our main results are as follows.}
\begin{enumerate}
    \item At any finite $S_0$, level repulsion makes the
imaginary part of the ensemble-averaged throat correlator vanish as $\omega^3$ \footnote{For a more general universality class it scales as
$\omega^{\beta_{\rm D}+1}$ for a Dyson index $\beta_{\rm D}$; we assume the unitary class
$\beta_{\rm D}=2$ in the rest, see Section~\ref{sec:mm}.}, and the DC viscosity
\emph{vanishes}. This is the transport counterpart of the ramp and plateau of the spectral
form factor \cite{CotlerEtAl2017}, the linear growth and late-time saturation of
$|\operatorname{Tr}e^{-(\beta+it)H}|^2$ that are produced by the same level correlations
(Section~\ref{sec:repulsion}). It is a property of the infinite-time limit of a chaotic
system with a discrete spectrum, and not a statement about hydrodynamics. The order of
limits matters: if the transverse volume is taken to infinity first, the spectrum is continuous
and the semiclassical viscosity is recovered.

\item A finite and universal viscosity is obtained in a \emph{bulk double-scaling limit}, in
which $S_0\to\infty$ and $\omega\to0$ with the ratio
\begin{equation}
r=\frac{e^{-S_0}}{C\omega}
\end{equation}
held fixed. The parameter $r$ compares the bare level-spacing scale $e^{-S_0}/C$ with the probe
frequency. In this limit
\begin{equation}
\frac{\eta_{\rm DS}}{\etasc}=\frac{\vev{k\,R_r(k)}_T}{2\pi CT},\qquad
R_r(k)=1-\sinc^2\Bigl(\frac{\sinh2\pi k}{2\pi r}\Bigr).
\label{eq:intro-DS}
\end{equation}
Here the energy is parametrized as $E=k^2/2C$, and $\vev{\cdot}_T$ is the average over the
disk thermal weight, which is proportional to $k\,e^{-k^2/2CT}\sinh(2\pi k)$. The factor
$R_r$ is the probability, relative to an uncorrelated spectrum, of finding a second level at
distance $\omega$. For $R_r=1$, i.e.\ for $r\to0$, equation \eqref{eq:intro-DS} reproduces
\eqref{eq:intro-disk}. At $CT\ll1$ the scaling function has a closed form in terms of Dawson's
function. At fixed $\omega$, the $(CT)^{-1/2}$ growth of the viscosity saturates at
\begin{equation}
\eta_{\max}\simeq0.174\,e^{S_0}C\omega\;\etasc
\end{equation}
(Section~\ref{sec:DS-asymptotics}), and the viscosity starts decreasing when the level spacing at the thermal energy exceeds $\omega$.

\item We show that this result is correctly normalized. To do so we include the
non-perturbative one-level effects of the same origin: the oscillations of the mean density of
states \cite{SaadShenkerStanford2019}, the Airy behaviour of the density near the edge of
the spectrum, the sum over higher-genus surfaces (expressed through Weil--Petersson volumes),
and eigenvalue instantons, i.e.\ single eigenvalues tunnelling into the classically forbidden
region. These are described in Section~\ref{sec:normalization}. At fixed $CT$ they
change the partition function only at order $e^{-2S_0}$ (in the soft-edge completion; a
completion with a hard wall adds a correction of order $e^{-4S_0/3}$, see point 5 below). For the
Airy kernel all
corrections to the disk density integrate \emph{exactly} to the exponentiated genus series
\cite{OkuyamaSakai2020}.
Equation \eqref{eq:intro-DS} is therefore correctly normalized, and can also be used to compute
$\eta/s$. We also clarify the meaning of a novel $\omega$-dependent normalization that
suggests itself in this context. It is a \emph{pair} partition function, and it defines a
viscosity per resolved transition, which stays finite as $r\to\infty$ but is not a
thermodynamic ratio.

\item At $CT\sim e^{-2S_0/3}$ three things happen together: the genus expansion of $Z$
breaks down, the one-loop entropy vanishes, and the level spacing at thermal energies becomes
of order $T$. Numerator and normalization are then both governed by the Airy kernel, and
\begin{equation}
\frac{\eta}{\etasc}=\frac{2^{2/3}}{\pi^{3/2}}\,e^{S_0/3}\,\cE(\kappa\omega,\kappa T),\qquad
\kappa=2^{1/3}Ce^{2S_0/3},
\label{eq:intro-edge}
\end{equation}
with a universal scaling function $\cE$, defined in \eqref{eq:E-def}, that we compute in the annealed and quenched
ensembles. The annealed entropy becomes negative for $\kappa T<0.967$. The quenched entropy
stays positive, and at low temperature it vanishes as \cite{JanssenMirbabayi2021}
\begin{equation}
S_{\rm q}\simeq\frac{7\pi^4}{180}\,(\kappa T)^3 ,\qquad \kappa T\ll1 .
\end{equation}
The resulting $\eta/s$ is finite where the one-loop expression diverges. At the lowest
temperatures and frequencies, $\omega,T\ll1/\kappa$, the response is that of the two lowest
levels, and $\eta/\etasc\simeq\frac12\,e^{S_0}C\omega\tanh(\omega/2T)$.

\item We repeat the analysis for the explicit non-perturbative completions constructed
from the string equation in \cite{Johnson2020,Johnson2020c}, in which the spectrum is
bounded below by a hard wall. In the edge limit such a completion is the Airy process
conditioned to have no level below the wall. The bulk results are unchanged. The partition
function acquires a correction of relative order $e^{-4S_0/3}$, proportional to the density of
states at the wall, which we compute in closed form. In the edge regime the
quenched quantities change only through the members of the soft-edge ensemble that have
a level below the wall. Their probability is $1-F_2(0)\simeq3\%$ when the wall is placed at
the classical edge of the spectrum, as in \cite{Johnson2020}; here $F_2$ is the Tracy--Widom
distribution of the lowest level. Numerically, the quenched entropy and viscosity change
by a few per cent. The annealed quantities, by contrast, change at order
one. They do not become physical, however: the annealed entropy is still negative at low
temperature (for $\kappa T<0.840$ when the wall sits at the classical edge), although it now
diverges only logarithmically as $T\to0$, instead of $-1/T^3$
(Section~\ref{sec:string-edge-regime}). A wall therefore does not remove the need for the
quenched prescription.
\end{enumerate}
Figure~\ref{fig:scales} summarizes the hierarchy of temperature scales, and
Tables~\ref{tab:summary} and~\ref{tab:summary-edge} in Section~\ref{sec:discussion} collect the results for $\eta$ and
$\eta/s$ in each regime.

It is worth stressing what these results mean physically. The vanishing of $\eta_{\rm DC}$ at
finite $S_0$ does not mean that a macroscopic fluid stops dissipating. It is the statement that
the Kubo formula of an isolated system with a discrete spectrum has no smooth DC limit, a
familiar fact in mesoscopic physics
\cite{Kubo1962,GorkovEliashberg1965,Efetov1983,Halperin1986,AltshulerShklovskii1986,SivanImry1987,Wilkinson1988,GuhrEtAl1998,Mirlin2000}.
The physics is contained in
the scaling functions, which describe the response at frequencies comparable to the level
spacing, and in their normalization. Both $S_0$ and $C$ are proportional to the transverse
area of the black brane, eq.~\eqref{eq:Ssc}. In the strict planar limit, in which this area is
infinite at fixed $T$, one has $CT\to\infty$ and $e^{-S_0}\to0$: the brane is semiclassical, its
spectrum is continuous, and none of the new scales survives. The effects discussed in this paper
are therefore finite-volume effects. They set in at temperatures $T\lesssim1/C$, which are
inversely proportional to the transverse volume, and at frequencies of order $e^{-S_0}/C$, which
are exponentially small in it. They are relevant whenever the transverse space is compact: for a
brane wrapped on a torus, as considered here, for near-extremal black holes whose horizons
are compact (such as $S^2$), and more generally for holographic descriptions of finite systems at very low
temperature.

The regime in which a response function resolves individual energy levels has a long
history, and it is useful to place our results in this context. In mesoscopic physics the
electromagnetic absorption of a small metallic grain at frequencies below its mean
level spacing is controlled by the two-level correlation function of random-matrix theory
\cite{GorkovEliashberg1965,Efetov1983,SivanImry1987}, exactly as the viscosity is here, and the
dissipation produced by a slow drive depends on the symmetry class of the level statistics
\cite{Wilkinson1988}. For chaotic many-body systems the same low-frequency regime has been
analysed within the eigenstate thermalization hypothesis, where the question is at which
frequency the matrix elements of local operators become those of a random matrix, and how
this scale is related to the slowest hydrodynamic mode
\cite{DAlessioEtAl2016,Dymarsky2022,RichterEtAl2020}. In two-dimensional gravity the imprint
of level repulsion on real-time observables has been studied for late-time correlation
functions \cite{Saad2019,BlommaertMertensVerschelde2019clocks,BlommaertMertensVerschelde2021eigenbranes,AltlandSonner2021,AltlandEtAl2023}
and for Hawking--Unruh radiation, whose spectrum is depleted at energies below the level
spacing \cite{BlommaertMertensVerschelde2021unruh}. The latter effect is the closest
analogue of the vanishing of $\eta_{\rm DC}$ found here.

\paragraph{Outline.} Section~\ref{sec:setup} reviews the holographic
computation of the shear viscosity and the near-extremal black brane, derives the Kubo
formula in the AdS$_2$ throat, formulates it in the matrix-model completion of JT gravity,
and reviews the disk result. Section~\ref{sec:repulsion} introduces level repulsion and shows
that the DC viscosity vanishes. Section~\ref{sec:DS} analyses the bulk double-scaling limit.
Section~\ref{sec:normalization} determines the partition function that normalizes it and
computes $\eta/s$. Section~\ref{sec:airy} treats the edge regime exactly.
Section~\ref{sec:string} repeats the analysis for explicit completions defined by the string
equation and compares them with the soft-edge one. Section~\ref{sec:discussion} summarizes the results, discusses their physical meaning and
lists open problems. Five appendices contain closed-form moments, Weil--Petersson data, a derivation of the oscillatory term of the density of states, a remark on bulk--edge matching, and our numerical methods.

\section{The viscosity of near-extremal black branes and its matrix-model completion}
\label{sec:setup}

In this section we first review, briefly, how the shear viscosity is computed in holography
and how the near-extremal black brane reduces the computation to a correlator in the AdS$_2$
throat. We follow \cite{KanargiasKiritsisMurthyPapadoulakiPorfyriadis2025}, to which we refer
for details. We then formulate this correlator in the matrix-model completion of JT gravity
and review the disk result.

\subsection{Shear viscosity in holography}
\label{sec:holo-visc}

Consider a relativistic fluid in $2+1$ dimensions. The shear viscosity $\eta$ controls the
dissipation of shear flows, and linear response theory relates it to the retarded correlator
of the transverse, traceless component of the stress tensor,
\begin{equation}
G^R_{xy,xy}(\omega,\vec p)=-i\!\int\!\dd t\,\dd^2x\;e^{i\omega t-i\vec p\cdot\vec x}\,
\theta(t)\,\vev{[T_{xy}(t,\vec x),T_{xy}(0)]}.
\end{equation}
At zero spatial momentum hydrodynamics predicts the low-frequency expansion
\cite{BaierEtAl2008}
\begin{equation}
G^R_{xy,xy}(\omega,\vec 0)=P-i\eta\,\omega+\cO(\omega^2),
\label{eq:hydro-GR}
\end{equation}
with $P$ the pressure. This gives the Kubo formula \cite{Kubo1957}
\begin{equation}
\eta=-\lim_{\omega\to0}\frac{1}{\omega}\,\operatorname{Im}G^R_{xy,xy}(\omega,\vec 0).
\label{eq:Kubo-def}
\end{equation}
The imaginary part of $G^R$ is proportional to the spectral density of $T_{xy}$. It measures the
rate at which the system absorbs energy from a homogeneous shear deformation of frequency
$\omega$, so $\eta$ is the low-frequency limit of this absorption rate per unit frequency.

In holography, $T_{xy}$ is dual to the boundary value of the metric perturbation
$h^x{}_y$. At zero spatial momentum this perturbation decouples from the other fluctuations
and obeys the equation of a minimally coupled massless scalar
\cite{PolicastroSonStarinets2002,KovtunSonStarinets2005},
\begin{equation}
\partial_M\bigl(\sqrt{-g}\,g^{MN}\partial_N\phi\bigr)=0,\qquad \phi=h^x{}_y .
\end{equation}
The retarded correlator is obtained from the solution that is ingoing at the horizon
\cite{SonStarinets2002}. For $\omega\to0$ the radial flux of this solution is conserved, so it
can be evaluated at the horizon rather than at the boundary. This is the membrane paradigm \cite{IqbalLiu2009}. If $a_h$ denotes the area of the horizon per unit boundary area, it gives
\begin{equation}
\eta=\frac{a_h}{16\pi G},\qquad s=\frac{a_h}{4G},\qquad\text{hence}\qquad\frac{\eta}{s}=\frac{1}{4\pi}.
\end{equation}
This result was first obtained for $\mathcal N=4$ super-Yang--Mills theory in
\cite{PolicastroSonStarinets2001,PolicastroSonStarinets2002}, derived from the membrane picture
in \cite{KovtunSonStarinets2003}, and shown to hold for all two-derivative gravitational duals in
\cite{KovtunSonStarinets2005,BuchelLiu2004,Buchel2005,IqbalLiu2009}; see \cite{SonStarinets2007}
for a review. Equivalently, $\eta$ is proportional to the low-frequency absorption cross section
of the massless scalar, which equals the area of the horizon
\cite{DasGibbonsMathur1997,KovtunSonStarinets2005}. The only input is a regular horizon
described by a classical geometry. The question of this paper is what replaces this argument
when the horizon itself fluctuates strongly.

\subsection{Near-extremal black branes and the \texorpdfstring{AdS$_2$}{AdS2} throat}
\label{sec:brane}

The simplest holographic model of a strongly coupled system at finite density is
Einstein--Maxwell theory in AdS$_4$ \cite{ChamblinEmparanJohnsonMyers1999},
\begin{equation}
S=\frac{1}{16\pi G}\int\!\dd^4x\,\sqrt{-g}\,\Bigl(R[g]-F^2+\frac{6}{L^2}\Bigr),
\end{equation}
dual to a three-dimensional CFT at chemical potential $\mu$. Here $R[g]$ is the Ricci
scalar and $L$ the AdS radius. Following
\cite{KanargiasKiritsisMurthyPapadoulakiPorfyriadis2025}, we take the spatial directions of
the boundary to be a torus of area $4\pi V_2$, so that the system has a finite volume. The metric of the
charged black brane is
\begin{equation}
\dd s^2=-f(R)\,\dd t^2+\frac{\dd R^2}{f(R)}+\frac{R^2}{L^2}\,\dd\vec x^{\,2},\qquad
f(R)=\frac{R^2}{L^2}-\frac{2GM}{R}+\frac{GQ^2}{R^2},
\end{equation}
where $R$ is the radial coordinate and $M$ and $Q$ are
the mass and charge parameters. The gauge
field is $A=A_t\,\dd t$ with $A_t=\mu L-\sqrt G\,Q/R$. The temperature and the chemical
potential are $T=f'(R_+)/4\pi$ and $\mu=A_t(\infty)/L=\sqrt G\,Q/(LR_+)$, where $R_+$ is the outer
horizon and $A_t$ vanishes there. At extremality the inner and outer horizons merge at
$R_\star=\mu L^2/\sqrt3$, $f(R)$ has a double zero, and the near-horizon geometry is
AdS$_2\times T^2$ with AdS$_2$ radius $L/\sqrt6$.
In the near-extremal regime $T\ll\mu$ the Bekenstein--Hawking entropy is
\cite{KanargiasKiritsisMurthyPapadoulakiPorfyriadis2025}
\begin{equation}
\begin{gathered}
S^{\rm s.c.}(\mu,T)=\bigl(S_0+4\pi^2CT\bigr)\bigl(1+\cO(T/\mu)\bigr),\\
S_0=\frac{\pi}{3}\,V_2\,\mu^2\mathcal N,\qquad
C=\frac{1}{E_{\rm gap}}=\frac{\mu V_2\mathcal N}{3\sqrt3},
\end{gathered}
\label{eq:Ssc}
\end{equation}
where $\mathcal N=L^2/G$ counts the degrees of freedom of the boundary theory. The
coefficient of the linear term defines the coupling $C$ of the Schwarzian mode. Both $S_0$ and
$C$ are proportional to $V_2\mathcal N$, and their ratio is $S_0/C=\sqrt3\,\pi\mu$. Hence
$E_{\rm gap}/\mu=\sqrt3\,\pi/S_0\ll1$: the quantum regime $T\lesssim E_{\rm gap}$ lies deep inside
the near-extremal regime $T\ll\mu$. In the planar limit $V_2\to\infty$ at fixed $T$, on the
other hand, $CT\to\infty$ and the brane is always semiclassical.

The origin of the Schwarzian mode is the following
\cite{AlmheiriPolchinski2015,MaldacenaStanfordYang2016,NayakShuklaSoniTrivediVishal2018,MoitraTrivediVishal2019,IliesiuTuriaci2020,IliesiuMurthyTuriaci2022}.
The leading deviation of the near-horizon geometry from AdS$_2\times T^2$ is a mode that changes
the size of the torus and grows linearly towards the boundary of the throat. In
two-dimensional language the area of the torus plays the role of the dilaton of JT gravity,
and this mode is the familiar dilaton profile of nearly-AdS$_2$. Its strength is set by $C$,
and its value at the horizon produces the linear term $4\pi^2CT$ in \eqref{eq:Ssc}. The
boundary dynamics of the throat is then the Schwarzian theory with coupling $C$. Its
fluctuations have an action of order $CT$, so they are small for $CT\gg1$ and strong for
$CT\lesssim1$. The other fields in the throat, including the shear mode discussed below,
behave as ordinary matter fields on AdS$_2$, whose contributions have a smooth limit as
$T\to0$. The only other light mode of the charged brane is a $U(1)$ phase mode, which
decouples from the neutral correlators studied here
\cite{KanargiasKiritsisMurthyPapadoulakiPorfyriadis2025}.

The throat is also where all low-frequency dissipation takes place. For $\omega,T\ll\mu$ the
bulk wave equation can be solved by matching the solution in the AdS$_2$ throat to the solution
in the exterior region \cite{FaulknerLiuMcGreevyVegh2011,EdalatiJottarLeigh2010c,EdalatiJottarLeigh2010}.
The exterior region contributes only real coefficients that are analytic in $\omega$, while all
non-analytic low-frequency behaviour comes from the throat. At zero spatial momentum
$h^x{}_y$ sources an operator $\cO$ of dimension $\Delta_\cO=1$ in the boundary quantum mechanics
of the throat, and one finds, up to real terms that are analytic in $\omega$ (such as the
pressure in \eqref{eq:hydro-GR})
\cite{EdalatiJottarLeigh2010c,EdalatiJottarLeigh2010,KanargiasKiritsisMurthyPapadoulakiPorfyriadis2025},
\begin{equation}
G^R_{xy,xy}(\omega,\vec 0)=-\frac{\mu^2\mathcal N}{48\pi}\,
\frac{\cG^R(\omega)}{1+\xi L^2\,\cG^R(\omega)}
=-\frac{\mu^2\mathcal N}{48\pi}\,\cG^R(\omega)\bigl(1+\cO(\omega/\mu)\bigr).
\label{eq:UV-IR}
\end{equation}
Here $\cG^R(\omega)$ is the retarded Green's function of $\cO$ in the throat and $\xi$ is a
constant of order $1/R_\star$ \cite{KanargiasKiritsisMurthyPapadoulakiPorfyriadis2025}. Following
\cite{KanargiasKiritsisMurthyPapadoulakiPorfyriadis2025}, $\cG^R$ is defined with the opposite
overall sign to $G^R_{xy,xy}$, so that $\operatorname{Im}\cG^R(\omega)>0$ for $\omega>0$. This
is the origin of the minus sign in \eqref{eq:UV-IR}. The prefactor of \eqref{eq:UV-IR} is
written at leading order in $T/\mu$; see the comment below \eqref{eq:Kubo-IR}.
Computing it in the classical AdS$_2$ geometry, one obtains the semiclassical result
\begin{equation}
\cG^R_{\rm s.c.}(\omega)=i\omega ,
\label{eq:grsc}
\end{equation}
and the Kubo formula \eqref{eq:Kubo-def} gives, at extremality,
\begin{equation}
\eta^{\rm s.c.}_0=\frac{\mu^2\mathcal N}{48\pi}=\frac{s_0}{4\pi},\qquad s_0=\frac{S_0}{4\pi V_2}=\frac{\mu^2\mathcal N}{12}.
\label{eq:eta0}
\end{equation}

\subsection{Kubo formula and \texorpdfstring{$\eta/s$}{eta/s}}
\label{sec:kubo}

Equation \eqref{eq:UV-IR} reduces the computation of the viscosity to a computation in the
throat. Inserting it into \eqref{eq:Kubo-def} gives
\cite{KanargiasKiritsisMurthyPapadoulakiPorfyriadis2025}
\begin{equation}
\eta=\etasc\,\lim_{\omega\to0}\frac{\operatorname{Im}\cG^R(\omega)}{\omega},
\label{eq:Kubo-IR}
\end{equation}
where now $\etasc(\mu,T)=s^{\rm s.c.}(\mu,T)/4\pi$, with $s^{\rm s.c.}=S^{\rm s.c.}/4\pi V_2$,
includes the classical growth of the horizon, $R_\star\to R_+$, at finite temperature
\cite{IqbalLiu2009,KanargiasKiritsisMurthyPapadoulakiPorfyriadis2025}. This last step does
not follow from \eqref{eq:UV-IR} as written, whose prefactor
$\mu^2\mathcal N/48\pi=\eta^{\rm s.c.}_0$ is the leading term in $T/\mu$. It uses the fact that in
the classical geometry $\eta=s/4\pi$ holds at any temperature \cite{IqbalLiu2009}, so that at
finite temperature the prefactor is $\etasc(\mu,T)$. In the quantum theory the
classical propagator in AdS$_2$ is replaced by the exact thermal correlator of $\cO$ in the
throat theory. It is convenient to work with the Wightman spectral function
\begin{equation}
\cG^>(\omega)=\frac1\pi\int_{-\infty}^{\infty}\!\dd t\;e^{i\omega t}\,\vev{\cO(t)\cO(0)}_\beta .
\end{equation}
The fluctuation--dissipation theorem relates it to the retarded function,
\begin{equation}
\operatorname{Im}\cG^R(\omega)=\frac12\bigl(1-e^{-\beta\omega}\bigr)\,\cG^>(\omega).
\label{eq:FDT}
\end{equation}
The factor $1-e^{-\beta\omega}$ expresses the Kubo--Martin--Schwinger (KMS) condition of the
thermal state: the Wightman function with the opposite operator ordering equals
$e^{-\beta\omega}\cG^>(\omega)$, and $\operatorname{Im}\cG^R$ is proportional to the difference of
the two. We refer to it as the \emph{KMS factor}.

\paragraph{Notation.} Several correlators appear in this paper, and we use the following
conventions throughout. $G^R_{xy,xy}$ is the retarded correlator of $T_{xy}$ in the boundary
theory. $\cG^R$ and $\cG^>$ are the retarded and Wightman correlators of the throat operator
$\cO$, related by \eqref{eq:FDT}. A subscript on a throat correlator states how it is computed:
\begin{itemize}
\item $\cG^R_{\rm s.c.}$: classical AdS$_2$ geometry, eq.~\eqref{eq:grsc};
\item $\cG^>_{\rm disk}$: exact Schwarzian (disk) correlator, which treats the spectrum as
continuous (Section~\ref{sec:disk});
\item $\cG^>_{\rm a}$ and $\cG^>_{\rm q}$: annealed and quenched averages over the
matrix-model ensemble (Section~\ref{sec:mm});
\item $\cG^>_{\rm cut}$: the sharp-cutoff approximation of Section~\ref{sec:cut}.
\end{itemize}
The same labels are used for the corresponding viscosities, e.g.\ $\eta_{\rm disk}$, and
partition functions, e.g.\ $Z_{\rm disk}$. Unlabelled symbols refer to a single member of the
ensemble or to statements that hold for all of the above.

The DC limit in \eqref{eq:Kubo-IR} will turn out to be subtle. We therefore keep the frequency
finite and define the frequency-dependent viscosity
\begin{equation}
\begin{gathered}
\eta(\omega)\equiv-\frac{\operatorname{Im}G^R_{xy,xy}(\omega,\vec 0)}{\omega}
=\etasc\,\frac{\operatorname{Im}\cG^R(\omega)}{\omega}
=\etasc\,\frac{1-e^{-\beta\omega}}{2\omega}\,\cG^>(\omega),\\
\eta_{\rm DC}=\lim_{\omega\to0}\eta(\omega).
\end{gathered}
\label{eq:Kubo}
\end{equation}

At finite $\omega$, $\eta(\omega)$ is the dissipative part of the response to a homogeneous
shear of frequency $\omega$. The corrections of relative order $\omega/\mu$ in
\eqref{eq:UV-IR} are negligible for all the frequencies considered in this paper, which lie far
below $E_{\rm gap}\ll\mu$.\footnote{The correction $\xi L^2\cG^R$ in the denominator of
\eqref{eq:UV-IR} is of order $(\omega/\mu)\,\eta(\omega)/\etasc$. It remains small even in the
edge regime of Section~\ref{sec:airy}, where $\eta/\etasc\sim e^{S_0/3}$ but
$\omega\sim e^{-2S_0/3}E_{\rm gap}$.} The normalization of $\cG^>$ is such that
$\eta\to\etasc$ in the semiclassical regime $CT\gg1$, where $\cG^R\to\cG^R_{\rm s.c.}$.

The semiclassical entropy is the Bekenstein--Hawking entropy \eqref{eq:Ssc}. If $S(T)$ is the
thermodynamic entropy of the same quantum ensemble that is used to compute $\cG^>$, then
\cite{KanargiasKiritsisMurthyPapadoulakiPorfyriadis2025}
\begin{equation}
4\pi\,\frac{\eta}{s}=\frac{\eta/\etasc}{S(T)/S^{\rm s.c.}(T)} .
\label{eq:eta-over-s}
\end{equation}
This combination measures the departure from the semiclassical value $\eta/s=1/4\pi$. It
receives contributions from the numerator (the dissipative response) and from the
denominator (the entropy). As we will see, both are modified in the quantum regime.
For an ensemble of Hamiltonians, \eqref{eq:eta-over-s} is the ratio of the ensemble average of
the viscosity to the ensemble average of the entropy. As long as both are self-averaging this
coincides with the average of the ratio computed for each member of the ensemble. In the edge
regime of Section~\ref{sec:airy} it does not, and we discuss the difference in
Section~\ref{sec:edge-etas}.

\subsection{Spectral representation in the matrix-model completion}
\label{sec:mm}

The Schwarzian computation of $\cG^>$ treats the spectrum of the throat as continuous. It was shown in \cite{SaadShenkerStanford2019} that the JT path integral, summed over all topologies, is
equivalent, order by order in the genus expansion, to a double-scaled Hermitian matrix integral. In
this description the boundary Hamiltonian $H$ of the throat is a large random Hermitian matrix,
and the JT path integral on surfaces with $n$ asymptotic boundaries computes the ensemble
average $\vev{\operatorname{Tr}e^{-\beta_1H}\cdots\operatorname{Tr}e^{-\beta_nH}}$. The disk
gives the leading density of eigenvalues. Surfaces of higher genus give corrections suppressed
by $e^{-2S_0}$ per handle, and there are in addition effects that are non-perturbative in
$e^{-S_0}$. Together they describe the discreteness of the spectrum. We now formulate the
Kubo formula in this language.

Two assumptions enter here, and we state them at the outset. The first concerns conserved
charges. The brane carries a $U(1)$ charge and, on the torus, two components of momentum, and the
Hamiltonian is block diagonal in them. Random-matrix statistics are expected within each block,
while different blocks are statistically independent
\cite{Iliesiu2019,KapecMahajanStanford2019}. The operator dual to $h_{xy}$ at zero spatial
momentum is neutral and carries no momentum, so it acts within these blocks. Throughout the paper $H$ denotes
the Hamiltonian of a single block, of fixed charge and momentum, and $S_0$, $C$ and the level
spacing are those of that block.
The second assumption concerns the symmetry class. A Hermitian matrix integral, dual to JT
gravity on orientable surfaces, describes Hamiltonians without antiunitary symmetry: the unitary
(GUE) class, with Dyson index $\beta_{\rm D}=2$ \cite{SaadShenkerStanford2019,StanfordWitten2019}.
Whether this is the class of the brane depends on the microscopic theory, and is not decided by
the bulk action. The Einstein--Maxwell action is invariant under time reversal, which leaves the
electric charge invariant. If the boundary theory has an antiunitary symmetry that acts within a
block of fixed charge and zero momentum, the statistics of that block are orthogonal or
symplectic, $\beta_{\rm D}=1$ or $4$, and the dual sum includes unorientable surfaces
\cite{StanfordWitten2019}. The block is in the unitary class if time reversal is broken in the
boundary theory, or if the only antiunitary symmetries reverse the charge: they then relate the
blocks of charge $Q$ and $-Q$, and each block with $Q\neq0$ is unitary
\cite{KapecMahajanStanford2019}. We assume $\beta_{\rm D}=2$ throughout. The qualitative
conclusions do not depend on this choice, but the scaling functions and some powers of
$\omega$ and $T$ do. We indicate the dependence on $\beta_{\rm D}$ where it is simple
(Sections~\ref{sec:DC} and~\ref{sec:edge-etas}) and return to it in
Section~\ref{sec:limitations}.

Let $|E_n\rangle$ be the energy eigenstates of the boundary theory. The Lehmann
representation of the spectral function is
\begin{equation}
\cG^>(\omega)=\frac{2}{Z(\beta)}\sum_{m,n}e^{-\beta E_m}\,|\cO_{mn}|^2\,\delta\bigl(\omega-E_n+E_m\bigr).
\label{eq:Lehmann}
\end{equation}
In JT gravity the eigenvalues are distributed as those of a double-scaled random matrix,
with leading (disk) density \cite{StanfordWitten2017,SaadShenkerStanford2019}
\begin{equation}
\rho_0(E)=\frac{e^{S_0}C}{2\pi^2}\sinh\bigl(2\pi\sqrt{2CE}\bigr),\qquad E=\frac{k^2}{2C}.
\label{eq:rho0}
\end{equation}
We parametrize energies by $k=\sqrt{2CE}$ throughout. With matter, the matrix elements
$\cO_{mn}$ are Gaussian random variables, independent of the eigenvalues, with variance
fixed by the Schwarzian vertex \cite{Saad2019,JafferisKolchmeyerMukhametzhanovSonner2022}:
\begin{equation}
\overline{|\cO_{mn}|^2}=\frac{N_\cO}{2}\,\Gamma(1\pm ik_m\pm ik_n),\qquad N_\cO=\frac{e^{-S_0}}{C^2}.
\label{eq:vertex}
\end{equation}
The product runs over the four sign choices. This is the $\Delta_\cO=1$ case of
$\Gamma(\Delta_\cO\pm ik_1\pm ik_2)/\Gamma(2\Delta_\cO)$. The normalization $N_\cO$ is fixed below by
matching to the disk correlator. Averaging over matrix elements and eigenvalues gives
the \emph{annealed} spectral function
\begin{equation}
\cG^>_{\rm a}(\omega)=\frac{N_\cO}{Z(\beta)}\int\!\dd E_1\,\dd E_2\;e^{-\beta E_1}\,
\Gamma(1\pm ik_1\pm ik_2)\;\vev{\rho(E_1)\rho(E_2)}\;\delta(\omega-E_2+E_1).
\label{eq:G-spectral}
\end{equation}
The \emph{quenched} spectral function instead averages the ratio sample by sample:
\begin{equation}
\cG^>_{\rm q}(\omega)=N_\cO\Bigl\langle\frac{\sum_{m\neq n}e^{-\beta E_m}\,\Gamma(1\pm ik_m\pm ik_n)\,\delta(\omega-E_n+E_m)}{\sum_le^{-\beta E_l}}\Bigr\rangle .
\label{eq:G-quenched}
\end{equation}
The diagonal terms $m=n$ contribute only at $\omega=0$. They are annihilated by the KMS
factor $1-e^{-\beta\omega}$ of \eqref{eq:FDT}, which vanishes at $\omega=0$, and never
enter $\operatorname{Im}\cG^R$.

The form \eqref{eq:vertex} of the matrix elements is an ansatz in the spirit of the
eigenstate thermalization hypothesis, and it is the main dynamical input of this paper. Two
remarks make clear what is actually used. First, $\cG^>$ is quadratic in $\cO$, and the
partition function does not depend on $\cO$. The annealed and quenched spectral functions
\eqref{eq:G-spectral} and \eqref{eq:G-quenched} therefore involve only the mean of
$|\cO_{mn}|^2$ at fixed eigenvalues, and the Gaussian form of the distribution plays no role.
What is assumed is that this mean is the smooth function \eqref{eq:vertex} of $E_m$ and $E_n$,
uncorrelated with the positions of the neighbouring levels. This is the structure found at
leading order in the matrix-model description of JT gravity with matter
\cite{Saad2019,JafferisKolchmeyerMukhametzhanovSonner2022}. Corrections to it are suppressed by
powers of $e^{-S_0}$ at fixed energies, and we assume that they are not enhanced in the scaling
limits considered below. Second, with this input the dependence of the response on $\omega$ at
the scale of the level spacing comes entirely from the eigenvalue statistics. The vanishing of
the DC limit (Section~\ref{sec:DC}), the level-repulsion factor $R_r$ of the bulk
double-scaling limit (Section~\ref{sec:DS}) and the edge scaling functions of
Section~\ref{sec:airy}, where the vertex reduces to a constant, follow from the sine and Airy
kernels alone. The Schwarzian form of the vertex and the density $\rho_0$ determine the thermal
weights with which these universal factors are averaged, and with them numerical coefficients
such as those of \eqref{eq:eta-disk} and \eqref{eq:eta-max}.

The eigenvalue statistics enter through $\vev{\rho(E_1)\rho(E_2)}$. For a double-scaled
Hermitian matrix integral the eigenvalues form a determinantal point process
\cite{Mehta2004,Forrester2010}: all $n$-point correlation functions are $n\times n$
determinants of a kernel $K(E_i,E_j)$. In particular,
\begin{equation}
\begin{gathered}
\vev{\rho(E)}=K(E,E),\\
\vev{\rho(E_1)\rho(E_2)}=K(E_1,E_1)\,K(E_2,E_2)-K(E_1,E_2)^2+\delta(E_1-E_2)\,K(E_1,E_1).
\end{gathered}
\label{eq:determinantal}
\end{equation}
The first term of the two-level density is the result for uncorrelated levels. The second,
$-K(E_1,E_2)^2$, is negative: it lowers the probability of finding a second level close to a
given one. In random-matrix theory this depletion of the two-level density near $E_1=E_2$ is
called the \emph{correlation hole}. The last term arises because the same level contributes to
both factors of $\rho(E_1)\rho(E_2)$. For the double-scaled matrix integrals considered here $K$
is the kernel of a projection operator, $\int\dd E'\,K(E,E')K(E',E'')=K(E,E'')$
\cite{Mehta2004,Johnson2025kernel}. In particular $\int\dd E_2\,K(E_1,E_2)^2=K(E_1,E_1)$: the
correlation hole around a level contains exactly one level, and integrating over $E_2$ it
cancels the contact term. The annealed partition function is
\begin{equation}
Z(\beta)=\int\!\dd E\;e^{-\beta E}K(E,E).
\label{eq:Z-kernel}
\end{equation}
Equations \eqref{eq:G-spectral}--\eqref{eq:Z-kernel} make precise what it means to
normalize the correlator by a partition function that contains the same effects. The kernel
$K$ enters both the numerator, through the two-level density, and the denominator, through
the one-level density. In the quenched case both come from the full distribution of the
same point process.

\subsection{The disk result}
\label{sec:disk}

At disk level the eigenvalues are uncorrelated, $\vev{\rho\rho}\to\rho_0\rho_0$, and
the partition function is
\cite{StanfordWitten2017,MertensTuriaciVerlinde2017,SaadShenkerStanford2019}
\begin{equation}
Z_{\rm disk}(T)=\int_0^\infty\!\dd E\,e^{-E/T}\rho_0(E)=\frac{e^{S_0}(CT)^{3/2}e^{2\pi^2CT}}{\sqrt{2\pi}} .
\end{equation}
To evaluate \eqref{eq:G-spectral}, write $E_1=k^2/2C$ and $E_2=E_1+\omega=k_2^2/2C$. Then
\begin{equation}
\rho_0(E_1)\,\dd E_1=\frac{e^{S_0}}{2\pi^2}\,k\sinh(2\pi k)\,\dd k,\qquad
\Gamma(1\pm ik\pm ik_2)=\frac{\pi^2(k_2^2-k^2)}{\sinh\pi(k_2+k)\,\sinh\pi(k_2-k)},
\end{equation}
using $|\Gamma(1+ix)|^2=\pi x/\sinh\pi x$. Since $k_2^2-k^2=2C\omega$, the factors combine into
the $\Delta_\cO=1$ Schwarzian correlator of
\cite{KanargiasKiritsisMurthyPapadoulakiPorfyriadis2025}:
\begin{equation}
\cG^>_{\rm disk}(\omega)=\frac{e^{S_0}\omega}{2\pi^2Z_{\rm disk}}\int_0^\infty\!\dd k\;g_T(k)\,f(k,\omega),
\qquad g_T(k)=k\,e^{-\frac{k^2}{2CT}}\sinh(2\pi k),
\label{eq:C1}
\end{equation}
\begin{equation}
f(k,\omega)=\frac{\sinh 2\pi k_2}{\sinh\pi(k_2+k)\,\sinh\pi(k_2-k)}
=\coth\pi(k_2+k)+\coth\pi(k_2-k),
\label{eq:f}
\end{equation}
where $k_2=\sqrt{k^2+2C\omega}$. Matching to \cite{KanargiasKiritsisMurthyPapadoulakiPorfyriadis2025} fixes $N_\cO=e^{-S_0}/C^2$.
With \eqref{eq:FDT} and \eqref{eq:Lehmann} this is the
value for which $\eta_{\rm disk}\to\etasc$ at $CT\gg1$, eq.~\eqref{eq:eta-disk}.%

\begin{figure}[!ht]
  \centering
  \includegraphics[width=\textwidth]{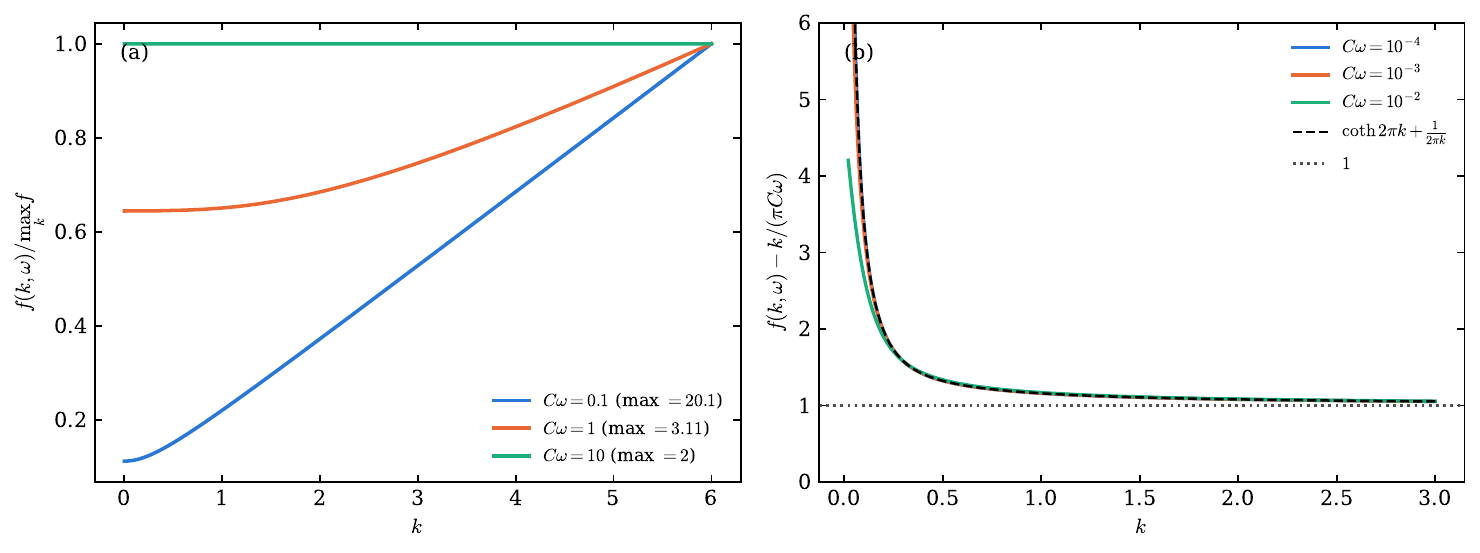}
  \caption{(a) The combination $f(k,\omega)$ of final-state density and Schwarzian vertex,
  eq.~\eqref{eq:f}, for three values of $C\omega$, normalized to its maximum on the
  interval. (b) The constant term $f-k/(\pi C\omega)$ at small $C\omega$. It converges to
  $\coth2\pi k+1/2\pi k$ (dashed), not to $1$ (dotted), as in eq.~\eqref{eq:f-expansion}.}
  \label{fig:f}
\end{figure}

The function $g_T(k)$ is the Boltzmann-weighted density of initial states. The function
$f$ combines the density of final states with the Schwarzian vertex. Figure~\ref{fig:f}(a)
shows $f$. At small $C\omega$ it grows linearly in $k$; at large $C\omega$ it is
essentially constant. The small-frequency expansion is controlled by the second
$\coth$, whose argument $\pi(k_2-k)\simeq\pi C\omega/k$ is small. Using
$1/(k_2-k)=(k_2+k)/2C\omega$, one finds, at fixed $k>0$,
\begin{equation}
f(k,\omega)=\frac{k}{\pi C\omega}+\coth(2\pi k)+\frac{1}{2\pi k}+\cO(C\omega).
\label{eq:f-expansion}
\end{equation}
The pole is what survives in the DC limit. The constant term
$a_0(k)=\coth2\pi k+1/2\pi k$ is not $1$ (Figure~\ref{fig:f}(b)). The two agree only for
$k\gg1$. This matters for the $\cO(\omega)$ part of $\cG^>$, and hence for the normalization
discussed in Section~\ref{sec:pair}.

\paragraph{DC viscosity.} With the moments
\begin{equation}
\begin{aligned}
I_1(CT)&=\int_0^\infty\!\dd k\,g_T(k)=\pi\sqrt{2\pi}\,(CT)^{3/2}e^{2\pi^2CT},\\
I_2(CT)&=\int_0^\infty\!\dd k\,k\,g_T(k)=2\pi (CT)^2+\sqrt{\tfrac\pi2}\,(CT)^{3/2}(1+4\pi^2CT)\,e^{2\pi^2CT}\erf\bigl(\pi\sqrt{2CT}\bigr),
\end{aligned}
\label{eq:I12}
\end{equation}
and $Z_{\rm disk}=e^{S_0}I_1/2\pi^2$, the Kubo formula \eqref{eq:Kubo} gives
\cite{KanargiasKiritsisMurthyPapadoulakiPorfyriadis2025}
\begin{equation}
\frac{\eta_{\rm disk}}{\etasc}=\frac{\beta}{2}\,\frac{e^{S_0}}{2\pi^2Z_{\rm disk}}\,\frac{I_2}{\pi C}
=\frac{I_2(CT)}{2\pi CT\,I_1(CT)}
=\erf\Bigl(\sqrt{2\pi^2CT}\Bigr)\Bigl(1+\frac{1}{4\pi^2CT}\Bigr)+\frac{e^{-2\pi^2CT}}{\sqrt{2\pi^3CT}} .
\label{eq:eta-disk}
\end{equation}
This is the result \eqref{eq:intro-disk} quoted in the introduction. Only the pole of
\eqref{eq:f-expansion} contributes in the DC limit, and the KMS factor supplies
$1-e^{-\beta\omega}\simeq\beta\omega$. The middle expression has a transparent meaning:
$\eta_{\rm disk}/\etasc=\vev{k}_T/(2\pi CT)$, where $\vev{\cdot}_T$ is the average over the normalized
thermal weight $g_T(k)/I_1(CT)$. The viscosity is the mean momentum $k=\sqrt{2CE}$ of the
thermally occupied states, in units of its semiclassical value $k_{\rm th}\simeq2\pi CT$. For
$CT\gg1$ the weight is a Gaussian centred at $2\pi CT$, and
$\eta_{\rm disk}/\etasc=1+1/(4\pi^2CT)$ up to exponentially small terms. For $CT\ll1$ the weight
is concentrated at $k\sim\sqrt{CT}$, where $\rho_0\propto\sqrt E$, and
\begin{equation}
\frac{\eta_{\rm disk}}{\etasc}=\frac{\sqrt2}{\pi^{3/2}\sqrt{CT}}
\Bigl[1+\frac{2\pi^2}{3}CT-\frac{2\pi^4}{15}(CT)^2+\frac{4\pi^6}{105}(CT)^3+\cO\bigl((CT)^4\bigr)\Bigr].
\label{eq:eta-disk-small-q}
\end{equation}
The first three terms reproduce the second line of \eqref{eq:intro-disk-limits}. The growth at
small $CT$ is simply $\vev{k}_T/CT\sim 1/\sqrt{CT}$: the typical momentum of the occupied
states decreases more slowly than $CT$.

\begin{figure}[!ht]
  \centering
  \includegraphics[width=\textwidth]{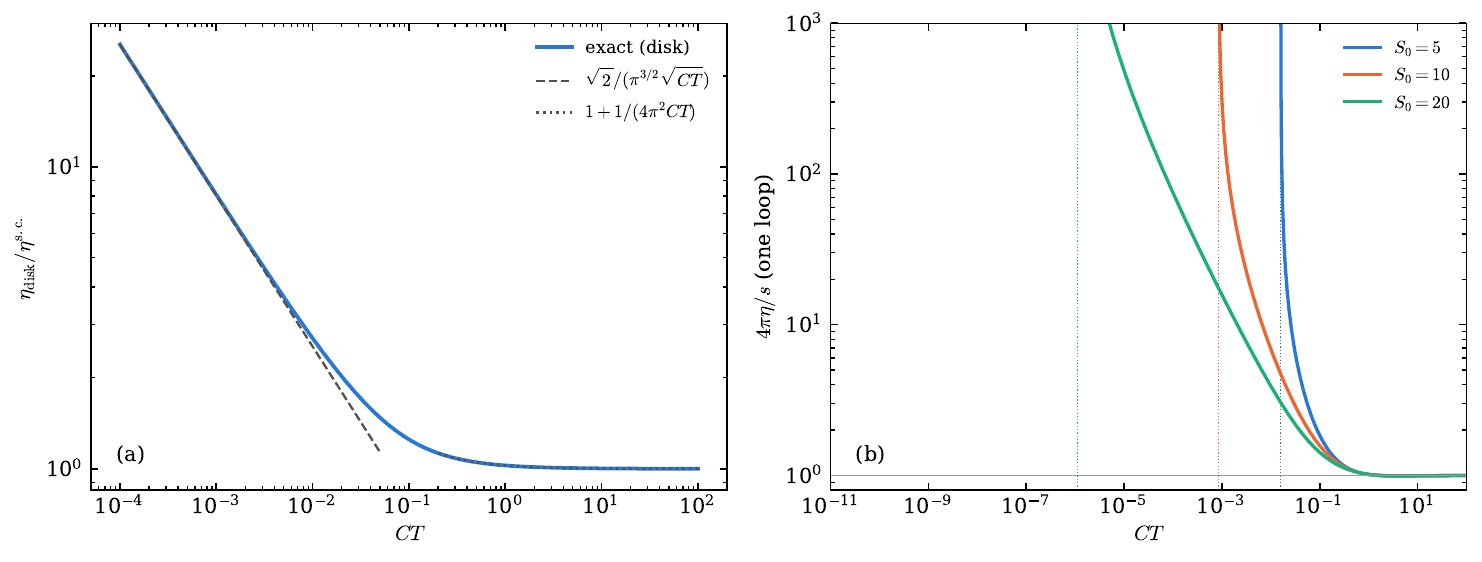}
  \caption{The disk (Schwarzian) results. (a) $\eta_{\rm disk}/\etasc$, eq.~\eqref{eq:eta-disk},
  with its small- and large-$CT$ asymptotics. (b) The one-loop ratio $4\pi\eta/s$ built from
  \eqref{eq:eta-disk} and \eqref{eq:S-disk}, for $S_0=5,10,20$. It dips below $1$ and then
  diverges where $S_{\rm disk}=0$ (vertical dotted lines), at $CT\sim e^{-2S_0/3}$.}
  \label{fig:disk}
\end{figure}

\paragraph{One-loop entropy.} The disk entropy is
\begin{equation}
S_{\rm disk}(T)=(1+T\partial_T)\log Z_{\rm disk}=S_0+4\pi^2CT+\frac32\log CT+\frac32-\frac12\log2\pi .
\label{eq:S-disk}
\end{equation}
The constant $\frac32$ comes from $T\partial_T\log T^{3/2}$. Like the other
temperature-independent terms it is often dropped, because it is negligible compared with
$S_0$, but it matters in Section~\ref{sec:airy}, where the entropy itself is of order one (see
also the comment below \eqref{eq:eta-s-DS}). The logarithm makes
$S_{\rm disk}$ vanish at
\begin{equation}
CT=(2\pi)^{1/3}e^{-1}\,e^{-2S_0/3}.
\end{equation}
There the one-loop $\eta/s$,
obtained by inserting \eqref{eq:eta-disk} and \eqref{eq:S-disk} into
\eqref{eq:eta-over-s}, diverges. Figure~\ref{fig:disk} shows both quantities. The
divergence of $\eta/s$ is the first indication that $CT\sim e^{-2S_0/3}$ is a genuinely new
scale, where the disk approximation cannot be trusted.

\section{Level repulsion and the fate of the DC viscosity}
\label{sec:repulsion}

\subsection{The sine kernel}

The spectrum of a quantum chaotic system is not a random sprinkling of points: neighbouring
levels repel each other \cite{Dyson1962,Mehta2004,BohigasGiannoniSchmit1984}. For the matrix
integral dual to JT gravity this follows from the joint distribution of the eigenvalues of the
$N\times N$ Hermitian matrix $H$,
\begin{equation}
p_N(\lambda_1,\dots,\lambda_N)\propto\prod_{i<j}|\lambda_i-\lambda_j|^{2}\;e^{-N\sum_iV(\lambda_i)} .
\label{eq:joint}
\end{equation}
The Vandermonde factor suppresses configurations in which two eigenvalues are close. The power
$2$ is the Dyson index of Hermitian matrices, appropriate for JT gravity on orientable surfaces
\cite{SaadShenkerStanford2019,StanfordWitten2019}. As discussed in Section~\ref{sec:mm}, the
unitary class is an assumption about the boundary theory; in the orthogonal and symplectic
classes the power is $\beta_{\rm D}=1$ and $4$.

The quantity that enters the Kubo formula is the two-level correlation function $R_2(x)$. It is
the probability of finding a level, not necessarily the nearest one, at distance $x\,\Delta(E)$
from a given level, relative to an uncorrelated spectrum, where $\Delta(E)=1/\rho(E)$ is the mean
level spacing. For Hermitian ensembles \cite{Mehta2004}
\begin{equation}
R_2(x)=1-\Bigl(\frac{\sin\pi x}{\pi x}\Bigr)^2 .
\label{eq:R2}
\end{equation}
It vanishes as $x^2$ for $x\to0$ and tends to $1$, up to decaying oscillations, for $x\gg1$:
nearby levels avoid each other, while distant levels are uncorrelated.

In the determinantal language of \eqref{eq:determinantal}, $R_2$ is encoded in the kernel.
Away from the spectral edge ($E\gg e^{-2S_0/3}/C$) and at separations small compared with $E$,
the kernel takes the universal sine form \cite{Mehta2004,SaadShenkerStanford2019}
\begin{equation}
K(E_1,E_2)\simeq\frac{\sin\bigl(\pi\int_{E_1}^{E_2}\rho_0\bigr)}{\pi(E_2-E_1)}+\dots
\end{equation}
The omitted terms oscillate in the mean energy $\bar E=\frac12(E_1+E_2)$ with a phase of
order $e^{S_0}$, and they average to zero against any smooth weight. The two-level density
at $\omega=E_2-E_1\neq0$ is therefore
\begin{equation}
\vev{\rho(E_1)\rho(E_2)}\simeq\rho_0(E_1)\rho_0(E_2)\Bigl[1-\sinc^2\bigl(\pi\rho_0(\bar E)\,\omega\bigr)\Bigr],
\qquad \sinc z=\frac{\sin z}{z}.
\label{eq:sine-kernel}
\end{equation}
The bracket is $R_2$ of \eqref{eq:R2}, evaluated at $x=\rho_0(\bar E)\,\omega$. It vanishes
quadratically for $\omega\ll\Delta$ and approaches $1$ for $\omega\gg\Delta$.

The $\sinc^2$ term is the correlation hole of \eqref{eq:determinantal}, and its total weight is
fixed by the projection property of the kernel stated below \eqref{eq:determinantal}:
\begin{equation}
\int\!\dd E_2\,\rho_0(E_1)\rho_0(E_2)\,\sinc^2\bigl(\pi\rho_0(\bar E)\,(E_2-E_1)\bigr)\simeq\rho_0(E_1).
\label{eq:hole-sum}
\end{equation}
The hole around each level thus contains exactly one level, and it cancels the contact term in
\eqref{eq:determinantal}. This has a simple consequence for the Kubo formula. Level repulsion
does not destroy spectral weight. The pairs of levels that are missing at frequencies
$|\omega|\lesssim\Delta$ are compensated by the self-pairs $m=n$ in the Lehmann sum
\eqref{eq:Lehmann}, which sit at $\omega=0$. These terms do not contribute to
$\operatorname{Im}\cG^R$, because the factor $1-e^{-\beta\omega}$ in \eqref{eq:FDT} vanishes at
$\omega=0$. Level repulsion can therefore remove low-frequency dissipation without violating any
sum rule. The cancellation is exact for the unweighted two-level density. In the Lehmann sum
the pairs are weighted by the Boltzmann factor and by the matrix elements \eqref{eq:vertex}. In
the bulk of the spectrum these weights are constant across the correlation hole, whose width is
the level spacing, up to corrections that are exponentially small in $S_0$, and the statement
holds with this accuracy. In the edge regime of Section~\ref{sec:airy} the hole is as wide as the
thermal window, and only the unweighted statement holds. 
Replacing $\bar E$ by $E_1$ changes the argument by a relative amount
$\cO(\omega\rho_0'/\rho_0)=\cO(e^{-S_0})$ when $\pi\rho_0\omega=\cO(1)$. We may therefore use
$\pi\rho_0(E_k)\,\omega=e^{S_0}C\omega\sinh(2\pi k)/2\pi$. Inserting \eqref{eq:sine-kernel} into
\eqref{eq:G-spectral} gives the level-repulsion-corrected correlator
\begin{equation}
\begin{aligned}
\cG^>_{\rm a}(\omega)=\frac{e^{S_0}\omega}{2\pi^2Z(T)}\int_0^\infty\!\dd k\;
&\underbrace{k\,e^{-\frac{k^2}{2CT}}\sinh(2\pi k)}_{g_T(k)}\;
\underbrace{\frac{\sinh 2\pi k_2}{\sinh\pi(k_2+k)\sinh\pi(k_2-k)}}_{f(k,\omega)}\\
&\times\Bigl[1-\sinc^2\bigl(\pi\rho_0(E_k)\,\omega\bigr)\Bigr].
\end{aligned}
\label{eq:G-corrected}
\end{equation}
Relative to the disk result \eqref{eq:C1}, two things have changed. The first is the last
factor, which accounts for level repulsion in the numerator. The second is the normalization.
In \eqref{eq:G-corrected}, $Z(T)$ is the partition function of the full matrix integral,
averaged over the ensemble as appropriate for the annealed correlator,
\begin{equation}
Z(T)=\int\!\dd E\;e^{-E/T}\,\vev{\rho(E)}=\int\!\dd E\;e^{-E/T}K(E,E),
\end{equation}
as in \eqref{eq:Z-kernel}, and not the disk partition function $Z_{\rm disk}$ that appears in
\eqref{eq:C1}. The one-level density $\vev{\rho(E)}$ differs from $\rho_0(E)$ by
higher-genus corrections and by non-perturbative oscillations that have the same origin as the
$\sinc^2$ term in the numerator. A consistent calculation must keep both. (In the two-level
density we have replaced $K(E_i,E_i)$ by $\rho_0(E_i)$. The difference is of relative order
$e^{-S_0}$ in the bulk of the spectrum, and it is negligible compared with the order-one change
produced by the $\sinc^2$ term.) We determine $Z(T)$ in Section~\ref{sec:normalization}. There we
show that, at fixed $CT$, $Z(T)=Z_{\rm disk}(T)\bigl[1+\cO\bigl(e^{-2S_0}/(CT)^3\bigr)\bigr]$
in the soft-edge completion, and that other completions change only the size of the correction.
In the bulk regime the change of the normalization is therefore negligible compared with the
order-one change of the numerator. This is no longer true at the exponentially small scale $CT\sim e^{-2S_0/3}$, which probes the edge of the spectrum. We treat this regime in Section~\ref{sec:airy}.

\begin{figure}[!ht]
  \centering
  \includegraphics[width=\textwidth]{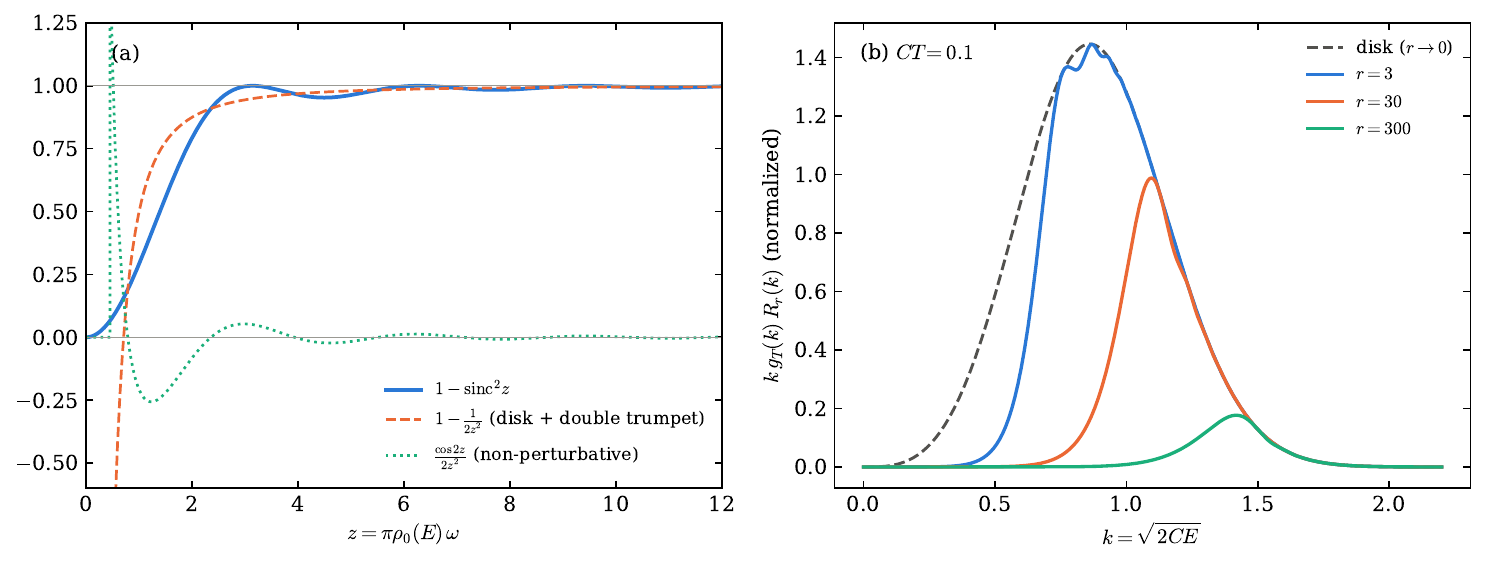}
  \caption{(a) The level-repulsion factor $1-\sinc^2z$ and its decomposition
  \eqref{eq:ramp-plateau} into the perturbative (disk plus double trumpet) and the
  non-perturbative (plateau) parts. (b) The integrand of the Kubo numerator at $CT=0.1$,
  $k\,g_T(k)R_r(k)$, normalized with respect to the disk. Here $R_r(k)=1-\sinc^2(\sinh2\pi k/2\pi r)$ is the
  bracket of \eqref{eq:G-corrected} with $r=e^{-S_0}/(C\omega)$.}
  \label{fig:repulsion}
\end{figure}

\paragraph{Perturbative and non-perturbative parts.} The level-repulsion factor contains
two physically distinct pieces. With $z=\pi\rho_0\omega$,
\begin{equation}
1-\sinc^2z=\underbrace{\Bigl[1-\frac{1}{2z^2}\Bigr]}_{\text{disk}\,+\,\text{double trumpet}}
+\underbrace{\frac{\cos2z}{2z^2}}_{\text{non-perturbative}} .
\label{eq:ramp-plateau}
\end{equation}
Multiplied by $\rho_0\rho_0$, the term $-1/2z^2$ gives $-1/(2\pi^2\omega^2)$. This is the
coincident limit of the genus-zero, two-boundary (double-trumpet) contribution to the
connected two-level density, i.e.\ the ramp of the spectral form factor
\cite{SaadShenkerStanford2019,CotlerEtAl2017}. It is perturbative, suppressed by $e^{-2S_0}$
relative to $\rho_0\rho_0$ at fixed $\omega$. The term $\cos(2z)/2z^2$ has a phase proportional
to $e^{S_0}$. It is the non-perturbative, plateau contribution
\cite{SaadShenkerStanford2019,SaadStanfordYangYao2022,BlommaertKruthoffYao2022}. Only the
sum of the two removes the spectral weight completely as $z\to0$. Both are needed as soon as
$z=\cO(1)$, see Figure~\ref{fig:repulsion}(a).

\paragraph{The weight at small $k$ is effectively removed.} Level repulsion acts at energies where the level
spacing exceeds $\omega$. Since the density grows with energy, this is the low-energy part
of the spectrum, $k\lesssim k_{\min}$, where $\pi\rho_0(E_{k_{\min}})\omega=1$:
\begin{equation}
k_{\min}(S_0,C\omega)=\frac{1}{2\pi}\operatorname{arsinh}\Bigl(\frac{2\pi e^{-S_0}}{C\omega}\Bigr).
\label{eq:k-min}
\end{equation}
Figure~\ref{fig:repulsion}(b) shows the effect on the thermal integrand. As the resolution
$r=e^{-S_0}/(C\omega)$ grows, the weight at small $k$ is removed, the remaining weight is
pushed to higher energies, and its total decreases.

\subsection{Sharp-cutoff approximation}
\label{sec:cut}

The smooth crossover of the level-repulsion factor can be modelled by a step function at
$k_{\min}$. With the linear-plus-constant form $f\simeq1+k/(\pi C\omega)$, which is adequate
for the thermal weight at $CT\gtrsim1$, this gives
\begin{equation}
\begin{gathered}
\cG^>_{\rm cut}(\omega)\simeq\frac{e^{S_0}\omega}{2\pi^2Z(T)}
\Bigl[\cI_1(CT;k_{\min})+\frac{\cI_2(CT;k_{\min})}{\pi C\omega}\Bigr],\\
\cI_n(CT;k_0)=\int_{k_0}^\infty\!\dd k\,k^ne^{-\frac{k^2}{2CT}}\sinh(2\pi k).
\end{gathered}
\label{eq:G-cut}
\end{equation}
The truncated moments are elementary (Appendix~\ref{app:moments}) and reduce to
$I_{1,2}$ at $k_0=0$. The approximation is accurate whenever the thermal weight sits at
$k\gg k_{\min}$ or when the crossover is narrow compared with the thermal width. An
improved version, with the step placed optimally, is given in
Section~\ref{sec:DS-asymptotics}.

\subsection{The DC limit at fixed entropy}
\label{sec:DC}

At fixed $S_0$, $k_{\min}\to\infty$ as $\omega\to0$. Every level eventually has its
neighbours further away than $\omega$. At fixed $k$ one has
$1-\sinc^2z=z^2/3+\cO(z^4)$. Using also the leading term $f\simeq k/(\pi C\omega)$ of the
small-frequency expansion \eqref{eq:f-expansion} of the vertex function $f(k,\omega)$ defined in
\eqref{eq:f}, the correlator \eqref{eq:G-corrected} becomes
\begin{equation}
\cG^>_{\rm a}(\omega)=\frac{e^{3S_0}C\omega^2}{24\pi^5Z(T)}\int_0^\infty\!\dd k\;k^2e^{-\frac{k^2}{2CT}}\sinh^3(2\pi k)+\cO(\omega^3),
\qquad \operatorname{Im}\cG^R(\omega)=\cO(\omega^3).
\label{eq:omega-cubed}
\end{equation}
Therefore
\begin{equation}
\eta_{\rm DC}=0\qquad(S_0<\infty)
\label{eq:eta-zero}
\end{equation}
for any normalization that is finite and non-zero at $\omega=0$. The frequency-resolved viscosity
behaves as $\eta(\omega)\propto e^{2S_0}\omega^2$ at small $\omega$. The exponent is the
Dyson index: in general $R_2(x)\propto x^{\beta_{\rm D}}$ at small $x$ \cite{Mehta2004}, so that
$\eta(\omega)\propto\omega^{\beta_{\rm D}}$, and the DC viscosity vanishes in all three symmetry
classes. The limits do not commute:
\begin{equation}
\lim_{\omega\to0}\,\lim_{S_0\to\infty}\eta(\omega)=\eta_{\rm disk},\qquad
\lim_{S_0\to\infty}\,\lim_{\omega\to0}\eta(\omega)=0 .
\end{equation}
The first order of limits is the one that defines a transport coefficient of the infinite
system. Both $S_0$ and $C$ are proportional to the transverse volume, so that in the planar
limit also $CT\to\infty$ and $\eta_{\rm disk}\to\etasc$.
The vanishing of the DC viscosity \eqref{eq:eta-zero} is a property of the infinite-time
limit of an isolated system with a discrete spectrum. For a single member of the ensemble
$\operatorname{Im}\cG^R(\omega)$ is a sum of delta functions at the level differences
$E_n-E_m$. After averaging over the ensemble these become a smooth function of $\omega$, which
vanishes at $\omega=0$ because pairs of nearly degenerate levels are suppressed by level
repulsion. The physical information is therefore carried by scaling limits in which $\omega$ is
kept comparable to the level spacing. We turn to these next.

\paragraph{A remark on $\omega$-dependent normalizations.} Because the numerator vanishes,
one may be tempted to divide by a quantity that vanishes as well. For instance, one could
use the coefficient of the $\cO(\omega)$ term of \eqref{eq:G-cut},
$Z_{\rm cut}(T,\omega)=\frac{e^{S_0}}{2\pi^2}\cI_1(CT;k_{\min})$. This gives
\begin{equation}
\frac{\eta_{\rm cut}}{\etasc}\simeq\frac{\omega}{2T}+\frac{\cI_2(CT;k_{\min})}{2\pi CT\,\cI_1(CT;k_{\min})}
\;\sim\;\frac{k_{\min}}{2\pi CT}\;\sim\;\frac{1}{4\pi^2CT}\log\frac{4\pi e^{-S_0}}{C\omega},
\end{equation}
which grows logarithmically as $\omega\to0$. The logarithm is an artifact of combining the
sharp step with an $\omega$-dependent normalization. $Z_{\rm cut}$ is not a thermal
partition function: it depends on the probe frequency. We return to the meaning of such
normalizations in Section~\ref{sec:pair}.

\section{The bulk double-scaling limit}
\label{sec:DS}

\subsection{The scaling function}

The unfolded frequency $\pi\rho_0(E_k)\omega=e^{S_0}C\omega\sinh(2\pi k)/2\pi$ depends on
$S_0$ and $\omega$ only through the combination $e^{S_0}C\omega$. This suggests taking the following double-scaling limit
\begin{equation}
r=\frac{e^{-S_0}}{C\omega},\qquad S_0\to\infty,\quad C\omega\to0,\quad r\ \text{and}\ CT\ \text{fixed}.
\label{eq:DS-limit}
\end{equation}
The parameter $r$ is the ratio of the bare level-spacing scale $e^{-S_0}/C$ to the probe
frequency. It measures how finely the probe resolves the spectrum. It is important to
stress that $r$ is not a parameter of the model alone. The gravitational data fix $S_0$ and $C$,
and hence the level-spacing scale $e^{-S_0}/C$, but $r$ also depends on how the system is
probed: on the frequency of the applied shear or, for a nominally static measurement, on the
frequency resolution that is actually achieved, which is limited by the duration of the
observation and by any broadening of the levels due to a coupling to an environment (see the
discussion around \eqref{eq:smeared}). Within our framework $r$ is therefore a free,
tunable parameter, which we cannot fix, and our results are functions of it. The disk result
of \cite{KanargiasKiritsisMurthyPapadoulakiPorfyriadis2025} corresponds to $r\to0$, and the
strict DC limit at fixed $S_0$ to $r\to\infty$. In the double-scaling limit
\eqref{eq:DS-limit} the
unfolded frequency $\sinh(2\pi k)/(2\pi r)$ stays finite, while $\omega/T\to0$ and
$C\omega\to0$. The KMS factor in \eqref{eq:Kubo} becomes $\beta\omega$, and only the pole of
\eqref{eq:f-expansion} survives in $f$. Anticipating Section~\ref{sec:normalization}, where
we show that the correct normalization reduces to $Z_{\rm disk}$ in this limit, we obtain (we denote by $\eta_{\rm DS}$ the viscosity in the double-scaling limit)
\begin{equation}
\begin{gathered}
\boxed{\;\frac{\eta_{\rm DS}(r,CT)}{\etasc}=\frac{\cJ_2(CT,r)}{2\pi CT\,I_1(CT)}=\frac{\vev{k\,R_r(k)}_T}{2\pi CT}\;},\\
\cJ_n(CT,r)=\int_0^\infty\!\dd k\;k^ne^{-\frac{k^2}{2CT}}\sinh(2\pi k)\,R_r(k),
\end{gathered}
\label{eq:eta-DS}
\end{equation}
with $R_r(k)=1-\sinc^2\bigl(\sinh(2\pi k)/2\pi r\bigr)$. Compared with the disk result
\eqref{eq:eta-disk}, each thermally occupied state now contributes with the probability
$R_r(k)$ that a partner level sits at distance $\omega$. Since $0\le R_r\le1$, the
double-scaled viscosity is bounded by the disk value. It is finite for every $r$,
approaches $\eta_{\rm disk}$ as $r\to0$, and vanishes as $r\to\infty$.

\begin{figure}[!ht]
  \centering
  \includegraphics[width=\textwidth]{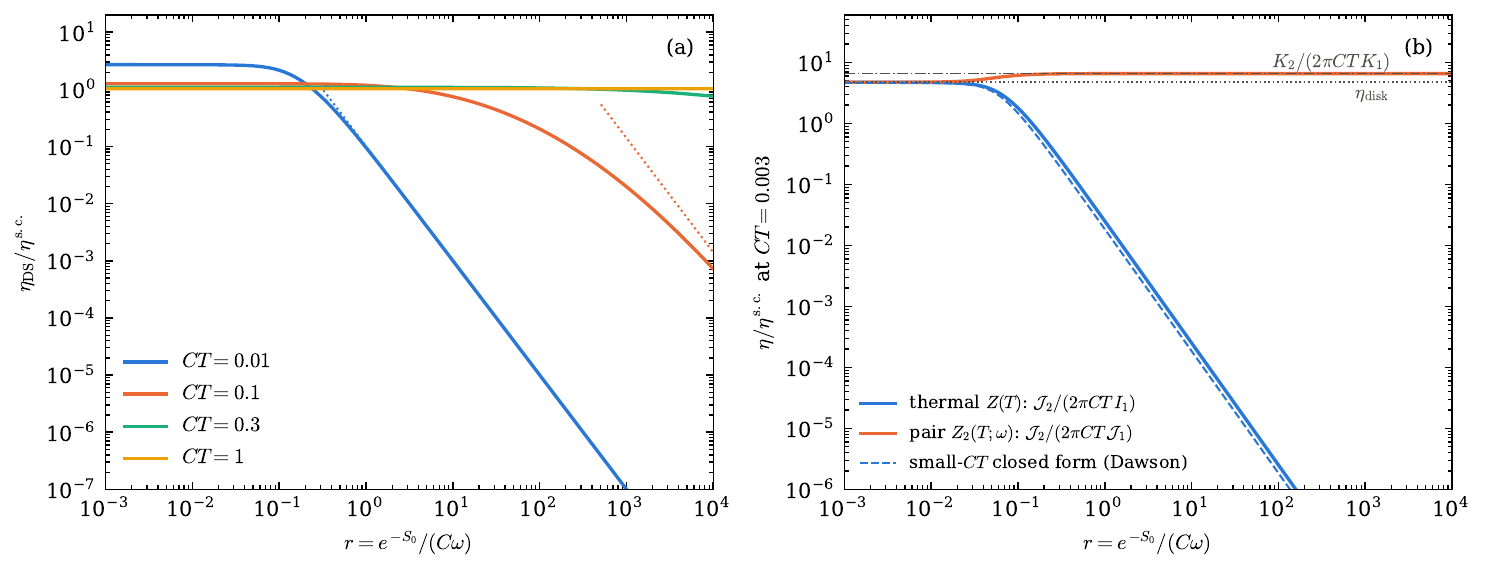}
  \caption{(a) The double-scaled viscosity \eqref{eq:eta-DS} as a function of
  $r=e^{-S_0}/(C\omega)$ for several $CT$, with the large-$r$ law \eqref{eq:DS-large-r}
  (dotted). (b) At $CT=0.003$: \eqref{eq:eta-DS} (blue), the small-$CT$ closed form
  \eqref{eq:DS-small-q} (dashed), and the pair-normalized ratio of Section~\ref{sec:pair}
  (orange). The latter tends to $K_2/(2\pi CT\,K_1)$ instead of zero.}
  \label{fig:DS-vs-r}
\end{figure}

This is the scaling limit in which the Kubo viscosity has a finite, non-trivial value.
Figures~\ref{fig:DS-vs-r} and \ref{fig:DS-map} display it. At fixed $CT$ the viscosity
decreases monotonically with $r$. The value of $r$ at which the decrease sets in grows
extremely fast with $CT$, because the thermally occupied density of states grows as
$e^{4\pi^2CT}$.

\subsection{Structure and closed forms}
\label{sec:DS-asymptotics}

\paragraph{Weak resolution: the double trumpet.} For $r\to0$ the argument of the sinc is
large over the thermal region. Using \eqref{eq:ramp-plateau}, the leading correction
comes from the perturbative term $-1/2z^2$, with $z=\sinh(2\pi k)/2\pi r$:
\begin{equation}
\frac{\eta_{\rm DS}}{\etasc}=\frac{\eta_{\rm disk}}{\etasc}-\frac{\pi r^2\mathcal L_2(CT)}{CT\,I_1(CT)}+\cO(r^4),
\qquad \mathcal L_n(CT)=\int_0^\infty\!\dd k\,\frac{k^n\,e^{-k^2/2CT}}{\sinh2\pi k}.
\label{eq:DS-small-r}
\end{equation}
The second term on the right-hand side of \eqref{eq:DS-small-r}, which is of order
$r^2=e^{-2S_0}/(C\omega)^2$, is the ramp contribution of the
double trumpet. It is the first correction in the genus-like expansion of the two-level
density at fixed $\omega$. The non-perturbative plateau term only enters at higher order,
through the endpoint $k\to0$.

The size of this correction depends strongly on temperature. Because of the factor
$1/\sinh2\pi k$, the moment $\mathcal L_2$ is dominated by $k\lesssim1$ and saturates at
$\mathcal L_2(\infty)=7\zeta(3)/16\pi^3$ for $CT\gtrsim1$, whereas $I_1$ grows as
$e^{2\pi^2CT}$. At high temperature the ramp correction is therefore suppressed by the
exponentially large number of thermally accessible levels, and level repulsion is invisible
unless $r$ is exponentially large. At low temperature $\mathcal L_2\simeq CT/2\pi$, and the
relative correction is $-r^2/(4CT)$. This is the first term of the large-$y$ expansion of the
closed form \eqref{eq:DS-small-q} below, with $y=\sqrt{CT}/r$.

\paragraph{Strong resolution.} When the argument of the sinc is small over the whole region
that dominates the integral, $R_r\simeq\sinh^2(2\pi k)/(12\pi^2r^2)$ and
\begin{equation}
\frac{\eta_{\rm DS}}{\etasc}\simeq\frac{K_2(CT)}{24\pi^3CT\,I_1(CT)}\,\frac{1}{r^2},\qquad
K_n(CT)=\int_0^\infty\!\dd k\;k^ne^{-k^2/2CT}\sinh^3(2\pi k).
\label{eq:DS-large-r}
\end{equation}
The moments $K_n$ are elementary (Appendix~\ref{app:moments}). Since the expansion of $R_r$
shifts the weight to $\sinh^3(2\pi k)$, the relevant region is the peak of the integrand of
$K_2$. We denote the position of this peak by $k_3$, where the subscript recalls the third
power of $\sinh$ in $K_2$. It lies at $k_3\simeq6\pi CT$ for $CT\gtrsim1$ and
$k_3\simeq\sqrt{5CT}$ for $CT\ll1$. The
condition of validity is therefore
\begin{equation}
r\gg\frac{\sinh(2\pi k_3)}{2\pi}.
\label{eq:strong-res-cond}
\end{equation}
For $CT\ll1$ this condition reads $r\gg\sqrt{CT}$, i.e.\ $\omega\ll\Delta(T)$: the probe
frequency must be small compared with the level spacing at the thermal energy. For $CT\gtrsim1$
it is much stronger than the corresponding condition at the thermal saddle
$k_{\rm th}\simeq2\pi CT$, which is $\pi\rho_0(E_{\rm th})\,\omega\ll1$, because the weight
$\sinh^3(2\pi k)$ moves the relevant momenta from $2\pi CT$ to $6\pi CT$. The approach to
\eqref{eq:DS-large-r} is then slow. In terms of the frequency, \eqref{eq:DS-large-r} reads
$\eta\propto e^{2S_0}(C\omega)^2$, in agreement with \eqref{eq:omega-cubed}.

Equation \eqref{eq:DS-large-r} is the first term of an expansion in inverse powers of $r^2$,
which is generated by the Taylor series of the level-repulsion factor,
\begin{equation}
R_r(k)=1-\sinc^2z=\sum_{m\ge1}\frac{(-1)^{m+1}\,2^{2m+1}}{(2m+2)!}\,z^{2m}
=\frac{z^2}{3}-\frac{2z^4}{45}+\dots,\qquad z=\frac{\sinh2\pi k}{2\pi r}.
\label{eq:R-series}
\end{equation}
Inserted into \eqref{eq:eta-DS}, the term $z^{2m}$ gives a contribution proportional to
$r^{-2m}$, whose coefficient is the moment
$\int_0^\infty\dd k\,k^2e^{-k^2/2CT}\sinh^{2m+1}(2\pi k)$. The term $m=1$ is
\eqref{eq:DS-large-r}. These moments grow as $e^{2\pi^2(2m+1)^2CT}$, faster than the factorial in
\eqref{eq:R-series} decreases. The expansion in powers of $1/r^2$ is therefore asymptotic and
not convergent.

\paragraph{Low temperature: a closed form.} For $CT\ll1$ the thermal weight lives at
$k\sim\sqrt{CT}$, where $\sinh2\pi k\simeq2\pi k$ and $\pi\rho_0\omega\simeq k/r$. Setting
$k=\sqrt{CT}\,u$,
\begin{equation}
\begin{aligned}
\frac{\eta_{\rm DS}}{\etasc}&\simeq\frac{1}{2\pi CT}\,\frac{\int_0^\infty\dd k\,k^3e^{-k^2/2CT}\bigl[1-\sinc^2(k/r)\bigr]}{\int_0^\infty\dd k\,k^2e^{-k^2/2CT}}\\
&=\frac{\sqrt2}{\pi^{3/2}\sqrt{CT}}\cdot\frac12\int_0^\infty\!\dd u\,u^3e^{-u^2/2}\bigl[1-\sinc^2(yu)\bigr],
\end{aligned}
\end{equation}
with $y=\sqrt{CT}/r$. The remaining integral follows from
$\int_0^\infty u^3e^{-u^2/2}\sinc^2(yu)\,\dd u=\frac{1}{2y^2}\bigl[1-\int_0^\infty u\,e^{-u^2/2}\cos(2yu)\,\dd u\bigr]$
and $\int_0^\infty u\,e^{-u^2/2}\cos(bu)\,\dd u=1-\sqrt2\,b\,F(b/\sqrt2)$. Here
$F(x)=e^{-x^2}\int_0^xe^{t^2}\dd t$ is Dawson's function. The result is
\begin{equation}
\boxed{\;\frac{\eta_{\rm DS}}{\etasc}\simeq\frac{\sqrt2}{\pi^{3/2}\sqrt{CT}}
\Bigl[1-\frac{F(\sqrt2\,y)}{\sqrt2\,y}\Bigr],\qquad
y=\frac{\sqrt{CT}}{r}=e^{S_0}C\omega\sqrt{CT}=\frac{\pi\rho_0(T)\,\omega}{\sqrt2}\;}
\label{eq:DS-small-q}
\end{equation}
The variable $y$ is, up to a numerical factor, the probe frequency in units of the level
spacing $\Delta(T)=1/\rho_0(T)$ at the thermal energy $E\simeq T$. The two limits of
\eqref{eq:DS-small-q} are
\begin{equation}
\frac{\eta_{\rm DS}}{\etasc}\simeq\frac{\sqrt2}{\pi^{3/2}\sqrt{CT}}\times
\begin{cases}
1-\dfrac{1}{4y^2}+\dots,& y\gg1,\\[8pt]
\dfrac43\,y^2+\dots,& y\ll1 .
\end{cases}
\label{eq:DS-small-q-limits}
\end{equation}
For $y\gg1$ we recover the $(CT)^{-1/2}$ law \eqref{eq:eta-disk-small-q} of the disk, together
with the ramp correction $-r^2/(4CT)$ found above. For $y\ll1$ we find
$\eta_{\rm DS}/\etasc\to\frac{4\sqrt2}{3\pi^{3/2}}\sqrt{CT}/r^2$, which is the small-$CT$ limit of
\eqref{eq:DS-large-r}.

The corrections to \eqref{eq:DS-small-q} are of relative order $CT$. To obtain them we
expand the exact result \eqref{eq:eta-DS} to the next order in $CT$ at fixed $y$. This means
keeping the next term of $\sinh2\pi k=2\pi k\,\bigl(1+\frac{2\pi^2}{3}k^2+\dots\bigr)$, both in
the thermal weight and in the argument of the sinc, and the factor $e^{2\pi^2CT}$ of $I_1$. The
result is
\begin{equation}
\frac{\eta_{\rm DS}}{\etasc}=\frac{\sqrt2}{\pi^{3/2}\sqrt{CT}}\;G(y)\,
\Bigl[1+c(y)\,CT+\cO\bigl((CT)^2\bigr)\Bigr],\qquad
G(y)\equiv1-\frac{F(\sqrt2\,y)}{\sqrt2\,y},
\label{eq:DS-small-q-corr}
\end{equation}
where the factor in front of the square bracket is the closed form \eqref{eq:DS-small-q}, and
\begin{equation}
c(y)=\frac{2\pi^2}{3}\Bigl[1+\frac{6\,y\,G'(y)+y^2G''(y)}{G(y)}\Bigr]
\;\longrightarrow\;
\begin{cases}
\dfrac{2\pi^2}{3}\simeq6.6,& y\gg1,\\[8pt]
10\pi^2\simeq98.7,& y\ll1 .
\end{cases}
\label{eq:c-of-y}
\end{equation}
The coefficient $c(y)$ is positive, so the closed form \eqref{eq:DS-small-q} underestimates the
exact result, and it decreases monotonically with $y$. Its limit for $y\gg1$ is the second
term of the disk expansion \eqref{eq:eta-disk-small-q}, and its limit for $y\ll1$ follows from
the expansion of \eqref{eq:DS-large-r} in $CT$. Numerically, $c\simeq31$ at $y=1$ and
$c\simeq37$ at the maximum $y_*$ discussed below. The correction is larger at small $y$ because level
repulsion pushes the thermal weight to larger $k$, where the approximation
$\sinh2\pi k\simeq2\pi k$ is less accurate. At $CT=10^{-3}$, for example, the closed form lies
$0.7\%$, $3\%$ and $8.5\%$ below the exact result at $y=10$, $1$ and $0.3$. The formula is
therefore quantitatively accurate for $CT\lesssim10^{-3}$ and qualitatively correct up to
$CT\sim10^{-2}$.

Equation \eqref{eq:DS-small-q} makes the effect of level repulsion on the quantum regime
transparent. At fixed probe frequency $\omega$ the viscosity is \emph{non-monotonic} in
temperature, because two effects compete as $T$ decreases. The first is the Schwarzian
enhancement, the prefactor $(CT)^{-1/2}$. The second is level repulsion: the density of thermally
accessible states decreases, so the level spacing $\Delta(T)$ at the thermal energy grows and
$y\propto\omega/\Delta(T)$ decreases. As long as $\Delta(T)\ll\omega$ the bracket in
\eqref{eq:DS-small-q} is close to one, and the viscosity grows as $(CT)^{-1/2}$, as in the
Schwarzian analysis. Once $\Delta(T)$ exceeds $\omega$, transitions at frequency $\omega$ become
rare, the bracket decreases as $y^2\propto CT$, and the viscosity falls as $(CT)^{1/2}$. The
maximum lies at $y_*=0.9198$, i.e.\ at
$CT_*=0.846\,r^2$, with
\begin{equation}
\frac{\eta_{\max}}{\etasc}=\frac{\sqrt2}{\pi^{3/2}}\,\max_y\frac{1-F(\sqrt2y)/\sqrt2y}{y}\;\frac1r
\simeq\frac{0.1736}{r}=0.1736\;e^{S_0}C\omega .
\label{eq:eta-max}
\end{equation}
The Schwarzian enhancement of
\cite{KanargiasKiritsisMurthyPapadoulakiPorfyriadis2025,CremoniniLiLiuNian2025} is
therefore capped at a value set by the ratio of the probe frequency to the bare
level-spacing scale $e^{-S_0}/C$. The derivation assumes $CT_*\ll1$, i.e.\ $r\ll1$ or
$C\omega\gg e^{-S_0}$. By \eqref{eq:DS-small-q-corr}, the leading correction multiplies
\eqref{eq:eta-max} by $1+c(y_*)\,CT_*=1+31.3\,r^2$, so that an accuracy of $3\%$ requires
$r\lesssim0.03$. It also requires
$CT_*\gg e^{-2S_0/3}$, i.e.\ $C\omega\ll e^{-2S_0/3}$. Otherwise the maximum lies in the edge
regime of Section~\ref{sec:airy}.

\begin{figure}[!ht]
  \centering
  \includegraphics[width=\textwidth]{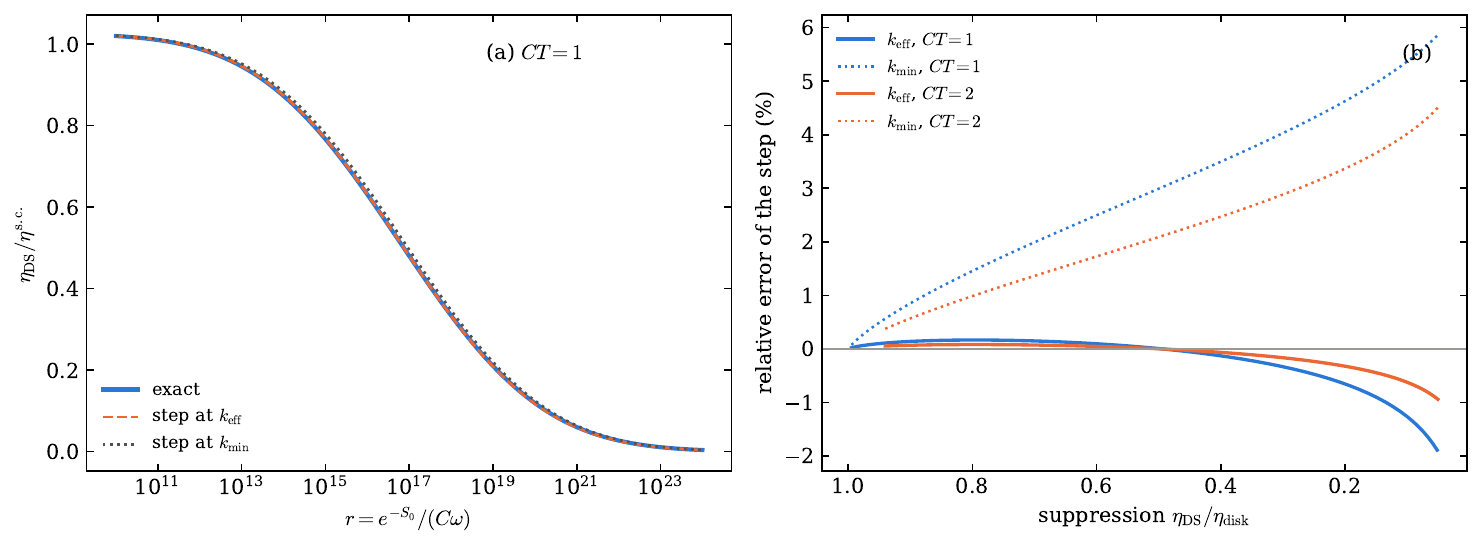}
  \caption{(a) At $CT=1$: the double-scaled viscosity \eqref{eq:eta-DS} and the two
  sharp-cutoff approximations, with the step at $k_{\rm eff}$ of \eqref{eq:k-eff} and at
  $k_{\min}$ of \eqref{eq:k-min}. (b) Relative error of the two approximations at $CT=1,2$,
  plotted against the suppression $\eta_{\rm DS}/\eta_{\rm disk}$.}
  \label{fig:cutoff}
\end{figure}

\paragraph{High temperature: an improved cutoff.} For $CT\gtrsim1$ the thermal weight is a
Gaussian of width $\sqrt{CT}$ centred at $k\simeq2\pi CT$, and level repulsion acts on it as a
filter in $k$. Since the unfolded frequency $\pi\rho_0(E_k)\omega=\sinh(2\pi k)/2\pi r$ grows
exponentially with $k$, the factor $R_r(k)$ rises from $0$ to $1$ over a range $\delta k\simeq1/2\pi$,
much narrower than the thermal width. To a good approximation $R_r$ is then a step function
$\Theta(k-k_{\rm eff})$, and the viscosity takes the sharp-cutoff form \eqref{eq:G-cut},
\begin{equation}
\frac{\eta_{\rm DS}}{\etasc}\simeq\frac{\cI_2(CT;k_{\rm eff})}{2\pi CT\,I_1(CT)} .
\end{equation}
We fix the position of the step by requiring that the step and $R_r$ have the same integral
over $k$. With $\sinh2\pi k\simeq e^{2\pi k}/2$, set $z=e^{2\pi k}/4\pi r$, so that
$R_r=1-\sinc^2z$ and $\dd k=\dd z/2\pi z$. The condition $\int\dd k\,[R_r-\Theta(k-k_{\rm eff})]=0$
becomes $\int_0^\infty\frac{\dd z}{z}\bigl[1-\sinc^2z-\Theta(z-z_{\rm eff})\bigr]=0$ and by performing this integral we find
$\log z_{\rm eff}=\frac32-\gamma_E-\log2$, i.e.
\begin{equation}
k_{\rm eff}=\frac{1}{2\pi}\Bigl[\log(2\pi r)+\frac32-\gamma_E\Bigr].
\label{eq:k-eff}
\end{equation}
This differs from the naive cutoff $k_{\min}$ of \eqref{eq:k-min}, defined by
$\pi\rho_0\omega=1$, by the constant $(\frac32-\gamma_E-\log2)/2\pi\simeq0.037$. As
Figure~\ref{fig:cutoff} shows, the step at $k_{\rm eff}$ reproduces \eqref{eq:eta-DS} to better
than $0.6\%$ ($0.3\%$ at $CT=2$) until the viscosity has dropped to a fifth of its disk value, whereas the step at
$k_{\min}$ is off by several per cent. The suppression sets in when $k_{\rm eff}$ reaches the
thermal peak $k\simeq2\pi CT$, i.e.\ at $r\sim\sinh(4\pi^2CT)/2\pi$, or
$C\omega\sim4\pi e^{-S_0-4\pi^2CT}=4\pi e^{-S^{\rm s.c.}}$. This is the level spacing at the
thermal energy, exponentially small in the Bekenstein--Hawking entropy.

\begin{figure}[!ht]
  \centering
  \includegraphics[width=0.72\textwidth]{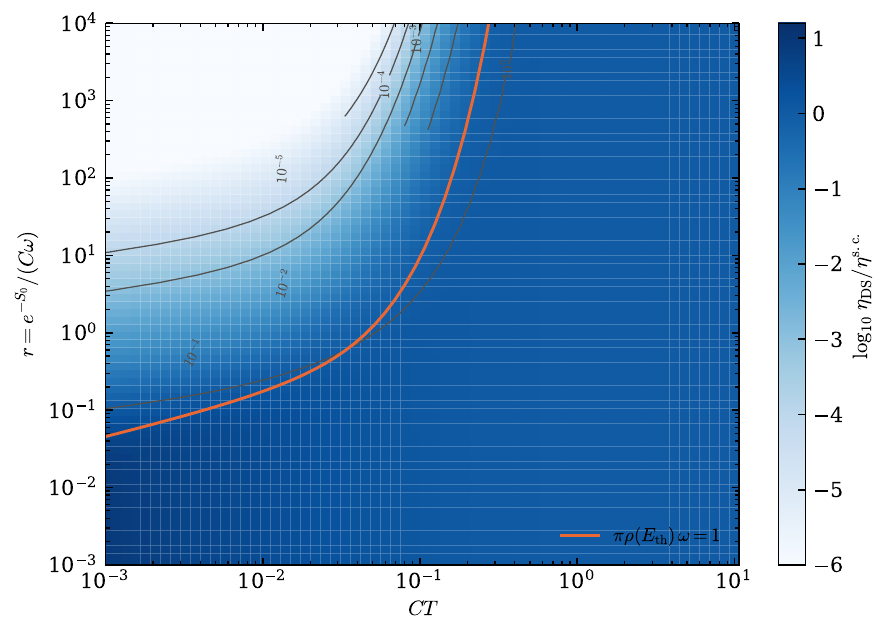}
  \caption{$\log_{10}\eta_{\rm DS}/\etasc$ in the $(CT,r)$ plane. The orange line marks
  $\pi\rho_0(E_{\rm th})\omega=1$ at the thermal saddle $k_{\rm th}$, defined by
  $k/CT=1/k+2\pi\coth2\pi k$. Below and to the right of it level repulsion is invisible;
  above it the viscosity is suppressed as $r^{-2}$.}
  \label{fig:DS-map}
\end{figure}

\paragraph{Summary.} The picture that emerges is simple. The quantum viscosity is
controlled by the ratio of the probe frequency to the level spacing at the thermal energy.
At high temperature this spacing is $\propto e^{-S^{\rm s.c.}}$, and level repulsion is
invisible for all practical purposes. At low temperature it is
$\Delta(T)=1/\rho_0(T)\simeq\pi e^{-S_0}/(C\sqrt{2CT})$. It grows as $T$ decreases
and eventually cuts off the Schwarzian enhancement. The line $\pi\rho_0(E_{\rm th})\omega=1$
in Figure~\ref{fig:DS-map} separates the two behaviours.

\section{The partition function of the matrix-model completion}
\label{sec:normalization}

The double-scaled result \eqref{eq:eta-DS} was normalized by the disk partition function. In
this section we show that this is correct. By \eqref{eq:Z-kernel}, the annealed correlator must
be normalized by the Laplace transform of the exact one-level density $\vev{\rho(E)}=K(E,E)$ of
the same matrix integral. This density differs from $\rho_0$ in two ways. It receives
perturbative corrections from surfaces of higher genus, and it contains non-perturbative
oscillations of the same origin as the plateau term in the numerator. Since the numerator is
modified at order one by non-perturbative effects in the double-scaling limit, one must check
whether the denominator is modified as well. We find that it is not. For the soft-edge
completion of the matrix integral, in which the eigenvalues can tunnel into the forbidden
region, at fixed $CT$,
\begin{equation}
Z(T)=Z_{\rm disk}(T)\Bigl[1+\cO\Bigl(\frac{e^{-2S_0}}{(CT)^3}\Bigr)\Bigr].
\label{eq:Z-result}
\end{equation}
The non-perturbative structure of the one-level density only resums the genus series of the
partition function. It becomes important only at $CT\sim e^{-2S_0/3}$, in the edge regime of
Section~\ref{sec:airy}. In Section~\ref{sec:string} we show that other completions change the
correction in \eqref{eq:Z-result}, but not the conclusion.

\subsection{The one-level density}
\label{sec:one-level}

\paragraph{Perturbative corrections.} The genus expansion of the one-level density reads
\begin{equation}
\vev{\rho(E)}=\rho_0(E)+\sum_{g\ge1}e^{(1-2g)S_0}\rho_g(E)+\dots,
\end{equation}
where $\rho_g$ is the inverse Laplace transform of the genus-$g$ partition function
$Z_{g,1}$, given below in terms of Weil--Petersson volumes. At small energy
$\rho_g\propto E^{1/2-3g}$, so the corrections grow without bound towards the edge of the
spectrum. This signals the breakdown of the genus expansion at energies of the order of the edge
level spacing, $E\sim e^{-2S_0/3}/C$.

\paragraph{Non-perturbative oscillations.} In the classically allowed region $E>0$ the leading
non-perturbative correction is an oscillation
\cite{SaadShenkerStanford2019,Johnson2021consistency,Johnson2022microstates},
\begin{equation}
\vev{\rho(E)}=\rho_0(E)-\frac{1}{4\pi E}\cos\bigl(2\pi N(E)\bigr)+\dots,\qquad
N(E)=\int_0^E\!\dd E'\,\rho_0(E') .
\label{eq:sss-density}
\end{equation}
Here $N(E)$ counts the levels below $E$, so each eigenvalue leaves a ripple in the average
density. The ripple has an amplitude of order one, while $\rho_0=\cO(e^{S_0})$. Its amplitude
depends only on the distance from the edge of the spectrum, for any leading density $\rho_0$
\cite{SaadShenkerStanford2019}. A short derivation of this statement, and a numerical check,
are given in Appendix~\ref{app:osc}. For the JT density, with
$E=k^2/2C$,
\begin{equation}
\vev{\rho(E_k)}=\rho_0(E_k)-\frac{C\cos\Phi(k)}{2\pi k^2},\qquad
\Phi(k)=2\pi N(E_k)=\frac{e^{S_0}}{4\pi^3}\bigl[2\pi k\cosh2\pi k-\sinh2\pi k\bigr].
\label{eq:sss-density-JT}
\end{equation}
Formally, the oscillation contributes to the partition function
\begin{equation}
\delta Z_{\rm osc}(CT;k_*)=-\frac{1}{2\pi}\int_{k_*}^\infty\!\dd k\;\frac{e^{-k^2/2CT}}{k}\cos\Phi(k) .
\label{eq:dZosc}
\end{equation}
Near the edge $\Phi\simeq\frac23e^{S_0}k^3$, and the integral diverges logarithmically as
$k_*\to0$. The divergence is not physical. It signals the breakdown of the WKB approximation
behind \eqref{eq:sss-density} in the edge region $k\lesssim e^{-S_0/3}$, where the exact edge
kernel must be used. Near the edge, therefore, both the perturbative and the non-perturbative
corrections must be resummed. We do this next.

\subsection{The Airy edge resums the genus series}
\label{sec:airy-resum}

Near the edge the kernel is the Airy kernel \cite{BowickBrezin1991,Forrester1993,TracyWidom1994},
\begin{equation}
\begin{gathered}
K(E_1,E_2)\simeq\kappa\,K_{\Ai}(-\kappa E_1,-\kappa E_2),\\
K_{\Ai}(x,y)=\frac{\Ai(x)\Ai'(y)-\Ai'(x)\Ai(y)}{x-y}=\int_0^\infty\!\dd s\,\Ai(x+s)\Ai(y+s),
\end{gathered}
\label{eq:airy-kernel}
\end{equation}
with $\kappa=2^{1/3}Ce^{2S_0/3}$. The corresponding density is
\begin{equation}
\rho_{\Ai}(E)=\kappa\bigl[\Ai'(-\kappa E)^2+\kappa E\,\Ai(-\kappa E)^2\bigr]\;\xrightarrow{\;\kappa E\gg1\;}\;
\frac{\kappa^{3/2}\sqrt E}{\pi}-\frac{1}{4\pi E}\cos\Bigl(\frac43(\kappa E)^{3/2}\Bigr)+\dots
\label{eq:rho-airy}
\end{equation}
Two different limits are involved here, and they should not be confused. The Airy form
\eqref{eq:airy-kernel} of the kernel holds close to the edge, for $CE\ll1$. The expansion
displayed in \eqref{eq:rho-airy} holds for $\kappa E\gg1$, that is, for energies large compared
with the level spacing $1/\kappa\propto e^{-2S_0/3}/C$ at the edge. The two conditions are
compatible, and both descriptions are valid in the window $e^{-2S_0/3}\ll CE\ll1$. In this window
the first term of \eqref{eq:rho-airy} coincides with the JT disk density \eqref{eq:rho0}
expanded for $CE\ll1$, where $\sinh(2\pi\sqrt{2CE})\simeq2\pi\sqrt{2CE}$:
$\kappa^{3/2}\sqrt E/\pi=e^{S_0}C\sqrt{2CE}/\pi$. The second term is \eqref{eq:sss-density} for the
Airy density, with phase $\frac43(\kappa E)^{3/2}=\frac23e^{S_0}k^3$. For $E<0$ the density
decays as $\exp(-\frac43|\kappa E|^{3/2})$: eigenvalues can tunnel into the classically
forbidden region, with exponentially small probability.

The Laplace transform of $\rho_{\Ai}$ can be computed exactly
\cite{OkuyamaSakai2020}. Using the integral representation in \eqref{eq:airy-kernel} and the
identity $\int_{-\infty}^\infty\dd x\,e^{tx}\Ai(x+s)^2=(4\pi t)^{-1/2}e^{t^3/12-st}$, valid for
$t>0$, we find
\begin{equation}
\begin{aligned}
Z_{\Ai}(T)&=\int_{-\infty}^{\infty}\!\dd E\;e^{-E/T}\rho_{\Ai}(E)
=\int_0^\infty\!\dd s\,\frac{e^{t^3/12-st}}{\sqrt{4\pi t}}\bigg|_{t=1/\kappa T}\\
&=\frac{(\kappa T)^{3/2}}{2\sqrt\pi}\,e^{\frac{1}{12(\kappa T)^3}}
=\frac{e^{S_0}(CT)^{3/2}}{\sqrt{2\pi}}\exp\Bigl(\frac{e^{-2S_0}}{24(CT)^3}\Bigr).
\end{aligned}
\label{eq:Z-airy}
\end{equation}
Expanded in $e^{-2S_0}$,
\begin{equation}
Z_{\Ai}=\sum_{g\ge0}e^{(1-2g)S_0}Z^{\Ai}_{g,1}(CT),\qquad Z^{\Ai}_{g,1}(CT)=\frac{(CT)^{3/2-3g}}{\sqrt{2\pi}\,24^g\,g!} .
\label{eq:Z-airy-genus}
\end{equation}
This reproduces, genus by genus, the most singular term of the Weil--Petersson answer
\cite{Mirzakhani2007,SaadShenkerStanford2019}
\begin{equation}
Z_{g,1}(CT)=\sqrt{\frac{CT}{2\pi}}\int_0^\infty\!b\,\dd b\,V_{g,1}(b)\,e^{-CT\,b^2/2}.
\end{equation}
For example, $V_{1,1}(b)=(b^2+4\pi^2)/48$ gives
\begin{equation}
Z_{1,1}(CT)=\frac{1}{\sqrt{2\pi}}\Bigl(\frac{\pi^2}{12\sqrt{CT}}+\frac{1}{24(CT)^{3/2}}\Bigr),
\label{eq:Z11}
\end{equation}
whose most singular term is $Z^{\Ai}_{1,1}$. The genus-two check is in
Appendix~\ref{app:WP}. Unlike the genus expansion of the full JT partition function, the series
\eqref{eq:Z-airy-genus} converges.

This exact result is what we need. Everything in $\rho_{\Ai}$ beyond $\rho_0$
integrates to \emph{exactly} the resummed genus series: the smooth higher-genus terms
$\propto E^{1/2-3g}$, the oscillations in \eqref{eq:rho-airy} and the exponentially small tail at
$E<0$. No additional non-perturbative term in $Z(\beta)$ arises. At fixed $CT$ the exponent
$e^{-2S_0}/[24(CT)^3]$ is small, so the non-perturbative structure of the density leaves only
perturbatively small corrections in $Z$. They become of order one when $CT\sim e^{-2S_0/3}$.

\paragraph{The exponent as a saddle point.} The exponent in \eqref{eq:Z-airy} has a
transparent origin. Write the oscillatory term of \eqref{eq:sss-density} as
\begin{equation}
\vev{\rho(E)}_{\rm osc}=-\frac{1}{4\pi E}\operatorname{Re}e^{2\pi iN(E)} .
\end{equation}
Its contribution to $Z(\beta)$ is $-\operatorname{Re}\int\dd E\,e^{-\beta E+2\pi iN(E)}/(4\pi E)$.
After continuation to $E<0$, the exponent $-\beta E+2\pi iN(E)$ is stationary where
$\beta=2\pi i\rho_0(E_*)$. For the edge density $\rho_0=\kappa^{3/2}\sqrt E/\pi$ this gives
$E_*=-e^{-2S_0}/[8C(CT)^2]$, in the forbidden region, where
\begin{equation}
-\beta E_*+2\pi i\!\int_0^{E_*}\!\rho_0=\frac{e^{-2S_0}}{24(CT)^3}.
\label{eq:saddle}
\end{equation}
So $Z_{\Ai}$ is dominated, when the exponent is large, by configurations with one eigenvalue
pulled into the forbidden region. The low temperature makes this favourable, since the
Boltzmann factor $e^{-\beta E_*}$ rewards negative energies. For the full JT density the saddle
condition reads $\sin v_*=\pi e^{-S_0}/(CT)$, with $v=2\pi\sqrt{2C|E|}$. Its value is
\begin{equation}
\mathcal F(CT)=\frac{v_*^2}{8\pi^2CT}-\frac{e^{S_0}}{4\pi^3}\bigl(\sin v_*-v_*\cos v_*\bigr)
=\frac{e^{-2S_0}}{24(CT)^3}+\frac{\pi^2e^{-4S_0}}{120(CT)^5}+\dots
\end{equation}
Together with the disk factor $e^{2\pi^2CT}$, $e^{\mathcal F}$ reproduces the two most singular
terms of both $Z_{1,1}$ and $Z_{2,1}$ (Appendix~\ref{app:WP}). The saddle exists only for
$CT\ge\pi e^{-S_0}$. At lower temperatures it runs into the region of the JT effective
potential whose extremum, at $E=-1/8C$, carries the eigenvalue-instanton weight
$e^{-e^{S_0}/4\pi^2}$ \cite{SaadShenkerStanford2019}. Enhanced by $e^{\beta/8C}$, this weight
competes with $Z_{\rm disk}$ only for $CT\lesssim\frac{\pi^2}{2}e^{-S_0}$. In this regime the annealed
partition function depends on the non-perturbative completion
\cite{Johnson2020,Johnson2020c}, which we discuss in Section~\ref{sec:string}.

\paragraph{A caveat on bulk--edge matching.} One might try to build a uniform density by
adding to the Airy result the difference between the oscillatory term of the JT density,
\eqref{eq:sss-density-JT}, and that of the Airy density, \eqref{eq:rho-airy}. This difference
is integrable at $k=0$ and
gives a finite correction to $Z$. The correction, however, is dominated by
$k\sim e^{-S_0/3}$, where neither WKB form is valid. It equals $0.1239\,e^{-2S_0/3}$,
independently of $CT$ (Appendix~\ref{app:matching}). In the soft-edge completion this region is
described exactly by the Airy kernel, which is already included in \eqref{eq:Z-airy}. The term is
therefore an artifact of the matching, and it must be discarded. In Section~\ref{sec:string} we will see that corrections from the edge region with fractional powers of $e^{-S_0}$ do arise in completions with a hard bound on the spectrum. They are then fixed by the exact edge kernel of that completion and not by a simple WKB matching.

\subsection{The normalized viscosity and the entropy}
\label{sec:DS-entropy}

\begin{figure}[!ht]
  \centering
  \includegraphics[width=\textwidth]{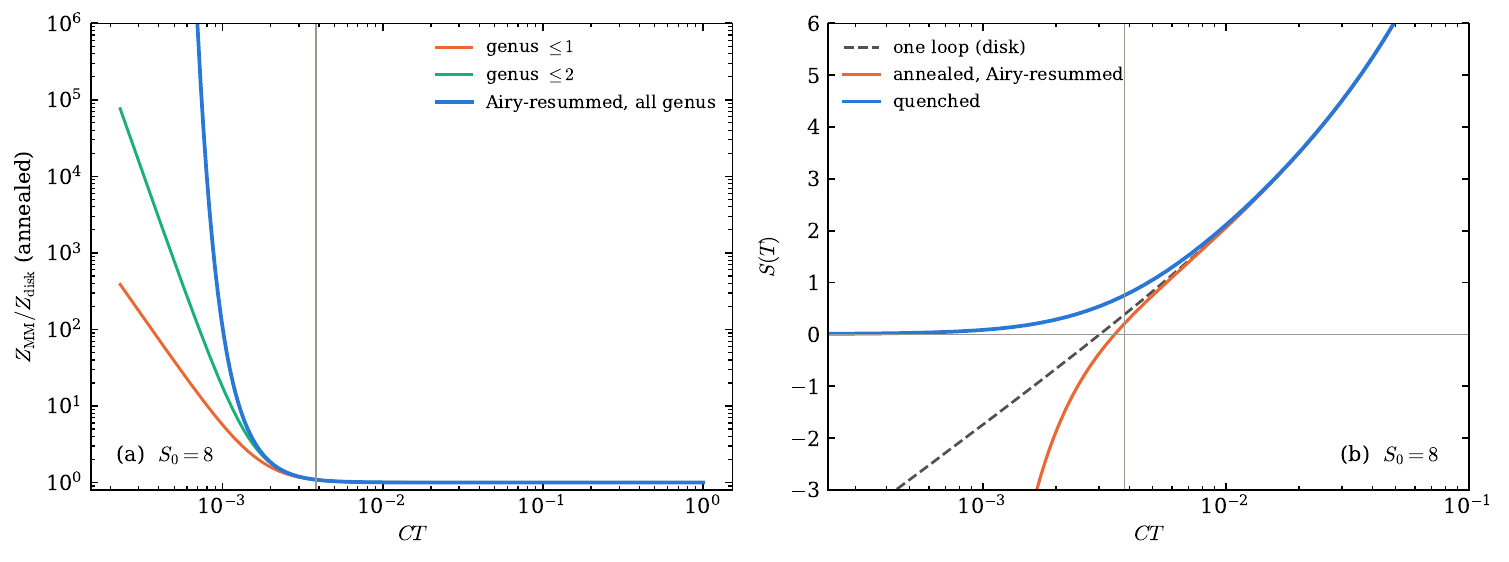}
  \caption{(a) $Z/Z_{\rm disk}$ at $S_0=8$: genus expansion truncated at $g=1,2$, and the
  Airy-resummed partition function \eqref{eq:Z-MM}. The corrections become of order one at
  $CT\sim e^{-2S_0/3}$ (vertical line: $\kappa T=1$). (b) The corresponding entropies: one loop
  \eqref{eq:S-disk}, annealed from \eqref{eq:Z-MM}, and quenched (Section~\ref{sec:airy}). The
  annealed entropy becomes negative; the quenched one does not.}
  \label{fig:genus}
\end{figure}

In the soft-edge completion, up to corrections that are doubly exponentially small in $S_0$,
the partition function
containing the one-level effects of the same origin, to all orders of the genus
expansion, is
\begin{equation}
Z_{\rm MM}(T)=\underbrace{Z_{\rm disk}(T)+Z_{\Ai}(T)-Z_{\Ai,0}(T)}_{\text{Airy-resummed}}
+\sum_{g\ge1}e^{(1-2g)S_0}\bigl[Z_{g,1}(CT)-Z^{\Ai}_{g,1}(CT)\bigr].
\label{eq:Z-MM}
\end{equation}
Here $Z_{\Ai,0}=e^{S_0}(CT)^{3/2}/\sqrt{2\pi}$ is the overlap of the disk and Airy answers. The
first three terms combine into the explicit composite
\begin{equation}
Z_{\rm disk}+Z_{\Ai}-Z_{\Ai,0}=\frac{e^{S_0}(CT)^{3/2}}{\sqrt{2\pi}}\Bigl[e^{2\pi^2CT}+e^{\frac{e^{-2S_0}}{24(CT)^3}}-1\Bigr],
\label{eq:Z-composite}
\end{equation}
which resums the most singular term at every genus. The remaining sum in \eqref{eq:Z-MM}
contains only less singular terms and is well behaved at small $CT$ term by term. Like
the genus expansion of JT gravity itself, it is an asymptotic series in $e^{-2S_0}$.
Equation \eqref{eq:Z-MM} is therefore a composite of the exact Airy resummation and of the
asymptotic genus expansion of JT gravity, with their overlap subtracted, and not an exact
expression at finite $S_0$. It is uniformly accurate in $CT$, order by order in the remaining
sum, which is what is needed below. At fixed $CT$,
\begin{equation}
\begin{gathered}
\frac{Z_{\rm MM}}{Z_{\rm disk}}=1+e^{-2S_0}a_1(CT)+e^{-4S_0}a_2(CT)+\dots,\\
a_1(CT)=e^{-2\pi^2CT}\Bigl(\frac{\pi^2}{12(CT)^2}+\frac{1}{24(CT)^3}\Bigr),
\end{gathered}
\label{eq:Z-ratio}
\end{equation}
with $a_2$ given in \eqref{eq:a2}. This proves \eqref{eq:Z-result}. Let $\eta^{(\rm MM)}$
denote the viscosity normalized by the partition function $Z_{\rm MM}$ of the same ensemble,
instead of $Z_{\rm disk}$. It differs from the double-scaled viscosity $\eta_{\rm DS}$ of
\eqref{eq:eta-DS} only by the ratio of the two partition functions, and both are measured in
units of the same semiclassical value $\etasc$:
\begin{equation}
\frac{\eta^{(\rm MM)}(r,CT)}{\etasc}=\frac{\cJ_2(CT,r)}{2\pi CT\,I_1(CT)}\,\frac{Z_{\rm disk}}{Z_{\rm MM}}
=\frac{\eta_{\rm DS}(r,CT)}{\etasc}\,\Bigl[1-e^{-2S_0}a_1(CT)+\cO\bigl(e^{-4S_0}\bigr)\Bigr].
\label{eq:eta-MM}
\end{equation}
The first factor in the middle expression is $\eta_{\rm DS}/\etasc$ by the definition
\eqref{eq:eta-DS}, and the second tends to one as $S_0\to\infty$ at fixed $r$ and $CT$. Hence
$\eta^{(\rm MM)}\to\eta_{\rm DS}$ in the double-scaling limit.
This justifies \eqref{eq:eta-DS}. The two-level numerator is modified at order one in the
double-scaling limit, while the one-level normalization is modified only at relative order
$e^{-2S_0}/(CT)^3$. The normalization therefore matters only when $CT\lesssim e^{-2S_0/3}$. There
the genus expansion itself breaks down (Figure~\ref{fig:genus}), and the problem has to be
treated exactly. We do so in Section~\ref{sec:airy}.

\paragraph{Entropy.} The same reasoning applies to the entropy. At fixed $CT$ the annealed
entropy is $S_{\rm disk}+(1+T\partial_T)\log\bigl(1+e^{-2S_0}a_1+\dots\bigr)$. The quenched
entropy, defined from $\vev{\log Z}$, differs from it through the connected two-boundary
amplitude. Expanding $\vev{\log Z}=\log\vev Z-\vev{Z^2}_{\rm c}/2\vev Z^2+\dots$ and using the
double trumpet \cite{SaadShenkerStanford2019}, $\vev{Z(\beta_1)Z(\beta_2)}_{\rm c}=\sqrt{\beta_1\beta_2}/[2\pi(\beta_1+\beta_2)]$,
which equals $1/4\pi$ at $\beta_1=\beta_2$, we find
\begin{equation}
\begin{gathered}
\vev{\log Z}-\log Z_{\rm MM}=-\frac{1}{8\pi Z_{\rm disk}^2}=-\frac{e^{-2S_0}e^{-4\pi^2CT}}{4(CT)^3},\\
S_{\rm q}-S_{\rm a}=\frac{e^{-2S_0}e^{-4\pi^2CT}}{2(CT)^3}\bigl(1+2\pi^2CT\bigr)+\dots
\end{gathered}
\label{eq:quenched-DT}
\end{equation}
These are replica-wormhole corrections \cite{EngelhardtFischettiMaloney2020}. They are
positive: sample-to-sample fluctuations of $Z$ increase the typical entropy relative to the
annealed one. In the double-scaling limit, $S/S^{\rm s.c.}\to1$ and
\begin{equation}
\boxed{\;4\pi\,\frac{\eta}{s}\Big|_{\rm DS}=\frac{\cJ_2(CT,r)}{2\pi CT\,I_1(CT)}\;}
\label{eq:eta-s-DS}
\end{equation}
At large but finite $S_0$ this is multiplied by the one-loop factor
$S^{\rm s.c.}/S_{\rm disk}=[1+(\frac32\log CT+\frac32-\frac12\log2\pi)/S^{\rm s.c.}]^{-1}$.
The same factor appears in \cite{KanargiasKiritsisMurthyPapadoulakiPorfyriadis2025}, but
without the constant $\frac32$. There the quantum entropy was taken to be
$S^{\rm s.c.}+\log\bigl[(CT)^{3/2}/\sqrt{2\pi}\bigr]$, i.e.\ the Bekenstein--Hawking entropy plus
the logarithm of the one-loop factor of $Z_{\rm disk}$. The thermodynamic entropy
$(1+T\partial_T)\log Z_{\rm disk}$ of \eqref{eq:S-disk} contains in addition the term
$T\partial_T\log(CT)^{3/2}=\frac32$. The two expressions differ by a temperature-independent
constant of order one, which is negligible for $S_0\gg1$ at the temperatures considered there.
We keep it here for two reasons: in the matrix model $e^{S_0}$ is fixed by the normalization of
$\rho_0$ in \eqref{eq:rho0}, so the constant is unambiguous, and it matters in the edge regime,
where the entropy itself is of order one. The one-loop factor diverges where
$S_{\rm disk}=0$, the first sign
of the edge regime.

\subsection{What an \texorpdfstring{$\omega$}{omega}-dependent normalization measures}
\label{sec:pair}

Level repulsion suppresses the integral $\cJ_2(CT,r)$ in the numerator of the Kubo formula
\eqref{eq:eta-DS}. One may be tempted to divide it by a quantity
that is suppressed in the same way, so that the ratio stays finite as $\omega\to0$ (see also the
remark at the end of Section~\ref{sec:repulsion}). Here we explain what such a ratio measures,
and why it is not the viscosity that enters $\eta/s$.

The natural candidate comes from the small-frequency form of the correlator. With
$f\simeq1+k/(\pi C\omega)$ in \eqref{eq:G-corrected}, the term of order $\omega$ in
$\cG^>_{\rm a}$ is proportional to
\begin{equation}
Z_2(T;\omega)=\frac{e^{S_0}}{2\pi^2}\,\cJ_1(CT,r)=\int\!\dd E\;e^{-\beta E}\,\frac{\vev{\rho(E)\rho(E+\omega)}}{\rho_0(E+\omega)} .
\label{eq:pair-Z}
\end{equation}
The second expression shows the meaning of $Z_2$. By \eqref{eq:sine-kernel}, the ratio in the
integrand is $\rho_0(E)R_2$: the density of levels at $E$ that have a partner at $E+\omega$. Hence
$Z_2$ is a Boltzmann-weighted count of pairs of levels separated by $\omega$, a \emph{pair}
partition function. For $\omega\gg\Delta$ it reduces to $Z_{\rm disk}$. For $\omega\lesssim\Delta$ it
is suppressed by level repulsion, exactly as the numerator. Unlike $Z(T)$, it depends on the
probe frequency and is not the Laplace transform of any one-level density. Even its motivation
is only approximate, since the true constant term of \eqref{eq:f-expansion} is
$\coth2\pi k+1/2\pi k$ rather than $1$.

\begin{figure}[!ht]
  \centering
  \includegraphics[width=0.66\textwidth]{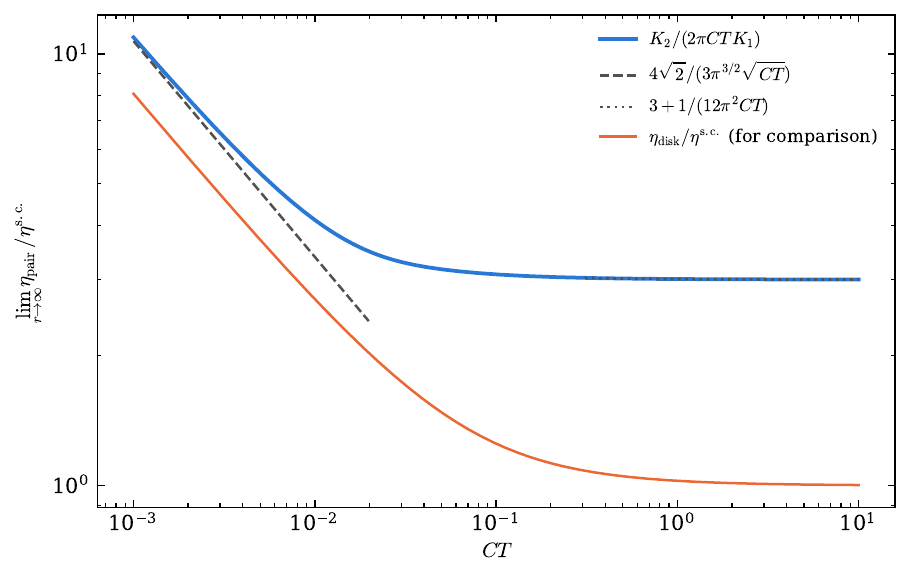}
  \caption{The $r\to\infty$ limit of the pair-normalized ratio \eqref{eq:eta-pair},
  $K_2/(2\pi CT\,K_1)$, with its asymptotics \eqref{eq:pair-limit}. The disk viscosity is
  shown for comparison. The Kubo viscosity itself vanishes in this limit.}
  \label{fig:pair}
\end{figure}

Dividing by $Z_2$ instead of $Z_{\rm disk}$ defines
\begin{equation}
\frac{\eta_{\rm pair}(r,CT)}{\etasc}=\frac{\cJ_2(CT,r)}{2\pi CT\,\cJ_1(CT,r)}=\frac{\vev{k}_{T,r}}{2\pi CT},
\label{eq:eta-pair}
\end{equation}
where $\vev{\cdot}_{T,r}$ is the average over the weight $g_T(k)R_r(k)$. Compared with
\eqref{eq:eta-DS}, only the normalization of the weight has changed. The quantity
$\eta_{\rm pair}$ is the mean momentum of the levels that have a partner at distance $\omega$,
rather than the mean over all thermally occupied levels. It is a conditional quantity: the
dissipation \emph{per resolved transition}. The two viscosities are related by
\begin{equation}
\eta_{\rm DS}=\eta_{\rm pair}\,\frac{Z_2(T;\omega)}{Z_{\rm disk}(T)} .
\end{equation}
All of the level-repulsion suppression of $\eta_{\rm DS}$ sits in the ratio
$Z_2/Z_{\rm disk}$, which vanishes as $r\to\infty$. The pair ratio therefore stays finite
(Figure~\ref{fig:pair}):
\begin{equation}
\lim_{r\to\infty}\frac{\eta_{\rm pair}}{\etasc}=\frac{K_2(CT)}{2\pi CT\,K_1(CT)}
=\begin{cases}3+\frac{1}{12\pi^2CT}+\dots,& CT\gg1,\\[2pt]
\frac{4\sqrt2}{3\pi^{3/2}\sqrt{CT}}\,\bigl(1+\cO(CT)\bigr),& CT\ll1 .\end{cases}
\label{eq:pair-limit}
\end{equation}
The factor $3$ at high temperature has a simple origin. At strong resolution only nearly
degenerate pairs contribute, with a weight $\propto\rho_0^3\propto e^{6\pi k}$. The thermal peak
therefore moves from $k=2\pi CT$ to $k=6\pi CT$, and the mean momentum triples. The thermodynamic
transport coefficient that enters $\eta/s$ is the Kubo viscosity normalized by the thermal
partition function, $\eta_{\rm DS}$. The pair ratio answers a different question: how
strongly the system absorbs \emph{per pair of levels that is available at the probe
frequency}, rather than per thermally occupied level. It characterizes the transitions that do
take place, but it carries no information on how rare they are.

We are not aware of an earlier use of a pair partition function of this kind in the
gravitational literature, but the underlying idea is standard in other contexts. In the
eigenstate thermalization hypothesis the variance of the off-diagonal matrix elements of an
operator is written as $e^{-S(\bar E)}|f(\bar E,\omega)|^2$, with $f$ a smooth function, and
dividing a spectral function by the density of pairs of levels at distance $\omega$ is
precisely what isolates $|f|^2$ from the level statistics
\cite{Srednicki1999,DAlessioEtAl2016}. In our problem $|f|^2$ is fixed by the Schwarzian vertex
\eqref{eq:vertex}, and $\eta_{\rm pair}$ is the average of the resulting absorption strength
over the pairs of levels that are present at the probe frequency. Similarly, in the theory of optical absorption the absorption
rate is factorized into a mean squared matrix element and the joint density of initial and
final states \cite{YuCardona2010,Fox2010}. In all these cases the quantity normalized per pair measures the strength of the
individual transitions, whereas the transport coefficient also counts how many of them
there are.

\section{The edge regime}
\label{sec:airy}

\subsection{Scaling limit and universality}
\label{sec:edge-scaling}

Introduce the edge variables
\begin{equation}
\theta=\kappa T=2^{1/3}e^{2S_0/3}\,CT,\qquad \nu=\kappa\omega,\qquad \zeta=\kappa E,
\label{eq:edge-variables}
\end{equation}
and consider $S_0\to\infty$ at fixed $(\theta,\nu)$. In this limit temperature, frequency and
energy are all measured in units of the level spacing near the edge of the spectrum,
$1/\kappa\propto e^{-2S_0/3}/C$. This scale is parametrically larger than the bulk level-spacing
scale $e^{-S_0}/C$, because the density of states vanishes at the edge. Five things happen at
once.
\begin{enumerate}
\item The expansion parameter of the partition function, $e^{-2S_0}/[24(CT)^3]=1/(12\theta^3)$,
is of order one. The genus series must be summed, and \eqref{eq:Z-airy} does this exactly.
\item The one-loop entropy becomes $S_{\rm disk}=\frac32\log\theta+\frac32-\log2\sqrt\pi+\cO(e^{-2S_0/3})$.
It is of order one and vanishes at $\theta_0=(4\pi)^{1/3}/e=0.8553$.
\item The level spacing at thermal energies is $\Delta(T)\simeq\pi/(\kappa\sqrt\theta)$, so
$\Delta/T\simeq\pi\theta^{-3/2}$. Only $\cO(1)$ levels are thermally accessible, and the
relevant frequencies are $\omega\sim T$. There is no hydrodynamic window.
\item The Schwarzian vertex becomes constant, $\Gamma(1\pm ik_1\pm ik_2)\to1$, since
$k\sim e^{-S_0/3}$.
\item The kernel is the Airy kernel \eqref{eq:airy-kernel}. The corrections to it are of
relative order $CE=2^{-1/3}e^{-2S_0/3}\zeta$.
\end{enumerate}
As a result, the problem becomes universal. In the scaling limit the details of JT gravity and
of the operator drop out, except for an overall normalization, and both the numerator and the
normalization of the Kubo formula are built from the Airy kernel alone. Physically, the black brane at
these temperatures is a quantum system with a handful of thermally accessible levels at the
bottom of a random-matrix spectrum. Its lowest level, $\zeta_0$ in the variables
\eqref{eq:edge-variables}, fluctuates from sample to sample according
to the Tracy--Widom distribution \cite{TracyWidom1994}:
\begin{equation}
{\rm Prob}\bigl(\zeta_0>\sigma\bigr)=F_2(-\sigma),\qquad
F_2(s)=\det\bigl(1-K_{\Ai}\bigr)_{L^2(s,\infty)}=\exp\Bigl(-\int_s^\infty\!\dd x\,(x-s)\,q(x)^2\Bigr).
\label{eq:TW}
\end{equation}
Here the determinant is the Fredholm determinant of the Airy kernel \eqref{eq:airy-kernel}
restricted to the interval $(s,\infty)$, and $q$ is the Hastings--McLeod solution of the
Painlev\'e II equation \cite{HastingsMcLeod1980,TracyWidom1994}, $q''=xq+2q^3$ with $q(x)\simeq\Ai(x)$ for $x\to+\infty$. The mean and the
standard deviation of $\zeta_0$ are $1.771$ and $0.902$. The ground-state energy thus fluctuates
by an amount $0.90/\kappa$, which is comparable to $T$ in this regime.\footnote{All
energies, temperatures and frequencies of this regime scale with the single power
$e^{-2S_0/3}$, and the scaling functions are universal. In this sense the edge regime is
critical in the language of random-matrix theory, where it is the soft-edge universality
class. We nevertheless do not call it quantum critical, for three reasons. First, the scale
$1/\kappa$ remains in the problem: there is no scale invariance, and the response depends on
$\kappa\omega$ and $\kappa T$ separately. Second, the fluctuations referred to in the text are
sample-to-sample fluctuations over the ensemble, not critical fluctuations of a given system.
Third, for near-extremal branes the term is commonly used for the scale-invariant regime
governed by the AdS$_2$ throat \cite{FaulknerLiuMcGreevyVegh2011}. In the planar limit this
regime extends from $T\ll\mu$ down to $T=0$, and lower temperatures lie deeper inside it. At
finite transverse volume, however, it has a lower end, at $E_{\rm gap}=1/C$. The Schwarzian
mode results from the breaking of the conformal symmetry of the throat, and for
$T\lesssim E_{\rm gap}$ its fluctuations are strong. Response functions then depend on the scale
$C$, as in \eqref{eq:eta-disk}, and no longer have the scaling form that is fixed by the
operator dimensions of the AdS$_2$ fixed point
\cite{BagretsAltlandKamenev2016,MertensTuriaciVerlinde2017}. The temperatures of the edge
regime, $T\sim e^{-2S_0/3}E_{\rm gap}$, are indeed much lower than those of the AdS$_2$ regime.
They are, however, not deep inside the quantum critical window $E_{\rm gap}\ll T\ll\mu$, but
far below its lower end.}

\subsection{Annealed and quenched averages}
\label{sec:aq}

With only a few accessible levels, the way in which the ensemble average is taken matters. The matrix integral is an ensemble of Hamiltonians. Each member of the ensemble, or
\emph{sample}, is one Hamiltonian $H$, with its own complete set of discrete energy levels
$\{E_n\}$. By the spectrum of a sample we mean this whole set, not an individual level.
Thermodynamic quantities and response functions are properties of a single quantum system: the
partition function $Z=\sum_ne^{-\beta E_n}$, the entropy and the Lehmann sum \eqref{eq:Lehmann}
are all defined for one sample, in terms of all its levels. The \emph{quenched} prescription
computes such a quantity sample by sample and then averages the result over the ensemble. For
the viscosity this means averaging the ratio in the Lehmann representation \eqref{eq:Lehmann},
which gives \eqref{eq:G-quenched}; for the entropy it means using $\vev{\log Z}$. The
\emph{annealed} prescription \eqref{eq:G-spectral} instead averages the numerator and the
partition function separately, and uses $\log\vev{Z}$ for the
entropy.
The two agree when the partition function is
self-averaging, i.e.\ when its sample-to-sample fluctuations are small compared with its mean.
This is the case for $CT\gg e^{-2S_0/3}$, where the thermal window contains a very large number of
levels. There the difference is the replica-wormhole correction \eqref{eq:quenched-DT},
suppressed by $e^{-2S_0}$. In the edge regime the fluctuations of $Z$ are as large as $Z$ itself,
and the two prescriptions differ at order one.

We compute both, for different reasons. The annealed quantities are the ones that the
gravitational path integral computes most directly: $\vev{Z}$ and the numerator of
\eqref{eq:G-spectral} are each given by a sum over geometries with a single asymptotic boundary,
and they are the quantities whose genus expansion we used in
Sections~\ref{sec:DS}--\ref{sec:normalization}. The quenched quantities are ensemble averages of
observables of a single system. If the black brane is described by one Hamiltonian, its entropy
and its response are those of one sample, and their mean over the ensemble is the quenched
average. In particular, the quenched entropy is an average of non-negative numbers, whereas
$\log\vev{Z}$ is not the average of any single-system quantity and can be dominated by rare,
atypical samples (Section~\ref{sec:edge-entropy}). In this sense the quenched average is the
physically meaningful one. One clarification is necessary. In the edge regime single-sample
quantities are not self-averaging: with $\cO(1)$ levels in the thermal window, the entropy and
the response of an individual sample fluctuate at order one around their mean. The quenched
average is therefore the ensemble mean of the physical observable. It gives the correct,
universal order of magnitude for a typical sample, but it is not a sharp prediction for an
individual member of the ensemble, whose response depends on the microscopic details of its
spectrum. Comparing the two prescriptions shows precisely where
and why the annealed answer fails.

\subsection{Exact formulas}
\label{sec:edge-exact}

Inserting the Airy kernel \eqref{eq:airy-kernel} into
\eqref{eq:G-spectral}--\eqref{eq:G-quenched} and
\eqref{eq:Kubo}, and using the normalization $N_\cO=e^{-S_0}/C^2$ of \eqref{eq:vertex}, which
gives $N_\cO\kappa^2/\nu=2^{2/3}e^{S_0/3}/\nu$, we find
\begin{equation}
\boxed{\;\frac{\eta(\omega,T)}{\etasc}=\frac{2^{2/3}}{\pi^{3/2}}\,e^{S_0/3}\,\cE(\nu,\theta),\qquad
\cE(\nu,\theta)=\frac{\pi^{3/2}}{2}\,\frac{1-e^{-\nu/\theta}}{\nu}\,\cP(\nu,\theta)\;}
\label{eq:E-def}
\end{equation}
The spectral weights $\cP$ are, in the annealed and quenched ensembles,
\begin{align}
\cP_{\rm a}(\nu,\theta)&=\frac{1}{Z_{\Ai}(\theta)}\int_{-\infty}^{\infty}\!\dd\zeta\;e^{-\zeta/\theta}
\Bigl[\hat\rho(\zeta)\hat\rho(\zeta+\nu)-\hat K(\zeta,\zeta+\nu)^2\Bigr],\qquad
Z_{\Ai}(\theta)=\frac{\theta^{3/2}}{2\sqrt\pi}\,e^{1/12\theta^3},
\label{eq:P-annealed}\\
\cP_{\rm q}(\nu,\theta)&=\Bigl\langle\frac{\sum_{m\neq n}e^{-\zeta_m/\theta}\,\delta(\nu-\zeta_n+\zeta_m)}{\sum_le^{-\zeta_l/\theta}}\Bigr\rangle_{\Ai},
\label{eq:P-quenched}
\end{align}
with $\hat K(\zeta_1,\zeta_2)=K_{\Ai}(-\zeta_1,-\zeta_2)$ and $\hat\rho(\zeta)=\hat K(\zeta,\zeta)$.
The average $\vev{\cdot}_{\Ai}$ is over the Airy point process.\footnote{The Airy point
process is the determinantal point process on the real line whose correlation functions are
determinants of the Airy kernel,
$\rho_n(\zeta_1,\dots,\zeta_n)=\det[\hat K(\zeta_i,\zeta_j)]_{i,j=1}^n$. It is the universal limit of
the eigenvalues of a large Hermitian random matrix near a soft (square-root) edge of its spectrum
\cite{Forrester1993,TracyWidom1994}. In our conventions its points $\zeta_0<\zeta_1<\dots$ are the
lowest energy levels of one member of the ensemble, in units of $1/\kappa$.}
The prefactor $e^{S_0/3}$ in \eqref{eq:E-def} is the value that the Schwarzian enhancement has
reached at the edge scale. In the Schwarzian regime the viscosity grows as the temperature is
lowered, $\eta/\etasc\simeq\sqrt2/(\pi^{3/2}\sqrt{CT})$, eq.~\eqref{eq:eta-disk-small-q}. In terms of
the edge variable, $(CT)^{-1/2}=2^{1/6}e^{S_0/3}\theta^{-1/2}$. By the time the temperature has
come down to the edge scale, $\theta=\cO(1)$, the viscosity has therefore grown to
$\eta/\etasc=\cO(e^{S_0/3})$. The function $\cE$ describes what happens to this growth at and below
the edge scale. With the normalization of \eqref{eq:E-def}, $\cE\to\theta^{-1/2}$ wherever the
Schwarzian result still applies, i.e.\ for $\theta\gg1$ and $\nu\sqrt\theta\gg1$ (see
\eqref{eq:E-overlap} below).

The annealed weight \eqref{eq:P-annealed} is a one-dimensional integral over Airy functions,
whereas the quenched weight requires the full distribution of the Airy process. It follows from two
standard facts about determinantal processes \cite{Forrester2010}. First, multiplicative
averages are Fredholm determinants:
\begin{equation}
\Bigl\langle\prod_l\bigl(1-\phi(\zeta_l)\bigr)\Bigr\rangle=\det\bigl(1-\phi\hat K\bigr).
\label{eq:fredholm-avg}
\end{equation}
Second, the process reweighted by $\prod_lw(\zeta_l)$ is again determinantal, with kernel
\begin{equation}
w\hat K\bigl(1-(1-w)\hat K\bigr)^{-1}.
\label{eq:reweight}
\end{equation}
This follows from the identity
\begin{equation}
\frac{\det\bigl(1-[1-w(1+g)]\hat K\bigr)}{\det\bigl(1-(1-w)\hat K\bigr)}
=\det\Bigl(1+g\,w\hat K\bigl(1-(1-w)\hat K\bigr)^{-1}\Bigr),
\label{eq:reweight-id}
\end{equation}
valid for an arbitrary function $g$. By \eqref{eq:fredholm-avg}, the left-hand side is the
average of $\prod_l[1+g(\zeta_l)]$ in the reweighted process, and the right-hand side is the same
average in a determinantal process with kernel \eqref{eq:reweight}.
To use these facts, write $1/Z=\int_0^\infty\dd x\,e^{-xZ}$ and note that
$e^{-xZ}=\prod_lw_x(\zeta_l)$ with $w_x(\zeta)=\exp(-xe^{-\zeta/\theta})$. The reweighted kernel
is $w_xM_x$ with $M_x=\hat K\bigl(1-(1-w_x)\hat K\bigr)^{-1}$. Hence
\begin{align}
\cP_{\rm q}(\nu,\theta)&=\int_0^\infty\!\dd x\;D(x)\!\int\!\dd\zeta\;e^{-\zeta/\theta}\,w_x(\zeta)\,w_x(\zeta')
\Bigl[M_x(\zeta,\zeta)M_x(\zeta',\zeta')-M_x(\zeta,\zeta')^2\Bigr]_{\zeta'=\zeta+\nu},
\label{eq:P-quenched-fredholm}\\
D(x)&=\vev{e^{-xZ}}=\det\bigl(1-(1-w_x)\hat K\bigr),\qquad
\vev{\log Z}=\int_0^\infty\frac{\dd x}{x}\bigl(e^{-x}-D(x)\bigr).
\label{eq:logZ-fredholm}
\end{align}
The mean energy follows in the same way,
$\vev{\beta E}=\theta^{-1}\int\dd x\,D(x)\int\dd\zeta\,\zeta\,e^{-\zeta/\theta}w_xM_x(\zeta,\zeta)$,
and the quenched entropy is $S_{\rm q}=\vev{\log Z}+\vev{\beta E}$. We evaluate the Fredholm
determinants and resolvents by Nystr\"om discretization \cite{Bornemann2010}. The method and its
checks are described in Appendix~\ref{app:numerics}. The checks include the Tracy--Widom
distribution and an exact sum rule, satisfied to $10^{-8}$--$10^{-12}$.

\subsection{Entropy}
\label{sec:edge-entropy}

The annealed entropy follows from \eqref{eq:Z-airy}:
\begin{equation}
S_{\rm a}(\theta)=\frac32\log\theta+\frac32-\log2\sqrt\pi-\frac{1}{6\theta^3}.
\label{eq:S-annealed}
\end{equation}
It is negative for $\theta<\theta_{\rm a}=0.9671$ and tends to $-\infty$ as $\theta\to0$. A
negative entropy cannot describe any single quantum system. The reason is that at low
temperature $Z_{\Ai}$ is dominated by rare samples with one eigenvalue at $\zeta_*=-1/4\theta^2$,
deep in the forbidden region. This is the saddle \eqref{eq:saddle}. Such samples have an
exponentially small probability, of order $\exp(-\frac43|\zeta_*|^{3/2})=e^{-1/6\theta^3}$, but
their Boltzmann weight $e^{-\zeta_*/\theta}=e^{1/4\theta^3}$ is exponentially large. The product
of the two is the factor $e^{1/12\theta^3}$ of \eqref{eq:Z-airy}. The average
$\vev{Z}$ is therefore not representative of a typical member of the ensemble, and its logarithm
is not a free energy. This is the edge version of the known pathology of the annealed free
energy \cite{EngelhardtFischettiMaloney2020,Johnson2020b,Johnson2021q,JanssenMirbabayi2021}.

\begin{figure}[!ht]
  \centering
  \includegraphics[width=\textwidth]{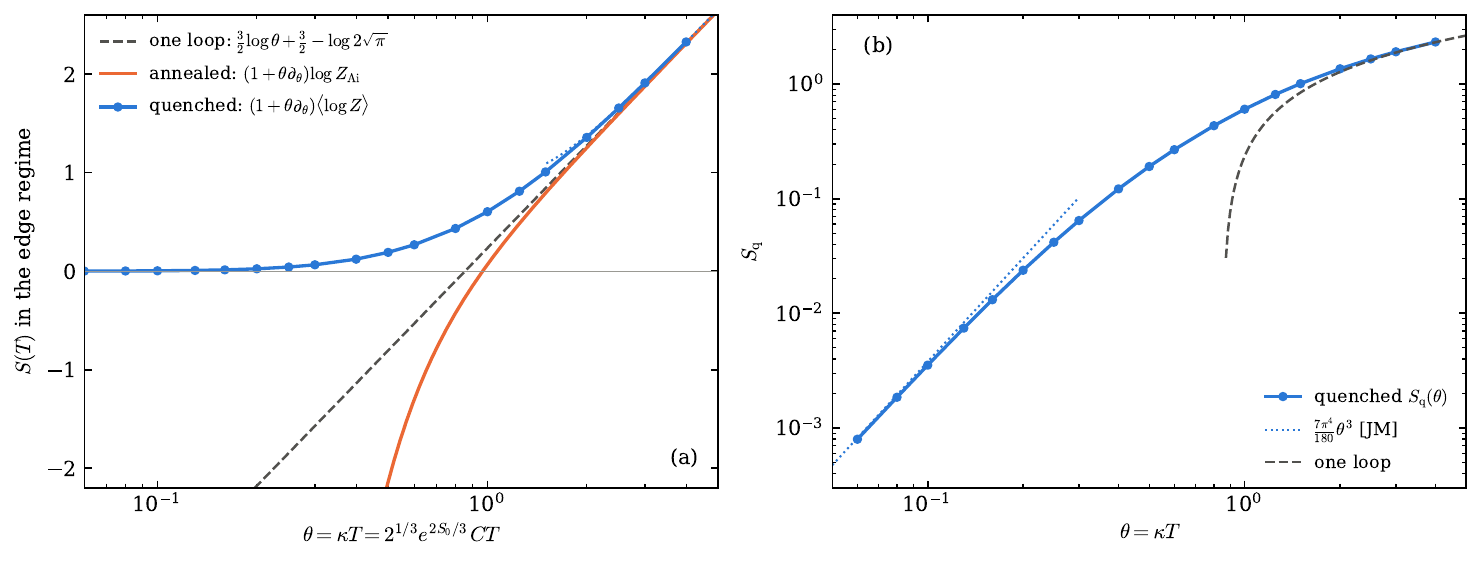}
  \caption{Entropy in the edge regime. (a) One-loop (Schwarzian), annealed
  \eqref{eq:S-annealed} and quenched (Fredholm, dots) entropy. The dotted line is the
  large-$\theta$ asymptotics \eqref{eq:S-quenched}. (b) The quenched entropy on
  logarithmic axes, compared with its low-temperature asymptotics
  $\frac{7\pi^4}{180}\theta^3$, the second line of \eqref{eq:S-quenched} (dotted). The label
  ``JM'' in the legend refers to \cite{JanssenMirbabayi2021}, where this law was
  derived.}
  \label{fig:airy-entropy}
\end{figure}

The quenched entropy is the physical one. It is positive and monotonic
(Figure~\ref{fig:airy-entropy}), with
\begin{equation}
S_{\rm q}(\theta)=
\begin{cases}
\frac32\log\theta+\frac32-\log2\sqrt\pi+\frac{5}{6\theta^3}+\cO(\theta^{-6}),&\theta\gg1,\\[3pt]
\frac{7\pi^4}{180}\,\theta^3\bigl(1+\cO(\theta^2)\bigr),&\theta\ll1.
\end{cases}
\label{eq:S-quenched}
\end{equation}
The coefficient $\frac56$ of $\theta^{-3}$ in the first line is the sum of two contributions.
The annealed entropy \eqref{eq:S-annealed} contributes $-\frac16$. The double-trumpet correction
\eqref{eq:quenched-DT} contributes $+1$: in the edge limit $CT\to0$ it reads
$S_{\rm q}-S_{\rm a}\simeq e^{-2S_0}/[2(CT)^3]=1/\theta^3$, because
$e^{-2S_0}/(CT)^3=2/\theta^3$ by \eqref{eq:edge-variables}. The Fredholm results confirm the first
line of \eqref{eq:S-quenched}
at $\theta=2,3,4$, with residuals $0.02$, $0.002$ and $0.0005$. The small-$\theta$ law reflects
the non-degenerate ground state and the quadratic repulsion of the first gap. Its coefficient
was derived in \cite{JanssenMirbabayi2021}. It can be understood as follows. For
$\theta\ll1$ only the two lowest levels of a sample are thermally populated, and only if their
gap $g=\zeta_1-\zeta_0$ is of order $\theta$. The density of small gaps in the Airy process is
\begin{equation}
p(g)\simeq\frac{g^2}{2}\qquad(g\ll1),
\label{eq:gap}
\end{equation}
as we have checked numerically from the joint distribution of the two lowest levels: we
find $p(g)/g^2=0.4995$ at $g=0.05$. A two-level system with gap $g$ has entropy
$\mathsf s(g/\theta)$, with $\mathsf s(x)=\log(1+e^{-x})+x/(e^x+1)$. Therefore
$S_{\rm q}\simeq\int_0^\infty\dd g\,p(g)\,\mathsf s(g/\theta)
=\frac12\theta^3\int_0^\infty\dd x\,x^2\,\mathsf s(x)=\frac{7\pi^4}{180}\,\theta^3$.
Our values $S_{\rm q}/\theta^3=3.53$, $3.62$, $3.69$
at $\theta=0.1$, $0.08$, $0.06$ approach $7\pi^4/180=3.788$ with a $\theta^2$ correction, an
independent check of the numerics. As $\theta\to0$, $\vev{\log Z}\to-\vev{\zeta_0}/\theta$ with
$\vev{\zeta_0}=1.7711$, the mean of the Tracy--Widom distribution \cite{TracyWidom1994}.
Table~\ref{tab:airy} in Appendix~\ref{app:numerics} collects selected values.

\subsection{The viscosity scaling function}
\label{sec:edge-visc}

\begin{figure}[!ht]
  \centering
  \includegraphics[width=\textwidth]{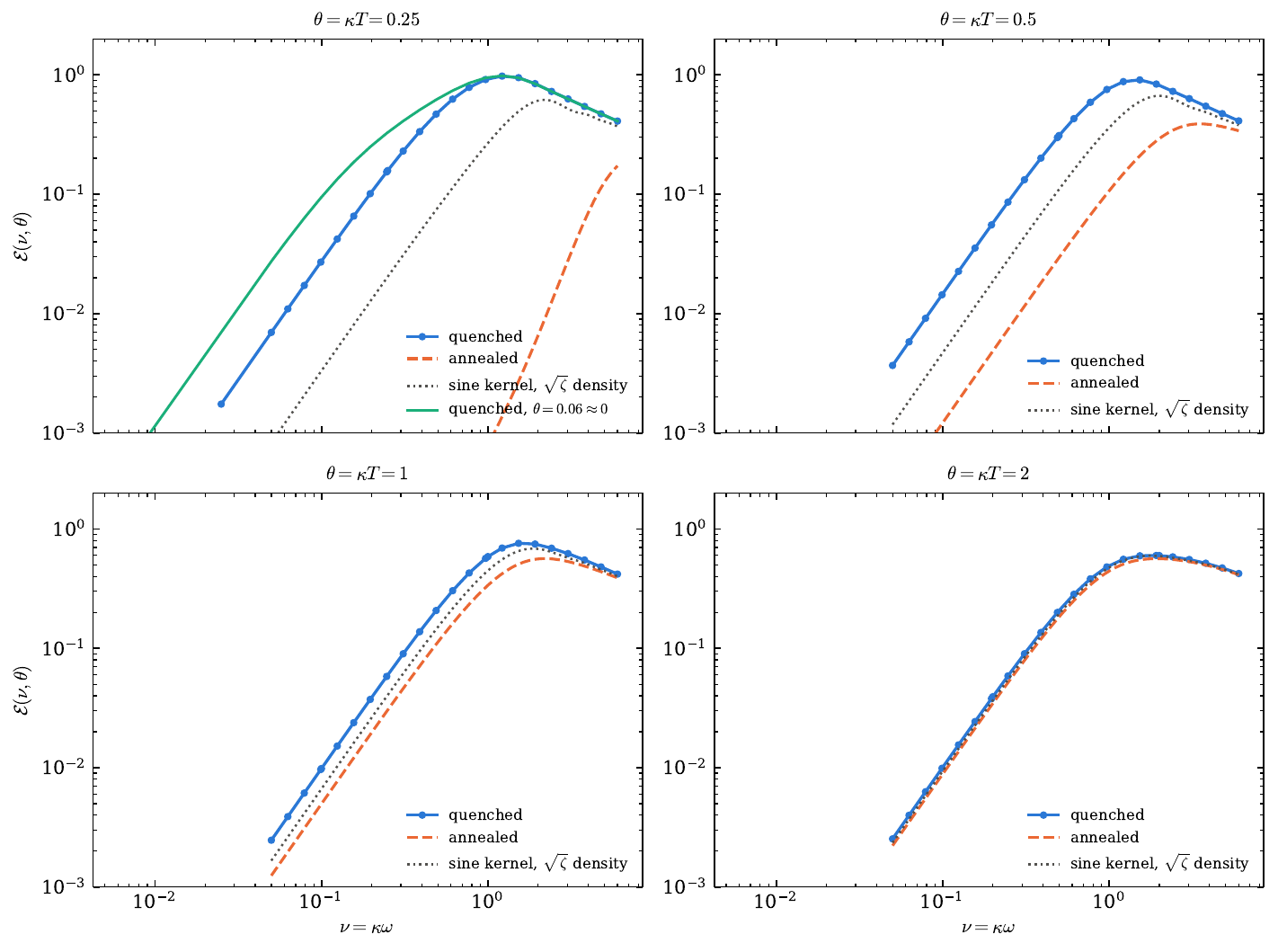}
  \caption{The edge scaling function $\cE(\nu,\theta)$ at $\theta=0.25,\,0.5,\,1,\,2$:
  quenched (dots), annealed (dashed), and the sine kernel with the density $\sqrt\zeta/\pi$
  (dotted), which is the edge limit of the bulk formula \eqref{eq:G-corrected}. In the first
  panel the quenched function at $\theta=0.06$ approximates the zero-temperature limit
  \eqref{eq:E-zeroT}.}
  \label{fig:airy-viscosity}
\end{figure}

Figure~\ref{fig:airy-viscosity} shows $\cE(\nu,\theta)$. At fixed $\theta$ it rises from zero at
small $\nu$, where level repulsion suppresses the response. It peaks at $\nu\sim1$--$2$, i.e.\ at
frequencies of the order of the spacing between the lowest levels, and it decays slowly at large
$\nu$, where the Kubo factor $(1-e^{-\nu/\theta})/\nu$ takes over. Five features are worth
noting.

\emph{Matching to the bulk.} For $\theta\to\infty$ at fixed $\nu\sqrt\theta$, so that
$\nu/\theta\to0$, both $\cE_{\rm a}$ and $\cE_{\rm q}$ approach
\begin{equation}
\cE\;\to\;\theta^{-1/2}\Bigl[1-\frac{F(\nu\sqrt\theta)}{\nu\sqrt\theta}\Bigr].
\label{eq:E-overlap}
\end{equation}
This is \eqref{eq:DS-small-q} written in edge variables ($\sqrt2y=\nu\sqrt\theta$). We have checked
it numerically to $2\times10^{-4}$ at $\theta=30$ and $\theta=60$. The bulk double-scaling
function and the edge function thus describe overlapping regions of a single crossover.

\emph{Level repulsion.} At fixed $\theta$ and $\nu\to0$, $\cP\propto\nu^2$ and
$\cE\propto\nu^2/\theta$. The DC viscosity vanishes in the edge regime as well.

\emph{Zero temperature.} As $\theta\to0$ at fixed $\nu$ a typical sample is in its ground state
$\zeta_0$, and the quenched function becomes the absorption spectrum of the ground state,
\begin{equation}
\cE_{\rm q}(\nu,0)=\frac{\pi^{3/2}}{2}\,\frac{\rho_{\rm exc}(\nu)}{\nu},\qquad
\rho_{\rm exc}(\nu)=\Bigl\langle\sum_{n\ge1}\delta\bigl(\nu-\zeta_n+\zeta_0\bigr)\Bigr\rangle_{\Ai}.
\label{eq:E-zeroT}
\end{equation}
It is linear in $\nu$ at small $\nu$ and approaches $\sqrt\pi/(2\sqrt\nu)$ at large $\nu$.

\emph{Low temperature and low frequency.} For $\theta,\nu\ll1$ the response is that of the two
lowest levels. A sample whose first gap equals $\nu$ absorbs at frequency $\nu$ with the
Boltzmann weight $1/(1+e^{-\nu/\theta})$ of its ground state. Hence
$\cP_{\rm q}\simeq p(\nu)/(1+e^{-\nu/\theta})$ with the gap density \eqref{eq:gap}, and
\begin{equation}
\cE_{\rm q}(\nu,\theta)\simeq\frac{\pi^{3/2}}{4}\,\nu\,\tanh\frac{\nu}{2\theta},\qquad\text{i.e.}\qquad
\frac{\eta}{\etasc}\simeq\frac12\,e^{S_0}C\omega\,\tanh\frac{\omega}{2T}\qquad(\omega,T\ll1/\kappa),
\label{eq:E-lowT}
\end{equation}
with relative corrections of order $\nu^2$. This contains the two previous limits,
$\cE_{\rm q}\simeq\frac{\pi^{3/2}}{8}\,\nu^2/\theta$ for $\nu\ll\theta$ and
$\cE_{\rm q}(\nu,0)\simeq\frac{\pi^{3/2}}{4}\,\nu$, and it agrees with the Fredholm results: at
$\theta=0.1$ it gives $\cE_{\rm q}(\theta,\theta)=0.0643$ and $\cE_{\rm q}(0.1\,\theta,\theta)=0.00070$,
against $0.0641$ and $0.00070$ in Table~\ref{tab:airy}. In the limit $T\to0$ the viscosity
becomes linear in the frequency, $\eta/\etasc\simeq\frac12e^{S_0}C\omega$: it is the absorption
rate of the ground state into the first excited level.

\emph{Annealed versus quenched.} For $\theta\lesssim1$ the annealed function is smaller than
the quenched one, and for $\theta\ll1$ it is exponentially smaller. The annealed normalization is dominated by single-eigenvalue
excursions into the forbidden region, $Z_{\Ai}\sim e^{1/12\theta^3}$. The annealed numerator
requires two eigenvalues in the forbidden region, at distance $\nu$, which gives only
$e^{1/48\theta^3}$. Hence $\log\cE_{\rm a}\simeq-1/16\theta^3$ for $\theta\ll1$, at leading
exponential order only. The
quenched function, which describes a typical sample, remains of order one.

\subsection{\texorpdfstring{$\eta/s$ across all temperatures}{eta/s across all temperatures}}
\label{sec:edge-etas}

Combining \eqref{eq:E-def} with the quenched entropy and $S^{\rm s.c.}\to S_0$ gives
\begin{equation}
\boxed{\;4\pi\,\frac{\eta}{s}=\frac{2^{2/3}}{\pi^{3/2}}\;S_0\,e^{S_0/3}\;\frac{\cE_{\rm q}(\kappa\omega,\kappa T)}{S_{\rm q}(\kappa T)}\;}
\label{eq:eta-s-edge}
\end{equation}
This is the ratio of two quenched averages, $\vev{\eta_H}/\vev{s_H}$, where $\eta_H$ and
$s_H$ are the viscosity and the entropy density of a single member $H$ of the ensemble. It is
the continuation to the edge of the bulk ratio \eqref{eq:eta-s-DS}, in which both quantities are
self-averaging. It is not the average $\vev{\eta_H/s_H}$ of the single-sample ratio, from which
it differs at order one in the edge regime; we return to this point at the end of this
section.
The ratio is large for two distinct reasons. The factor $e^{S_0/3}$ is the saturated
Schwarzian enhancement of $\eta$. The factor $S_0/S_{\rm q}$ is the collapse of the
thermal entropy relative to the Bekenstein--Hawking value that normalizes $\etasc$. At
fixed $\omega/T$ and $\theta\to0$ one has $\cE_{\rm q}\propto\theta$ by \eqref{eq:E-lowT}
and $S_{\rm q}\propto\theta^3$, so $4\pi\eta/s\propto S_0e^{S_0/3}/(\kappa T)^2$; explicitly,
\begin{equation}
4\pi\,\frac{\eta}{s}\simeq\frac{90}{7\pi^4}\,\frac{S_0\,e^{S_0}C\omega}{(\kappa T)^3}\,\tanh\frac{\omega}{2T}
\qquad(\omega,T\ll1/\kappa).
\label{eq:eta-s-lowT}
\end{equation}
The divergence
of the one-loop result at $\theta_0$ is gone. It is replaced by a smooth growth controlled
by the vanishing entropy of a system with a non-degenerate ground state. The power of $T$ at
fixed $\omega/T$ does not depend on the symmetry class. For Dyson index $\beta_{\rm D}$ the
density of small gaps is $p(g)\propto g^{\beta_{\rm D}}$, so that
$\cE_{\rm q}\propto\nu^{\beta_{\rm D}-1}\tanh(\nu/2\theta)$ and
$S_{\rm q}\propto\theta^{\beta_{\rm D}+1}$ \cite{JanssenMirbabayi2021}, and the ratio scales as
$\theta^{-2}$ at fixed $\nu/\theta$.

\begin{figure}[!ht]
  \centering
  \includegraphics[width=\textwidth]{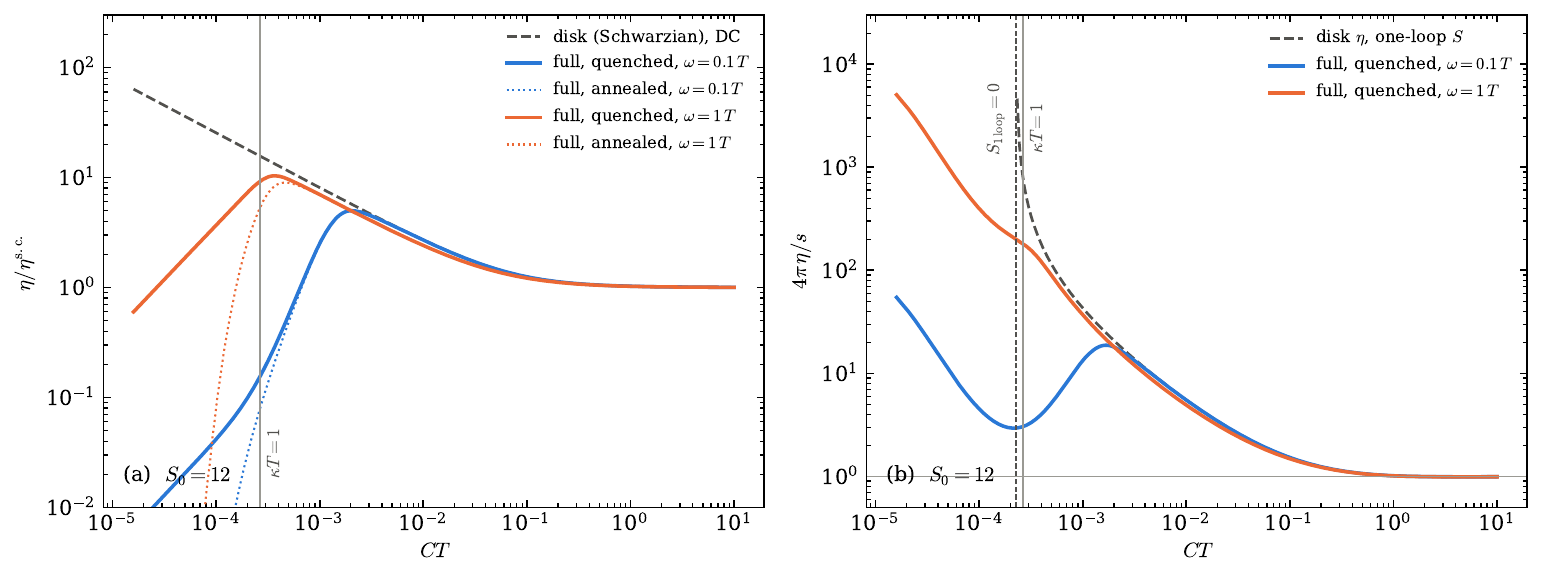}
  \caption{All regimes at $S_0=12$. (a) $\eta/\etasc$ at $\omega=0.1\,T$ and $\omega=T$: disk
  result (dashed), full quenched composite (solid) and annealed composite (dotted). (b)
  $4\pi\eta/s$: the one-loop result of \cite{KanargiasKiritsisMurthyPapadoulakiPorfyriadis2025}
  (dashed), which diverges where $S_{\rm disk}=0$, and the full quenched result. The annealed
  ratio is not shown because the annealed entropy is negative for $\kappa T<0.967$.}
  \label{fig:composite}
\end{figure}

Figure~\ref{fig:composite} assembles all regimes at $S_0=12$ for $\omega=0.1\,T$ and $\omega=T$.
We use the uniform composite
\begin{equation}
\eta=\eta_{\rm bulk}+\eta_{\rm edge}-\eta_{\rm overlap}.
\label{eq:composite}
\end{equation}
Here
$\eta_{\rm bulk}$ is \eqref{eq:G-corrected} at finite $S_0$ with the exact $f(k,\omega)$ and
KMS factor, $\eta_{\rm edge}$ is \eqref{eq:E-def}, and $\eta_{\rm overlap}$ is their common
limit (dotted curves in Figure~\ref{fig:airy-viscosity}). The entropy is composed in the
same way. From high to low temperature one sees, in turn:
\begin{itemize}
\item the semiclassical value;
\item the Schwarzian growth $(CT)^{-1/2}$;
\item its saturation once $\Delta(T)$ reaches $\omega$;
\item the edge regime.
\end{itemize}
For $\omega=0.1\,T$ the viscosity peaks at $\eta\simeq5.0\,\etasc$ at $CT\simeq2\times10^{-3}$.
For this frequency the quenched $4\pi\eta/s$ is finite at
$\theta_0$ and passes through a minimum of about $2.9$ near $\kappa T\simeq0.8$ before the final
$T^{-2}$ growth; for $\omega=T$ it grows monotonically as the temperature is lowered.

Two statements should be kept apart here, because they concern different quantities.
The viscosity itself is bounded. At fixed $\omega$ it reaches a maximum, given by
\eqref{eq:eta-max} in the bulk regime, and then decreases; as $T\to0$ at fixed
$\omega\ll1/\kappa$ it tends to the absorption rate of the ground state,
$\eta/\etasc\simeq\frac12e^{S_0}C\omega$. At fixed $\omega/T$, as in
Figure~\ref{fig:composite}(a), it also passes through a maximum, and then vanishes linearly in
$T$, eq.~\eqref{eq:E-lowT}. The ratio $\eta/s$ of
Figure~\ref{fig:composite}(b) nevertheless grows as $T\to0$, because the entropy,
$S_{\rm q}\propto T^3$, vanishes faster than the viscosity. By \eqref{eq:eta-s-lowT} this
growth is a power law, $4\pi\eta/s\propto T^{-2}$ at fixed $\omega/T$ and $\propto T^{-3}$ at
fixed $\omega$, and it sets in only below the edge scale $T\sim1/\kappa$. It does not signal
enhanced dissipation. It expresses the fact that a system with a non-degenerate ground state
has no entropy left at $T=0$, while it still absorbs energy at any non-zero frequency. This
should be contrasted with the one-loop ratio, which diverges at the non-zero temperature
$\theta_0$ and is negative below it. The quenched ratio is finite and positive at every
non-zero temperature.

\paragraph{Ratio of averages vs average of the ratio.} The power laws just quoted
are properties of the ratio of quenched averages \eqref{eq:eta-s-edge}. The average of the
single-sample ratio behaves differently, and the two-level regime shows why. For
$\theta,\nu\ll1$ a sample whose first gap is $g$ absorbs only at $\nu=g$, with
$\cE_H=\frac{\pi^{3/2}}{2}\,\nu^{-1}\tanh(\nu/2\theta)\,\delta(\nu-g)$, and its entropy is
$S_H=\mathsf s(g/\theta)$, with the function $\mathsf s$ defined below \eqref{eq:gap}. The
samples that absorb at frequency $\nu$ are thus those with $g=\nu$, and for $\nu\sim\theta$ their
entropy is of order one. Averaging with the gap density \eqref{eq:gap} gives
\begin{equation}
\Bigl\langle\frac{\cE_H}{S_H}\Bigr\rangle\simeq\frac{\pi^{3/2}}{4}\,\frac{\nu\,\tanh(\nu/2\theta)}{\mathsf s(\nu/\theta)},
\qquad\text{i.e.}\qquad
4\pi\Bigl\langle\frac{\eta_H}{s_H}\Bigr\rangle\simeq\frac{S_0}{2}\,e^{S_0}C\omega\,\frac{\tanh(\omega/2T)}{\mathsf s(\omega/T)}
\qquad(\omega,T\ll1/\kappa),
\label{eq:ratio-of-sample}
\end{equation}
up to corrections suppressed by powers of $\theta$ and $\nu$; those of relative order $\theta$
come from transitions between excited levels. We have checked \eqref{eq:ratio-of-sample} by
direct sampling of the lowest levels at the soft edge of the Gaussian unitary ensemble, with
$1.4\times10^6$ samples: at $\theta=0.1$ and $0.1\le\nu\le0.4$ the sampled average of the ratio
agrees with it to within about $5\%$, with statistical errors of $1$--$3\%$; the deviations
grow at larger $\theta$ and $\nu$, as expected. In \eqref{eq:eta-s-lowT} the same numerator is divided instead by the mean
entropy $\frac{7\pi^4}{180}\theta^3$, which is small because only a fraction of order $\theta^3$ of
the samples has a gap $g\lesssim\theta$, and hence an appreciable entropy. At fixed $\omega/T$ the
average of the ratio therefore vanishes linearly in $T$, whereas the ratio of averages grows as
$T^{-2}$; at fixed $\omega$ they grow as $T\,e^{\omega/T}$ and as $T^{-3}$, respectively. Neither
is a sharp prediction for a single system. As stressed in Section~\ref{sec:aq}, nothing is
self-averaging in this regime, and the response of one sample is a sum of delta functions. We
use the ratio of averages \eqref{eq:eta-s-edge} because it is built from two quantities that are
separately smooth and well defined after averaging, the mean absorption and the mean entropy,
and because it is continuous with the bulk result. The statements that do not depend on this
choice are that the ratio is finite and positive at every non-zero temperature, that the
divergence of the one-loop ratio at $\theta_0$ is absent, and the behaviour of the viscosity
itself. The power laws $T^{-2}$ and $T^{-3}$ are specific to the ratio of averages.

\section{Explicit non-perturbative completions}
\label{sec:string}

The results of Sections~\ref{sec:repulsion}--\ref{sec:airy} use two properties of the matrix
integral: the sine kernel in the bulk of the spectrum and the Airy kernel at its edge. Both
belong to the soft-edge completion of the matrix integral of \cite{SaadShenkerStanford2019}, in which eigenvalues can tunnel
into the classically forbidden region $E<0$. For JT gravity this completion is not
satisfactory. In the forbidden region the effective potential of a single eigenvalue is
proportional to $e^{S_0}(\sin v-v\cos v)$, with $v=2\pi\sqrt{2C|E|}$. Its first maximum, at
$E=-1/8C$, is the eigenvalue instanton of Section~\ref{sec:normalization}. Beyond it the
potential oscillates with growing amplitude and is unbounded below, so the eigenvalue integral
must be defined by a choice of contour \cite{SaadShenkerStanford2019}. Explicit completions
that avoid this problem, and in which the spectrum is bounded from below, have been constructed
in \cite{Johnson2020,Johnson2020c,Johnson2021consistency}. This line of work has also provided
the tools to compute their spectral statistics exactly at finite $e^{S_0}$: the numerical solution of the string
equation \cite{Johnson2020,Johnson2020c}, Fredholm determinants for the lowest levels and for
the quenched free energy \cite{Johnson2021fredholm,Johnson2021q,Johnson2022microstates}, a
differential equation for the distribution of the ground-state energy \cite{Johnson2022ground},
a differential equation that determines the kernel \cite{Johnson2025kernel}, and closed
formulae for the correlators of a broad class of such models \cite{Johnson2026universal}.

In this section we follow this approach and compute our observables in these completions. The
comparison with the previous sections can be summarized as follows.
\begin{itemize}
\item The bulk double-scaling results of Section~\ref{sec:DS} are unchanged, and so is their
normalization.
\item The partition function acquires a correction with a fractional power of $e^{-S_0}$,
which is absent in the soft-edge completion. We compute it in closed form,
\eqref{eq:Z-wall}. It is controlled by the density of states at the bottom of the spectrum,
and at fixed $CT$ it is suppressed by $e^{-4S_0/3}$.
\item In the edge regime the quenched entropy and viscosity change by a few per cent. The
change comes only from the samples of the soft-edge ensemble that have a level
below the bottom of the spectrum of the new completion. Their probability is about $3\%$ for the completion of
\cite{Johnson2020}, in which the wall sits at the classical edge of the spectrum, and the
size of the change is a numerical result.
\item The annealed edge quantities change at order one. The annealed entropy remains negative
at low temperature, but for a different reason, which is explained in
Section~\ref{sec:string-edge-regime} (see \eqref{eq:Sa-wall}): with a wall, the annealed
partition function is dominated by rare samples that have a level close to the wall, instead of
a level deep in the forbidden region.
\end{itemize}

\subsection{The string equation and the kernel}
\label{sec:string-eq}

In the double-scaling limit a Hermitian matrix integral is equivalent to the quantum mechanics
of a particle on a line
\cite{GrossMigdal1990,DouglasShenker1990,BrezinKazakov1990,BanksDouglasSeibergShenker1990},
with Hamiltonian
\begin{equation}
\cH=-\hbar^2\partial_x^2+u(x),\qquad\hbar=e^{-S_0}.
\label{eq:H-string}
\end{equation}
The coordinate $x$ is the continuum label of the orthogonal polynomials of the matrix integral,
and the half-line $x<0$ plays the role of the Fermi sea of its $N$ eigenvalues
\cite{BanksDouglasSeibergShenker1990}. If $\psi(E,x)$ are the eigenfunctions of $\cH$,
normalized as $\int\dd x\,\psi(E,x)\psi(E',x)=\delta(E-E')$, the kernel of the determinantal
process of Section~\ref{sec:repulsion} is
\begin{equation}
K(E,E')=\int_{-\infty}^0\!\dd x\;\psi(E,x)\,\psi(E',x),\qquad
\vev{\rho(E)}=\int_{-\infty}^0\!\dd x\;\psi(E,x)^2 .
\label{eq:K-string}
\end{equation}
Completeness of the $\psi(E,x)$ gives $\int\dd E'\,K(E,E')K(E',E'')=K(E,E'')$, which is the
projection property behind the sum rule \eqref{eq:hole-sum}.

The potential $u(x)$ is fixed by the string equation.\footnote{The string equation is
the non-linear ordinary differential equation obeyed by $u(x)$, which is, up to normalization,
the second derivative of the free energy of the matrix model with respect to $x$. It is the
double-scaling limit of the recursion relations of the orthogonal polynomials of the matrix
integral \cite{GrossMigdal1990,DouglasShenker1990,BrezinKazakov1990}. The potential of the matrix
integral enters through the couplings $t_\ell$. Different boundary conditions on $u(x)$ define
different non-perturbative completions with the same genus expansion.}
In units with $2C=1$ define
\begin{equation}
\cR=\sum_{\ell\ge1}t_\ell R_\ell[u]+x,\qquad t_\ell=\frac{\pi^{2\ell-2}}{2\,\ell!\,(\ell-1)!},
\label{eq:R-string}
\end{equation}
where $R_\ell[u]$ are the Gel'fand--Dikii polynomials, normalized so that $R_\ell=u^\ell$ plus
terms with derivatives \cite{Johnson2020}. At leading order in $\hbar$ the derivative terms drop
out and $\sum_\ell t_\ell u^\ell=\sqrt u\,I_1(2\pi\sqrt u)/2\pi$, where $I_1$ is the modified
Bessel function. The condition $\cR=0$ then determines
the classical potential $u_0(x)$ for $x<0$, and the WKB form of \eqref{eq:K-string},
\begin{equation}
\rho_0(E)=\frac{1}{2\pi\hbar}\int_{-\infty}^0\!\dd x\,\frac{\Theta\bigl(E-u_0(x)\bigr)}{\sqrt{E-u_0(x)}}
=\frac{\sinh(2\pi\sqrt E)}{4\pi^2\hbar},
\label{eq:rho0-WKB}
\end{equation}
is the disk density \eqref{eq:rho0} at $C=1/2$. Two completions share these perturbative data.

\paragraph{Soft-edge completion.} Imposing $\cR=0$ for all $x$ defines the matrix
integral of \cite{SaadShenkerStanford2019} over all Hermitian matrices. For $x>0$ the leading-order equation
$\sum_\ell t_\ell u^\ell=-x$ requires $u<0$, where
$\sum_\ell t_\ell u^\ell=-\sqrt{|u|}\,J_1(2\pi\sqrt{|u|})/2\pi$ oscillates with growing amplitude. This is the string-equation
counterpart of the instability described above \cite{Johnson2020}. The edge limit of this
completion is the Airy kernel used in Sections~\ref{sec:normalization} and~\ref{sec:airy}.

\paragraph{Walled completion.} The completion proposed in \cite{Johnson2020} is defined
instead by the string equation
\begin{equation}
(u-E_0)\,\cR^2-\frac{\hbar^2}{2}\,\cR\cR''+\frac{\hbar^2}{4}\,(\cR')^2=0,
\qquad u\to u_0(x)\ \ (x\to-\infty),\qquad u\to E_0\ \ (x\to+\infty),
\label{eq:string-sigma}
\end{equation}
with $E_0=0$. This is the string equation of complex matrix models and of type 0A
minimal strings \cite{DalleyJohnsonMorris1992,KlebanovMaldacenaSeiberg2003}, with a vanishing
right-hand side. For $x<0$ the first term dominates
and $\cR$ vanishes order by order in $\hbar$. The asymptotic expansion of $u(x)$, and with it
the genus expansion, is therefore the same as in the soft-edge completion. For $x>0$ the
potential tends to $E_0$. The spectrum of $\cH$ is bounded below by $E_0$, and $\rho(E)=0$ for
$E<E_0$ \cite{Johnson2020}: the completion has a hard wall at $E=E_0$. We keep $E_0$ as a free
parameter, since the position of the wall is not fixed by perturbation theory. It labels a
one-parameter family of completions, and the choice $E_0=0$ of \cite{Johnson2020}
puts the wall at the
classical edge.

\subsection{The edge limit: a conditioned Airy process}
\label{sec:string-edge}

We now take the edge limit of the two completions. This limit keeps only a neighbourhood of
$x=0$ of width $\hbar^{2/3}$, in which $u$ is small and only the term $\ell=1$ of
\eqref{eq:R-string} survives, $\cR\simeq\frac12u+x$. It is convenient to measure all quantities in
edge units. Since $2C=1$ in this section, $\kappa=(2\hbar)^{-2/3}$, and we set
\begin{equation}
E=\frac{\zeta}{\kappa},\qquad E_0=\frac{\sigma}{\kappa},\qquad u=\frac{\hat u}{\kappa},\qquad
x=\Bigl(\frac{\hbar^2}{2}\Bigr)^{1/3}\hat x,\qquad \cR=\Bigl(\frac{\hbar^2}{2}\Bigr)^{1/3}\hat\cR .
\label{eq:edge-rescaling}
\end{equation}
From now on we drop the hats. The energies $\zeta$ are the edge variables of
\eqref{eq:edge-variables}. The Hamiltonian \eqref{eq:H-string} becomes
$\kappa\,\cH=-\partial_x^2+u(x)$, so that effectively $\hbar=1$, and $\cR=u+x$. In these variables
the two completions are the following.
\begin{itemize}
\item \emph{Soft edge.} The condition $\cR=0$ gives $u=-x$ for all $x$. The eigenfunctions are
Airy functions, $\psi(\zeta,x)=\Ai(-x-\zeta)$, and \eqref{eq:K-string} is the Airy kernel
$\hat K(\zeta,\zeta')$ of Section~\ref{sec:airy}, in the integral representation of
\eqref{eq:airy-kernel}.
\item \emph{Wall at $\sigma$.} The string equation \eqref{eq:string-sigma} becomes
\begin{equation}
(u-\sigma)\,\cR^2-\tfrac12\,\cR\cR''+\tfrac14\,(\cR')^2=0,\qquad \cR=u+x,
\label{eq:P34}
\end{equation}
with the boundary conditions $u\simeq-x$ for $x\to-\infty$ and $u\to\sigma$ for $x\to+\infty$.
Written as an equation for $\cR$ alone, it is the Painlev\'e XXXIV equation
$\cR''=\cR'^2/2\cR+2\cR^2-2(x+\sigma)\cR$.
\end{itemize}

For the walled completion the solution can be written down explicitly. Substituting
$\cR=2q(s)^2$ with $s=-x-\sigma$ turns \eqref{eq:P34} into the Painlev\'e II equation
$q''=sq+2q^3$. The boundary conditions select the Hastings--McLeod solution \cite{HastingsMcLeod1980}, $q\simeq\Ai(s)$
for $s\to+\infty$ and $q\simeq\sqrt{-s/2}$ for $s\to-\infty$, which is the solution that
determines the Tracy--Widom distribution \eqref{eq:TW}. Hence
\begin{equation}
u_\sigma(x)=-x+2\,q(-x-\sigma)^2 .
\label{eq:u-HM}
\end{equation}
The position of the wall only shifts the argument of $q$.

The kernel of the walled completion follows from a simple observation. A hard wall at $\sigma$
restricts the eigenvalue integral to $\zeta_i\ge\sigma$. After normalization, the restricted
measure is the original one conditioned on the event that no eigenvalue lies below $\sigma$.
This holds at finite $N$, and it survives the edge limit because the probability of the event
stays finite. In the edge limit, the walled completion is therefore the Airy process
conditioned to have no points below $\sigma$. This is the reweighting \eqref{eq:reweight} with
$w=1-\chi_\sigma$, where $\chi_\sigma$ is the indicator function of $(-\infty,\sigma)$. The
conditioned kernel is
\begin{equation}
K_\sigma=\hat K+\hat K\chi_\sigma\bigl(1-\chi_\sigma\hat K\chi_\sigma\bigr)^{-1}\chi_\sigma\hat K
\quad\text{on }[\sigma,\infty),
\label{eq:K-cond}
\end{equation}
and the normalization is
\begin{equation}
F_2(-\sigma)=\det\bigl(1-\chi_\sigma\hat K\bigr),
\end{equation}
the probability that the lowest level of the Airy process lies above $\sigma$,
eq.~\eqref{eq:TW}. This kernel describes a hard edge that meets a soft edge. Kernels of
this type were obtained in \cite{ClaeysKuijlaars2008} in terms of Painlev\'e transcendents of the
Hastings--McLeod type, in line with \eqref{eq:u-HM}. The identification requires that the string equation has no constant on
its right-hand side, i.e.\ no factor $\prod_i(\lambda_i-E_0)^\Gamma$ in the eigenvalue measure.
This is the case for the completion of \cite{Johnson2020}.

Differentiating the restricted integral with respect to the position of the wall gives the
density at the wall,
\begin{equation}
\rho_\sigma(\sigma)=-\partial_\sigma\log F_2(-\sigma)=\frac{F_2'(-\sigma)}{F_2(-\sigma)} .
\label{eq:rho-wall}
\end{equation}
It is the probability density of the lowest Airy level at $\sigma$, divided by the probability
that this level lies above $\sigma$. The string equation gives the same quantity. By
\eqref{eq:u-HM} and the Tracy--Widom formula
$\partial_s\log F_2(s)=\int_s^\infty q(t)^2\,\dd t$,
\begin{equation}
\frac12\int_{-\infty}^0\!\dd x\;\cR(x)=\int_{-\sigma}^\infty\!\dd s\;q(s)^2=\rho_\sigma(\sigma).
\label{eq:R-wall}
\end{equation}
For a wall at the classical edge, $\sigma=0$, one has $F_2(0)=0.96937$ and $\rho_\sigma(\sigma)=0.06909$.
For $\sigma=-1$, $F_2(1)=0.99751$ and $\rho_\sigma(\sigma)=0.00704$.

We have also checked the equivalence numerically. We solved \eqref{eq:P34} as a boundary-value
problem, computed $\int_{-\infty}^0\psi^2\dd x$, and compared it with the diagonal of
\eqref{eq:K-cond}, evaluated by the Nystr\"om method of Appendix~\ref{app:numerics}. The two
agree to all six digits computed (Table~\ref{tab:string}), and so do the two sides of
\eqref{eq:R-wall}. Figure~\ref{fig:string}(a) compares the densities. The wall removes the tail
at $\zeta<\sigma$, which carries the probability $1-F_2(-\sigma)$ of finding a level there.
For $\sigma=0$, about half of this probability reappears within $\zeta<2$. The rest is spread
over a slowly decaying tail, which we determine in the next subsection.

\begin{table}[!ht]
\centering
\small
\setlength{\tabcolsep}{4.5pt}
\begin{tabular}{cccc@{\hspace{2.2em}}cccc}
\toprule
\multicolumn{4}{c}{$\sigma=0$} & \multicolumn{4}{c}{$\sigma=-1$}\\
\cmidrule(r{2.2em}){1-4}\cmidrule{5-8}
$\zeta$ & string eq. & $K_\sigma(\zeta,\zeta)$ & Airy & $\zeta$ & string eq. & $K_\sigma(\zeta,\zeta)$ & Airy\\
\midrule
0.3 & 0.117098 & 0.117097 & 0.113565 & $-0.7$ & 0.014930 & 0.014930 & 0.014893\\
0.5 & 0.159568 & 0.159568 & 0.154808 & $\phantom{-}0.0$ & 0.067150 & 0.067150 & 0.066987\\
1.0 & 0.295242 & 0.295242 & 0.286929 & $\phantom{-}0.5$ & 0.155168 & 0.155168 & 0.154808\\
1.5 & 0.429365 & 0.429365 & 0.418898 & $\phantom{-}1.5$ & 0.419630 & 0.419630 & 0.418898\\
3.0 & 0.529488 & 0.529488 & 0.529464 & $\phantom{-}3.0$ & 0.529476 & 0.529476 & 0.529464\\
5.0 & 0.723221 & 0.723221 & 0.722222 & $\phantom{-}5.0$ & 0.722329 & 0.722329 & 0.722222\\
\bottomrule
\end{tabular}
\caption{One-level density of the walled completion at the edge. ``String eq.'' is
$\int_{-\infty}^0\psi^2\,\dd x$ with the solution of \eqref{eq:P34}. $K_\sigma(\zeta,\zeta)$ is the
conditioned Airy kernel \eqref{eq:K-cond}. The last column of each block is the soft-edge
(Airy) density. The density at the wall is $\rho_\sigma(\sigma)=0.069091$ for $\sigma=0$ and
$0.007041$ for $\sigma=-1$, both from $\frac12\int_{-\infty}^0\cR\,\dd x$ and from
$-\partial_\sigma\log F_2(-\sigma)$.}
\label{tab:string}
\end{table}

\subsection{The partition function}
\label{sec:string-Z}

At temperatures large compared with the edge spacing the effect of the wall can be computed in
closed form. Write the partition function as a trace over the Fermi sea,
\begin{equation}
Z(\beta)=\int\!\dd E\,e^{-\beta E}\vev{\rho(E)}=\int_{-\infty}^0\!\dd x\;\bra{x}e^{-\beta\cH}\ket{x}.
\end{equation}
On the Fermi sea $x<0$ the two completions differ only in a transition region around $x=0$,
of width $\hbar^{2/3}$, where $u(x)$ interpolates between $u_0(x)$ and $E_0$. When the heat
kernel varies slowly across this region, i.e.\ when $\hbar\sqrt\beta\ll\hbar^{2/3}$ or
$\theta\gg1$, we can use $\bra{x}e^{-\beta\cH}\ket{x}\simeq e^{-\beta u(x)}/(2\hbar\sqrt{\pi\beta})$.
Expanding in the difference of the potentials of the walled and soft-edge completions gives
\begin{equation}
Z_\sigma-Z_{\rm soft}\simeq-\frac{\sqrt\beta}{2\hbar\sqrt\pi}\int_{-\infty}^0\!\dd x\,
\bigl(u_\sigma(x)-u_{\rm soft}(x)\bigr).
\end{equation}
In edge units $\hbar=1$, $\beta=1/\theta$, $u_{\rm soft}=-x$ and $Z_{\rm soft}=Z_{\Ai}$, so the
integrand is $\cR$. By \eqref{eq:R-wall},
\begin{equation}
Z_\sigma(\theta)=Z_{\Ai}(\theta)-\frac{\rho_\sigma(\sigma)}{\sqrt{\pi\theta}}
\bigl[1+\delta(\theta)\bigr],\qquad\theta\gg1 .
\label{eq:Z-wall-edge}
\end{equation}
For $\sigma=0$ the numerical Laplace transform of $K_\sigma(\zeta,\zeta)$ gives
$(Z_{\Ai}-Z_\sigma)\sqrt\theta=0.03918$, $0.03902$, $0.03899$ at $\theta=5$, $12$, $30$. The
prediction is $\rho_\sigma(\sigma)/\sqrt\pi=0.03898$. The deviations scale as $\theta^{-2}$,
$\delta(\theta)\simeq0.13/\theta^2$, which indicates that the term of order $\theta^{-1}$ in the
gradient expansion of the heat kernel cancels. In terms of the density, the same argument gives a
smooth tail above the wall, $\rho_\sigma(\zeta)-\rho_{\Ai}(\zeta)\simeq\rho_\sigma(\sigma)/(2\pi
\zeta^{3/2})$ after averaging over the oscillations. This tail carries the part of the
probability removed below the wall that does not reappear near it, and it ensures that
$Z_\sigma-Z_{\Ai}\to0$ as $\theta\to\infty$.

Since $Z_{\Ai}\simeq\theta^{3/2}/2\sqrt\pi$ and $\theta=2^{1/3}e^{2S_0/3}CT$, this is a
correction of relative order $e^{-4S_0/3}$. The derivation only uses the transition region, so
it holds for all $CT\gg e^{-2S_0/3}$, and the disk partition function provides the
normalization:
\begin{equation}
\boxed{\;Z_\sigma(T)=Z_{\rm disk}(T)\biggl[1-2^{1/3}\rho_\sigma(\sigma)\,
\frac{e^{-4S_0/3}\,e^{-2\pi^2CT}}{(CT)^2}+\cO\Bigl(\frac{e^{-2S_0}}{(CT)^3}\Bigr)\biggr]\;}
\label{eq:Z-wall}
\end{equation}
For $\sigma=0$ the coefficient is $2^{1/3}\rho_\sigma(\sigma)=0.0870$.

Three remarks are in order. First, the correction is a fractional power of $e^{-S_0}$. It is
not part of the genus expansion, which is generated by the asymptotic expansion of $u(x)$ away
from $x=0$ and is the same in both completions. It comes from the transition region near
$x=0$, where that expansion breaks down. In the soft-edge completion $x=0$ is not a special point for $u$, there
is no transition region, and the correction is absent, in agreement with \eqref{eq:Z-airy}.
Second, at fixed $CT$ the correction vanishes as $S_0\to\infty$. The normalization of the bulk
double-scaling limit by $Z_{\rm disk}$ in Section~\ref{sec:normalization} therefore holds for
both completions. The same is true of its numerator, since the sine kernel is fixed by the
local density of states and the two completions differ in the bulk only by the tail found
above. Third, the ratio of the wall term to the genus-one term $e^{-2S_0}/24(CT)^3$ of
\eqref{eq:Z-airy} is $24\rho_\sigma(\sigma)\,\theta$. The wall term therefore dominates
throughout the window $e^{-2S_0/3}\ll CT\ll1$, where both asymptotic forms are valid. (For
$\sigma=0$ the ratio formally reaches one at $\theta\simeq0.6$, where neither form applies.) The
leading finite-$S_0$ correction to the low-temperature thermodynamics thus distinguishes the
two completions. This complements the caveat on bulk--edge matching in
Section~\ref{sec:normalization}: in a completion with a hard wall, the edge region produces
fractional powers of $e^{-S_0}$, and they are fixed by the exact edge kernel of that
completion.

\subsection{The edge regime}
\label{sec:string-edge-regime}

All the formulas of Section~\ref{sec:airy} hold for the walled completion with
$\hat K\to K_\sigma$. Equivalently, one keeps $\hat K$, multiplies the weights $w_x$ of
Section~\ref{sec:airy} (where $x$ is the Laplace variable, not the string coordinate) by
$1-\chi_\sigma$ and divides by $F_2(-\sigma)$. For any quantity $A$ defined sample by sample,
the conditioning gives the exact relation
\begin{equation}
\vev{A}_\sigma-\vev{A}_{\Ai}=\bigl[1-F_2(-\sigma)\bigr]\Bigl(\vev{A}_{\zeta_0\ge\sigma}-\vev{A}_{\zeta_0<\sigma}\Bigr),
\label{eq:cond-diff}
\end{equation}
where $\zeta_0$ is the lowest level and the subscripts denote conditional averages in the Airy
process. It holds at any temperature. Quenched quantities are of this form, with $A$ the
entropy $\log Z+\beta E$ of a single sample, or the ratio in \eqref{eq:P-quenched} for the
viscosity. The walled completion therefore changes them only through the rare samples of the
soft-edge ensemble that have a level below the wall.
These samples have probability $1-F_2(0)=3.1\%$ for $\sigma=0$ and $1-F_2(1)=0.25\%$ for
$\sigma=-1$. Annealed quantities are not of this form, because $\vev{Z}$ is dominated by
atypical samples. 

\begin{figure}[!ht]
  \centering
  \includegraphics[width=\textwidth]{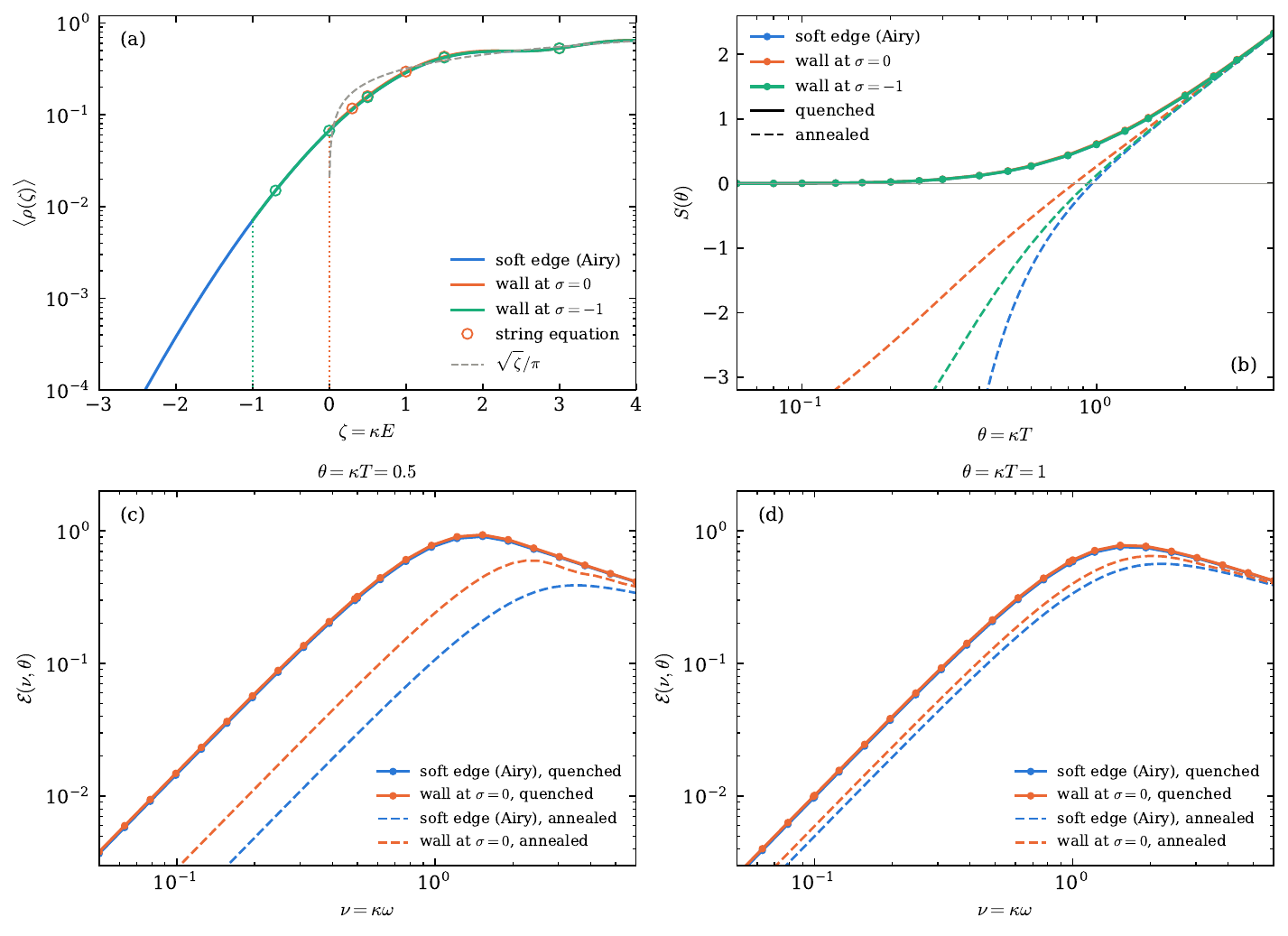}
  \caption{Soft-edge (Airy) and walled completions at the edge. (a) One-level density, on a
  logarithmic scale, for walls at $\sigma=0$ and $\sigma=-1$ (dotted lines). The wall removes
  the tail of the Airy density below $\sigma$. Open circles: solution of the string equation
  \eqref{eq:P34}. (b) Quenched (solid, Fredholm) and annealed (dashed) entropy. (c,d) The edge
  scaling function $\cE(\nu,\theta)$ at $\theta=0.5$ and $\theta=1$, quenched (solid) and
  annealed (dashed), for the soft-edge completion and the wall at $\sigma=0$.}
  \label{fig:string}
\end{figure}

Figure~\ref{fig:string} and Table~\ref{tab:string-edge} show the results.
\begin{itemize}
\item \emph{Quenched entropy.} At $\theta=1$, $S_{\rm q}=0.6034$ in the soft-edge completion,
$0.6154$ for $\sigma=0$ and $0.6046$ for $\sigma=-1$. For $\sigma=0$ the relative change
decreases from $3.6\%$ at $\theta=0.06$ to $0.3\%$ at $\theta=4$. For $\sigma=-1$ it is about
ten times smaller (Table~\ref{tab:string-edge}). The low-temperature law
$S_{\rm q}\propto\theta^3$ persists, since it only requires a non-degenerate ground state and
quadratic repulsion of the first gap. Its coefficient changes by a few per cent.
\item \emph{Quenched viscosity.} For $\sigma=0$, $\cE_{\rm q}(\theta,\theta)$ increases by $3.2\%$
at low temperature and by $0.4\%$ at $\theta=4$. The low-temperature value is close to
$[1-F_2(0)]/F_2(0)=3.2\%$, the value expected from \eqref{eq:cond-diff} if the rare samples
with a level below the wall contribute little to the absorption at small $\nu$. The
shape of $\cE_{\rm q}(\nu,\theta)$, with its $\nu^2$ suppression at small $\nu$ and its peak at
the spacing of the lowest levels, is unchanged (Figure~\ref{fig:string}(c,d)). Entropy and
viscosity change in the same direction, so the quenched $4\pi\eta/s$ of
\eqref{eq:eta-s-edge} at $\omega=T$ changes by at most $1.1\%$.
\item \emph{Annealed entropy.} The change is of order one at $\theta\lesssim1$. The wall
removes the saddle \eqref{eq:saddle}, which put an eigenvalue at $\zeta_*=-1/4\theta^2$, as soon
as $\zeta_*<\sigma$. At low
temperature $Z_\sigma$ is instead dominated by samples with a level within $\theta$ of the wall,
$Z_\sigma\simeq\theta\,\rho_\sigma(\sigma)\,e^{-\sigma/\theta}$. Hence
\begin{equation}
S_{\rm a}^{(\sigma)}(\theta)\simeq1+\log\bigl(\theta\,\rho_\sigma(\sigma)\bigr)
+2\theta\,\frac{\rho_\sigma'(\sigma)}{\rho_\sigma(\sigma)},\qquad\theta\to0 .
\label{eq:Sa-wall}
\end{equation}
For $\sigma=0$, $\rho_\sigma'/\rho_\sigma=1.88$ at the wall, and \eqref{eq:Sa-wall} gives
$-3.60$ at $\theta=0.1$, against the exact $-3.58$.
The annealed entropy still diverges to $-\infty$, but only logarithmically, instead of as
$-1/6\theta^3$. It is negative for $\theta<0.840$, $0.933$, $0.962$ at $\sigma=0,-1,-2$, and
the threshold approaches the soft-edge value $0.967$ as $\sigma\to-\infty$.
\item \emph{Annealed viscosity.} The annealed weight is now dominated by samples with a level at
the wall, and $\cP_{\rm a}$ tends to the density of a second level at distance $\nu$ from a
level at the wall. It is of order one instead of exponentially small, but it describes the
response of an atypical sample (Figure~\ref{fig:string}(c,d)).
\end{itemize}

\begin{table}[!ht]
\centering
\small
\begin{tabular}{c@{\quad}ccc@{\quad}ccc@{\quad}cc}
\toprule
& \multicolumn{3}{c}{$S_{\rm q}$} & \multicolumn{3}{c}{$\cE_{\rm q}(\theta,\theta)$} & \multicolumn{2}{c}{$S_{\rm a}$}\\
\cmidrule(r{0.8em}){2-4}\cmidrule(r{0.8em}){5-7}\cmidrule{8-9}
$\theta$ & Airy & $\sigma=0$ & $\sigma=-1$ & Airy & $\sigma=0$ & $\sigma=-1$ & Airy & $\sigma=0$\\
\midrule
0.10 & 0.0035 & 0.0036 & 0.0035 & 0.0641 & 0.0661 & 0.0642 & $-169.9$ & $-3.576$\\
0.20 & 0.0237 & 0.0245 & 0.0238 & 0.1269 & 0.1309 & 0.1272 & $-23.01$ & $-2.484$\\
0.30 & 0.0644 & 0.0664 & 0.0646 & 0.1883 & 0.1943 & 0.1888 & $-7.744$ & $-1.751$\\
0.50 & 0.1908 & 0.1961 & 0.1912 & 0.3102 & 0.3200 & 0.3110 & $-2.139$ & $-0.836$\\
0.80 & 0.4332 & 0.4431 & 0.4341 & 0.4897 & 0.5048 & 0.4910 & $-0.426$ & $-0.075$\\
1.00 & 0.6034 & 0.6154 & 0.6046 & 0.5863 & 0.6036 & 0.5877 & $\phantom{-}0.068$ & $\phantom{-}0.265$\\
1.50 & 1.0073 & 1.0212 & 1.0088 & 0.6566 & 0.6730 & 0.6580 & $\phantom{-}0.793$ & $\phantom{-}0.867$\\
2.00 & 1.3577 & 1.3707 & 1.3592 & 0.5977 & 0.6090 & 0.5988 & $\phantom{-}1.253$ & $\phantom{-}1.292$\\
3.00 & 1.9109 & 1.9202 & 1.9120 & 0.4957 & 0.4995 & 0.4963 & $\phantom{-}1.876$ & $\phantom{-}1.892$\\
4.00 & 2.3265 & 2.3329 & 2.3273 & 0.4307 & 0.4326 & 0.4309 & $\phantom{-}2.311$ & $\phantom{-}2.320$\\
\bottomrule
\end{tabular}
\caption{Edge-regime entropies and viscosity in the soft-edge completion (Airy) and in walled
completions with the wall at $\sigma=0$ (the choice of \cite{Johnson2020}) and $\sigma=-1$.}
\label{tab:string-edge}
\end{table}

\subsection{Summary of the comparison}
\label{sec:string-summary}

The comparison supports the picture of the previous sections and sharpens it. The quantities
that describe a typical black hole, namely the bulk double-scaling function, its
normalization and the quenched edge functions, are at most weakly sensitive to the
completion. The bulk
results are identical. The quenched edge results change only through the samples that
have a level below the wall, \eqref{eq:cond-diff}, and numerically by an amount of the order of
the probability $1-F_2(-\sigma)$ of such samples (Table~\ref{tab:string-edge}). The
annealed edge quantities depend on the
completion at order one, and in none of the completions we have considered do they give a
positive entropy at low temperature. The
partition function at fixed $CT$ distinguishes the completions only through the fractional
correction \eqref{eq:Z-wall}.

Our explicit computations are in the edge limit, where the walled completion is exactly the
conditioned Airy process. We have not solved the full JT string equation at finite $S_0$. The
statements about the bulk rest on the heat-kernel argument of Section~\ref{sec:string-Z} and
on the fact that the two completions share the asymptotic expansion of $u(x)$ for $x<0$. For
the full JT string equation the edge results receive corrections of relative order
$e^{-2S_0/3}$, as in Section~\ref{sec:airy}.

\section{Discussion}
\label{sec:discussion}

We have studied how the discreteness of the spectrum of a near-extremal black brane, as
encoded in the non-perturbative completion of JT gravity by a double-scaled matrix integral,
modifies its quantum shear viscosity and the ratio $\eta/s$. The main lesson is that, once
the extremal entropy $S_0$ is finite, the Kubo viscosity below the gap scale is no longer a
single number. It is a function of the probe frequency measured in units of the level
spacing. Three results summarize the picture. First, the strict DC limit vanishes, because
of level repulsion. Second, the Schwarzian growth of the viscosity found in
\cite{KanargiasKiritsisMurthyPapadoulakiPorfyriadis2025,CremoniniLiLiuNian2025} is cut off
when the level spacing at the thermal energy reaches the probe frequency. Third, at
temperatures $CT\sim e^{-2S_0/3}$, where the one-loop $\eta/s$ diverges, the response is
governed by the universal statistics of the lowest levels of the spectrum, and $\eta/s$ is
finite once the viscosity and the entropy are both computed for a typical member of the
ensemble, i.e.\ in the quenched prescription. These conclusions hold for the soft-edge
completion and for completions in which the spectrum is bounded from below. The explicit
formulas are collected in Table~\ref{tab:summary}, for the bulk of the spectrum, and in
Table~\ref{tab:summary-edge}, for the edge regime. In the rest of this section we discuss
their physical interpretation, the assumptions on which they rest, and some open
problems.

\begin{table}[!p]
\centering
\small
\renewcommand{\arraystretch}{1.2}
\setlength{\tabcolsep}{7pt}
\begin{tabularx}{\textwidth}{@{}>{\raggedright\arraybackslash}p{2.9cm}>{\raggedright\arraybackslash}p{4.15cm}>{\raggedright\arraybackslash}X>{\raggedleft\arraybackslash}p{1.1cm}@{}}
\toprule
\textbf{Regime} & \textbf{Conditions} & \textbf{Viscosity} $\eta/\etasc$ & \textbf{Eq.}\\
\midrule
\blockhead{\textbf{1.\ Continuous spectrum (disk)}, $\Delta(T)\ll\omega\ll T$. Here
$4\pi\eta/s=(\eta/\etasc)\,S^{\rm s.c.}/S_{\rm disk}$: it tends to $1$ for $CT\gg1$, grows as
$T$ decreases, and diverges where $S_{\rm disk}=0$.}\\
\addlinespace[5pt]
semiclassical & $CT\gg1$ & $1+\dfrac{1}{4\pi^2CT}$ & \eqref{eq:eta-disk}\\
\addlinespace[7pt]
Schwarzian & $e^{-2S_0/3}\ll CT\ll1$ & $\dfrac{\sqrt2}{\pi^{3/2}\sqrt{CT}}\Bigl[1+\dfrac{2\pi^2}{3}\,CT+\dots\Bigr]$ & \eqref{eq:eta-disk-small-q}\\
\addlinespace[4pt]
\midrule
\blockhead{\textbf{2.\ DC limit at finite $S_0$}: $\omega\to0$ at fixed $S_0$ and $T$. Both
$\eta$ and $\eta/s$ vanish.}\\
\addlinespace[5pt]
any temperature & $r\gg\sinh(2\pi k_3)/2\pi$ & $\dfrac{K_2(CT)}{24\pi^3CT\,I_1(CT)}\;e^{2S_0}(C\omega)^2$ & \eqref{eq:omega-cubed}\\
\addlinespace[4pt]
\midrule
\blockhead{\textbf{3.\ Bulk double-scaling limit}: $S_0\to\infty$ and $C\omega\to0$, with
$r=e^{-S_0}/(C\omega)$ and $CT$ fixed. In this limit $S/S^{\rm s.c.}\to1$, so that
$4\pi\eta/s=\eta_{\rm DS}/\etasc$ in every row, $\eta/s$ is given by eq.~\eqref{eq:eta-s-DS}.}\\
\addlinespace[5pt]
general & any $r$ and $CT$ & $\dfrac{\vev{k\,R_r(k)}_T}{2\pi CT}$,\quad
$R_r=1-\sinc^2\dfrac{\sinh2\pi k}{2\pi r}$ & \eqref{eq:eta-DS}\\
\addlinespace[7pt]
weak resolution & $r\to0$ & $\dfrac{\eta_{\rm disk}}{\etasc}-\dfrac{\pi r^2\,\mathcal L_2(CT)}{CT\,I_1(CT)}+\cO(r^4)$ & \eqref{eq:DS-small-r}\\
\addlinespace[7pt]
strong resolution & $r\gg\sinh(2\pi k_3)/2\pi$ & $\dfrac{K_2(CT)}{24\pi^3CT\,I_1(CT)}\;\dfrac{1}{r^2}$ & \eqref{eq:DS-large-r}\\
\addlinespace[7pt]
low temperature & $CT\ll1$;\quad $y=\sqrt{CT}/r$ & $\dfrac{\sqrt2}{\pi^{3/2}\sqrt{CT}}\Bigl[1-\dfrac{F(\sqrt2\,y)}{\sqrt2\,y}\Bigr]$ & \eqref{eq:DS-small-q}\\
\addlinespace[7pt]
maximum over $T$\newline at fixed $\omega$ & $e^{-S_0}\ll C\omega\ll e^{-2S_0/3}$;\newline reached at $CT_*=0.846\,r^2$ & $0.1736\;e^{S_0}C\omega$ & \eqref{eq:eta-max}\\
\addlinespace[7pt]
high temperature & $CT\gtrsim1$ & $\dfrac{\cI_2(CT;k_{\rm eff})}{2\pi CT\,I_1(CT)}$ & \eqref{eq:k-eff}\\
\addlinespace[3pt]
\bottomrule
\end{tabularx}
\caption{Summary of the results for the frequency-dependent viscosity $\eta(\omega)$ in the
bulk of the spectrum, $CT\gg e^{-2S_0/3}$. The edge regime is summarized in
Table~\ref{tab:summary-edge}. Each of the three blocks begins with a line that states the regime
and the corresponding ratio $4\pi\eta/s$. The rows below it list the special cases, the
conditions under which they hold, and the equation in which the result is derived. $\Delta(T)$
is the level spacing at the thermal energy, $\vev{\cdot}_T$ the average over the weight
$g_T(k)$ of \eqref{eq:C1}, $F$ Dawson's function, and $S_{\rm disk}$ the one-loop entropy
\eqref{eq:S-disk}. $k_3$ is the position of the peak of the integrand of $K_2$, with
$k_3\simeq6\pi CT$ for $CT\gtrsim1$ and $k_3\simeq\sqrt{5CT}$ for $CT\ll1$. The moments $I_n$,
$\cI_n$, $K_n$ and $\mathcal L_n$ are defined in \eqref{eq:I12}, \eqref{eq:G-cut},
\eqref{eq:DS-large-r} and \eqref{eq:DS-small-r}, and the cutoff $k_{\rm eff}$ in
\eqref{eq:k-eff}.}
\label{tab:summary}
\end{table}

\begin{table}[!p]
\centering
\small
\renewcommand{\arraystretch}{1.2}
\setlength{\tabcolsep}{7pt}
\begin{tabularx}{\textwidth}{@{}>{\raggedright\arraybackslash}p{3.3cm}>{\raggedright\arraybackslash}X>{\raggedright\arraybackslash}X>{\raggedleft\arraybackslash}p{1.1cm}@{}}
\toprule
\textbf{Regime} & \textbf{Viscosity} $\eta/\etasc$ & \textbf{Entropy and ratio} $4\pi\eta/s$ & \textbf{Eq.}\\
\midrule
\blockhead{\textbf{4.\ Edge regime}: $S_0\to\infty$ with $\theta=\kappa T$ and
$\nu=\kappa\omega$ fixed, where $\kappa=2^{1/3}Ce^{2S_0/3}$. Soft-edge (Airy) completion;
quenched averages, except in the last row.}\\
\addlinespace[5pt]
general $\nu$ and $\theta$ & $\dfrac{2^{2/3}}{\pi^{3/2}}\;e^{S_0/3}\,\cE_{\rm q}(\nu,\theta)$ &
$4\pi\dfrac{\eta}{s}=\dfrac{2^{2/3}}{\pi^{3/2}}\,S_0\,e^{S_0/3}\,\dfrac{\cE_{\rm q}}{S_{\rm q}}$ &
\eqref{eq:E-def}\newline\eqref{eq:eta-s-edge}\\
\addlinespace[7pt]
matching to the bulk,\newline $\theta\gg1$ & $\cE_{\rm q}\to\dfrac{1}{\sqrt\theta}\Bigl[1-\dfrac{F(\nu\sqrt\theta)}{\nu\sqrt\theta}\Bigr]$ &
$S_{\rm q}\to S_{\rm disk}+\dfrac{5}{6\theta^3}$ &
\eqref{eq:E-overlap}\newline\eqref{eq:S-quenched}\\
\addlinespace[7pt]
low temperature,\newline $\theta,\nu\ll1$ & $\dfrac12\,e^{S_0}C\omega\,\tanh\dfrac{\omega}{2T}$ &
$S_{\rm q}\simeq\dfrac{7\pi^4}{180}\,\theta^3$, so that\newline
$4\pi\eta/s\propto S_0\,e^{S_0/3}\,\theta^{-2}$ at fixed $\omega/T$ &
\eqref{eq:E-lowT}\newline\eqref{eq:eta-s-lowT}\\
\addlinespace[7pt]
annealed averages & $\log\cE_{\rm a}\simeq-\dfrac{1}{16\,\theta^3}$ for $\theta\ll1$ &
$S_{\rm a}<0$ for $\theta<0.967$:\newline the ratio is not defined &
\eqref{eq:S-annealed}\\
\addlinespace[4pt]
\midrule
\blockhead{\textbf{5.\ Completion with a hard wall} at $E_0=\sigma/\kappa$
(Section~\ref{sec:string}). The entries state the change relative to the soft-edge
completion.}\\
\addlinespace[5pt]
bulk (blocks 1--3) & unchanged & unchanged; $Z$ changes at relative order $e^{-4S_0/3}$ &
\eqref{eq:Z-wall}\\
\addlinespace[7pt]
edge, quenched & changes by about $3\%$ for $\sigma=0$ (numerical result; the samples
affected have probability $1-F_2(0)=3.1\%$)&
$4\pi\eta/s$ changes by at most $1.1\%$ (for $\sigma=0$ and $\omega=T$) &
\eqref{eq:cond-diff}\\
\addlinespace[7pt]
edge, annealed & changes at order one &
$S_{\rm a}<0$ for $\theta<0.840$ ($\sigma=0$):\newline the ratio is not defined &
\eqref{eq:Sa-wall}\\
\addlinespace[3pt]
\bottomrule
\end{tabularx}
\caption{Summary of the results in the edge regime, $CT\sim e^{-2S_0/3}$, and of their
dependence on the non-perturbative completion; the block numbers continue those of
Table~\ref{tab:summary}. $\cE_{\rm q}$ and $\cE_{\rm a}$ are the quenched and annealed edge
scaling functions of \eqref{eq:E-def}, $S_{\rm q}$ and $S_{\rm a}$ the quenched and annealed
entropies, $S_{\rm disk}=\frac32\log\theta+\frac32-\log2\sqrt\pi$ the one-loop entropy in edge
variables, $F$ Dawson's function, and $F_2$ the Tracy--Widom distribution
\eqref{eq:TW}. In the edge regime $4\pi\eta/s$ denotes the ratio of the quenched averages of
$\eta$ and $s$; the average of the single-sample ratio behaves differently,
eq.~\eqref{eq:ratio-of-sample}.}
\label{tab:summary-edge}
\end{table}

\subsection{\texorpdfstring{Physical interpretation}{Physical interpretation}}
\label{sec:interpretation}

\paragraph{The viscosity as a spectral observable.} Below $E_{\rm gap}$ the Kubo viscosity
of a near-extremal brane is a spectral statistic. It counts pairs of levels at distance
$\omega$, weighted by the Boltzmann factor and by the Schwarzian vertex, and it is
normalized by the one-level function of the same ensemble. Three energy scales compete:
the temperature $T$, the probe frequency $\omega$, and the level spacing $\Delta(T)$ at the
thermal energy. The regimes we have found are organized by their ratios.
\begin{itemize}
\item \emph{Disk.} Here $\Delta(T)\ll\omega\ll T$, and one obtains the Schwarzian result
\eqref{eq:eta-disk} with its $(CT)^{-1/2}$ growth.
\item \emph{Bulk double-scaling limit.} Here $\omega\sim\Delta(T)\ll T$. The answer
\eqref{eq:eta-DS} is a function of $CT$ and $e^{S_0}C\omega$ only. The normalizing
partition function is the disk one. Its non-perturbative corrections have the same origin as
the plateau term in the numerator: both come from the oscillating part of the eigenvalue
correlators, which is not captured by the sum over surfaces and which, in the language of
\cite{SaadShenkerStanford2019}, is computed by ``brane'' (determinant) insertions in the matrix
integral. In the partition function, however, these corrections only resum the genus series,
and they drop out of the double-scaling limit.
\item \emph{Edge regime.} Here $\omega\sim\Delta(T)\sim T$, and the Airy kernel controls both
the numerator and the normalization. The quenched ensemble is required, because the
annealed one is dominated by rare eigenvalues in the forbidden region.
\end{itemize}
In none of these regimes does level repulsion leave a finite DC coefficient at finite
$S_0$: the DC viscosity goes to zero, instead of diverging at low temperature as in the
Schwarzian approximation. This reflects the failure of a naive hydrodynamic description at
frequencies below the level spacing. What survives instead is a set of universal scaling functions. They describe how the
Schwarzian enhancement of dissipation is cut off by the discreteness of the spectrum.
Tables~\ref{tab:summary} and~\ref{tab:summary-edge} collect the explicit results for $\eta$ and $4\pi\eta/s$, together
with the ranges of $CT$, $C\omega$ and $r$ in which each of them applies.

\paragraph{Hydrodynamics near the level-spacing scale.} Hydrodynamics is a
coarse-grained effective theory. It assumes that a frequency bin contains many
microscopic transitions, i.e.\ that the observation time is shorter than the time needed
to resolve individual levels. A finite chaotic system has instead a discrete spectrum with
Wigner--Dyson correlations. The crossover between the two descriptions is controlled by
two scales, the Thouless energy $E_{\rm Thou}$ and the Heisenberg time
$t_{\rm H}=2\pi\rho$
\cite{AltshulerShklovskii1986,SivanImry1987,DAlessioEtAl2016,WinerSwingle2022}.\footnote{The Thouless time $t_{\rm Thou}$ is the time
that the slowest collective mode, for instance the diffusion of a conserved density, needs to
relax across the whole system; for a diffusive system of linear size $L$ and diffusion constant
$D$, $t_{\rm Thou}\sim L^2/D$. The Thouless energy is $E_{\rm Thou}=1/t_{\rm Thou}$. For energy
differences below $E_{\rm Thou}$ the spectral correlations are those of random-matrix theory. (We
write $E_{\rm Thou}$ to avoid confusion with the thermal energy $E_{\rm th}$.) The Heisenberg time $t_{\rm H}=2\pi\rho(E)=2\pi/\Delta(E)$
is the time needed to resolve individual energy levels. For $t\gtrsim t_{\rm H}$ the discreteness
of the spectrum becomes visible, and the spectral form factor reaches its plateau.}
They separate three regimes:
\begin{equation*}
\begin{array}{c|c|l}
\text{frequency} & \text{time} & \text{effective description}\\ \hline
E_{\rm Thou}\lesssim\omega\ll\tau_{\rm mic}^{-1} & \tau_{\rm mic}\ll t\lesssim t_{\rm Thou} & \text{spatial hydrodynamic modes}\\
\Delta\ll\omega\ll E_{\rm Thou} & t_{\rm Thou}\ll t\ll t_{\rm H} & \text{ergodic zero mode, random-matrix correlations}\\
\omega\lesssim\Delta & t\gtrsim t_{\rm H} & \text{discrete levels, plateau, recurrences}
\end{array}
\end{equation*}
Hydrodynamics does not end abruptly at $E_{\rm Thou}$. Fluctuating hydrodynamics controls
the approach of the connected spectral form factor to the random-matrix ramp
\cite{FriedmanEtAl2019,WinerSwingle2022}. Below $E_{\rm Thou}$ the theory is effectively
zero-dimensional, and the sine kernel is the universal two-level correlator. The ramp term
of \eqref{eq:ramp-plateau} is the part of the suppression that is still perturbative in
$e^{-S_0}$. The plateau term is what makes $\eta_{\rm DC}$ vanish. It is the transport
analogue of the Heisenberg time \cite{CotlerEtAl2017,BarbonRabinovici2003,Saad2019}.

For a finite system the macroscopic shear viscosity, that is, the transport coefficient
that appears in the hydrodynamic equations of the infinite system, is defined by the
order of limits
$\eta_{\rm hydro}=\etasc\lim_{\omega\to0}\lim_{V\to\infty}\operatorname{Im}\cG^R(\omega;V)/\omega$,
not by $\omega\to0$ at fixed volume. What one measures in a finite-time experiment or
simulation is instead a finite-resolution viscosity,
\begin{equation}
\eta_\Gamma=\etasc\int\!\dd\omega\,W_\Gamma(\omega)\,\frac{\operatorname{Im}\cG^R(\omega)}{\omega},\qquad
\int\!\dd\omega\,W_\Gamma(\omega)=1,
\label{eq:smeared}
\end{equation}
with $W_\Gamma$ a window of width $\Gamma$ set by the observation time or a bath-induced
linewidth. This is the standard way in which the Kubo formula is applied to a system with
a discrete spectrum. In the theory of disordered and mesoscopic conductors the width appears as
an imaginary part of the frequency, or as an inelastic broadening $\gamma$ of the levels, and
the response changes qualitatively between the continuous-spectrum limit $\gamma\gg\Delta$ and
the discrete-spectrum limit $\gamma\ll\Delta$
\cite{ThoulessKirkpatrick1981,SivanImry1987,TrivediBrowne1988,ReuletBouchiat1994,Imry2002}. In
numerical studies of transport in finite many-body systems the same role is played by the
broadening of the delta functions of the Lehmann sum, or by the finite time over which the
current correlator can be followed
\cite{Bulanchuk2020,BertiniEtAl2021}. The smeared viscosity $\eta_\Gamma$ follows from our scaling functions by
smearing over $r$ or $\nu$, and it reproduces the disk result when
$\Delta(T)\ll\Gamma$. For near-extremal branes the frequency below which level repulsion
suppresses the response, defined by $\pi\rho_0(E_{\rm th})\,\omega_{\rm lr}=1$ at the thermal
saddle (so that $\omega_{\rm lr}=\Delta(T)/\pi$ at $CT\ll1$), is
\begin{equation}
C\omega_{\rm lr}(T)\simeq
\begin{cases}4\pi\,e^{-S^{\rm s.c.}(T)},& CT\gtrsim1,\\[2pt]
e^{-S_0}/\sqrt{2CT},& e^{-2S_0/3}\ll CT\ll1.\end{cases}
\label{eq:omega-lr}
\end{equation}
In the semiclassical regime it is exponentially small in the Bekenstein--Hawking entropy,
which is why corrected and uncorrected curves agree over an enormous frequency range.
The continuous-spectrum description of the response requires a frequency window
$\omega_{\rm lr}(T)\ll\omega\ll T$. As $T$ is lowered, the lower end $\omega_{\rm lr}(T)$ of this
window rises, while its upper end $T$ decreases. The two meet, $\omega_{\rm lr}(T)=T$, at
$CT=2^{-1/3}e^{-2S_0/3}$, i.e.\ exactly at $\kappa T=1$. Below that
temperature the window is closed, and no hydrodynamic regime is left.

\paragraph{Finite volume and the planar limit.} Both $S_0$ and $C$ are extensive in the
transverse volume: by \eqref{eq:Ssc}, $S_0\propto\mu^2V_2\mathcal N$ and $C\propto\mu V_2\mathcal N$
\cite{KanargiasKiritsisMurthyPapadoulakiPorfyriadis2025}. In the strict planar limit at fixed
$T$ one has $CT\to\infty$, and the brane is semiclassical. The Schwarzian regime is itself a
finite-volume phenomenon, reached for $T\lesssim1/C\propto1/V_2$. Within the Schwarzian
regime, at fixed $CT$, the non-perturbative scales are suppressed by $e^{-S_0}$, which is
exponentially small in $V_2$, and \eqref{eq:intro-disk} becomes exact as $V_2\to\infty$. The
effects computed here are thus the leading finite-size effects in the low-temperature
transport of a finite holographic system: a brane on a compact transverse space, or a
near-extremal black hole. In the latter case the same $\Delta_\cO=1$ correlator governs the
low-frequency absorption of massless quanta. Our scaling functions then describe how the
Schwarzian-corrected greybody factors and quantum absorption cross sections
\cite{BrownIliesiuPeningtonUsatyuk2024,Emparan2025,Biggs2025,EmparanTrezzi2025,BetziosPapadoulakiZhou2025}
are modified by the discreteness of the spectrum; non-perturbative corrections to the
evaporation rate were studied in \cite{MohanThorlacius2024}, and the depletion of the
Hawking--Unruh spectrum at energies below the level spacing, which follows from the sine
kernel, was derived in \cite{BlommaertMertensVerschelde2021unruh}, building on the bulk
observables and on the treatment of random-matrix level statistics in JT gravity developed in
\cite{BlommaertMertensVerschelde2019clocks,BlommaertMertensVerschelde2021eigenbranes}.

The fluid mechanics of near-extremal branes was studied in the classical regime in
\cite{MoitraSakeTrivedi2020}; see also
\cite{EdalatiJottarLeigh2010sound,DavisonParnachev2013,AreanDavisonGouterauxSuzuki2021} for the
hydrodynamic modes of cold holographic matter and for the breakdown of hydrodynamics near
AdS$_2$ fixed points. Schwarzian corrections to the fluid/gravity correspondence were obtained
in \cite{NianPandoZayasYue2025}, and Schwarzian corrections to the conductivity and to transport
with momentum relaxation in
\cite{LiuNianPandoZayas2024,MaulikPandoZayasZhang2026,DasMaulikPandoZayasZhang2026}. Combining a fluid-dynamical description with the spectral effects
discussed here requires three ingredients that go beyond our analysis. First, we have only
computed the two-point function of $T_{xy}$ at zero spatial momentum, which reduces to quantum
mechanics in the throat. Fluid dynamics involves the correlators of all components of the
stress tensor and of the current at finite momentum $\vec p$, for which the throat operators
have momentum-dependent dimensions \cite{FaulknerLiuMcGreevyVegh2011,EdalatiJottarLeigh2010}.
Second, an operator of momentum $\vec p\neq0$ connects sectors of the spectrum with different
total momentum. One therefore needs the correlations between the levels of different momentum
sectors, which are not determined by the single matrix integral used here; if distinct sectors
are statistically independent, there is no level repulsion between them. Third, fluid dynamics
is non-linear, whereas the Kubo formula only captures linear response. It would be interesting
to develop these ingredients to obtain a more complete picture.

\paragraph{Annealed versus quenched.} With fixed boundary conditions, the Euclidean
gravitational path integral computes annealed quantities. At fixed $CT$ the difference from
the quenched ones is a replica-wormhole effect of order $e^{-2S_0}$
\cite{EngelhardtFischettiMaloney2020}, given in \eqref{eq:quenched-DT}. In the edge regime it is of
order one, and only the quenched entropy is physical
\cite{Johnson2020b,Johnson2021q,JanssenMirbabayi2021}. The Fredholm representation of
equations \eqref{eq:P-quenched-fredholm} and \eqref{eq:logZ-fredholm} gives a concrete
non-perturbative definition of the quenched Kubo formula. It is only weakly sensitive to the
non-perturbative completion. Section~\ref{sec:string} makes this precise for completions with a
hard wall: by \eqref{eq:cond-diff}, which holds at any temperature, quenched quantities change
only through the samples of the soft-edge ensemble that have a level below the wall, a
fraction $1-F_2(-\sigma)$ of them, about $3\%$ for a wall at the classical edge of the
spectrum; numerically they change by a few per cent
(Table~\ref{tab:string-edge}). Annealed quantities change at
order one. The reason is that a typical sample is described by the Airy process near its ground
state, whereas the completion affects only rare configurations, which dominate $\vev{Z}$ but
not $\vev{\log Z}$.

\subsection{\texorpdfstring{Assumptions and limitations}{Assumptions and limitations}}
\label{sec:limitations}

\paragraph{Matrix elements and symmetry class.} We have treated the shear mode as a $\Delta_\cO=1$ probe with
Gaussian matrix elements independent of the eigenvalues, as appropriate for JT gravity with
matter at leading order \cite{JafferisKolchmeyerMukhametzhanovSonner2022}; as explained in
Section~\ref{sec:mm}, only their variance \eqref{eq:vertex} is used. We have also
assumed a completion in the GUE class, either soft-edge or with a hard wall as in
Section~\ref{sec:string}. This is an assumption about the boundary theory, which does not
follow from the Einstein--Maxwell action (Section~\ref{sec:mm}). Non-Gaussianities of the matrix elements are
suppressed by further powers of $e^{-S_0}$ \cite{JafferisKolchmeyerMukhametzhanovSonner2022}
and do not affect the scaling limits. For time-reversal-invariant or supersymmetric
completions \cite{StanfordWitten2019,HeydemanIliesiuTuriaciZhao2020} the sine and Airy
kernels are replaced by their orthogonal, symplectic or hard-edge counterparts. The structure
of our arguments is unchanged, but the scaling functions differ. For example, the small-$T$
entropy scales as $T^{\beta_{\rm D}+1}$ \cite{JanssenMirbabayi2021}, the low-frequency viscosity as $\omega^{\beta_{\rm D}}$, and
numbers such as $0.1736$ in \eqref{eq:eta-max} and $7\pi^4/180$ in \eqref{eq:S-quenched} are
specific to $\beta_{\rm D}=2$. The vanishing of $\eta_{\rm DC}$, the cutoff of the Schwarzian
growth at $\omega\sim\Delta(T)$ and the $T^{-2}$ law \eqref{eq:eta-s-lowT} at fixed $\omega/T$
hold in all three classes; for $\beta_{\rm D}=1$ the viscosity saturates below the cutoff
instead of decreasing, because the bracket of \eqref{eq:DS-small-q} is replaced by a function
that vanishes as $y^{\beta_{\rm D}}\propto(CT)^{\beta_{\rm D}/2}$. We have also ignored the other
light fields of the throat, such as the $U(1)$ phase mode of the charged brane. At fixed
chemical potential we expect it to modify $\rho_0$ by smooth factors without changing the
random-matrix universality class of the edge.

\paragraph{Conserved charges.} The brane carries a $U(1)$ charge and, on
the torus, two components of momentum. The Hamiltonian is block diagonal in these charges.
Random-matrix statistics are expected within each block, while the levels of different blocks
are uncorrelated; for JT gravity coupled to gauge fields this structure was derived in
\cite{Iliesiu2019,KapecMahajanStanford2019}. The operator dual to $h_{xy}$ at zero spatial
momentum is neutral, so the Lehmann sum \eqref{eq:Lehmann} only involves pairs of levels in the
same block of charge and momentum. Our formulas therefore apply to each charge sector separately, with the values of
$S_0$ and $C$ of that sector, and the level spacing that enters $r$ and $\kappa$ is the spacing
within a sector, not that of the full spectrum. We have implicitly worked in a single sector. If
several sectors are thermally populated, as can happen at fixed chemical potential, the scaling
functions have to be averaged over the sectors with their thermal weights. This affects neither
the vanishing of $\eta_{\rm DC}$ nor the structure of the scaling limits, but it smears the
dependence on $r$ and on $\kappa T$, because $e^{S_0}$ differs between neighbouring sectors by
factors of order one.

\paragraph{Stability of the brane.} We have assumed that the Reissner--Nordstr\"om
brane dominates the ensemble of fixed charge down to zero temperature, so that its $e^{S_0}$
states persist, with the Schwarzian density, down to the edge of the spectrum. In
Einstein--Maxwell theory this is true by construction. In an ultraviolet-complete theory it is
an additional assumption. Charged matter can make the brane unstable towards a condensate
below a temperature that is a fixed fraction of $\mu$
\cite{Gubser2008,HartnollHerzogHorowitz2008,DenefHartnoll2009}, non-supersymmetric extremal
black branes are expected to be able to decay \cite{ArkaniHamedMotlNicolisVafa2007}, and the
generic ground state of a conformal field theory at large charge is a superfluid without
entropy \cite{HellermanOrlandoReffertWatanabe2015}. All these scales lie far above
$E_{\rm gap}=\sqrt3\,\pi\mu/S_0$. Our results apply when none of these instabilities occurs. The
edge regime, which requires the Schwarzian density down to the lowest levels, is the most
sensitive to this assumption. A related point is that a given boundary theory has a single
Hamiltonian, and not an ensemble. In the bulk double-scaling regime this is harmless, because
the thermal window contains exponentially many pairs of levels at every frequency and the
response is self-averaging. In the edge regime it is not: as stressed in
Section~\ref{sec:aq}, the averages computed there are statistical statements.

\paragraph{Limitations on the hydrodynamic interpretation.} Three remarks limit the hydrodynamic reading
of our results, and of the disk result that they extend. The first concerns the Thouless
energy. With $\eta=s/4\pi$ and the enthalpy density (energy density plus pressure) $w$, the shear diffusion
constant of the extremal brane is $D=\eta/w=\sqrt3/(12\mu)$, and the slowest hydrodynamic mode
on a square torus relaxes at the rate
\begin{equation}
E_{\rm Thou}=D\Bigl(\frac{2\pi}{L}\Bigr)^2=\frac{\sqrt3\,\pi}{12\,\mu V_2}=\frac{\pi\,\mathcal N}{36}\,E_{\rm gap}.
\label{eq:Thouless}
\end{equation}
For $\mathcal N\gg1$ the whole quantum regime $\omega,T\lesssim E_{\rm gap}$ lies below the
Thouless energy, in the ergodic regime of the table above. This supports the use of
random-matrix statistics. It also means that no hydrodynamic mode of the torus probes
$\eta(\omega)$ at $\omega\lesssim E_{\rm gap}$: the longest-wavelength shear mode decays at the
rate $E_{\rm Thou}$, where the viscosity is classical. The quantity computed here, and in
\cite{KanargiasKiritsisMurthyPapadoulakiPorfyriadis2025,CremoniniLiLiuNian2025}, is the
absorption of a homogeneous, slowly varying deformation of the shape of the torus, the
analogue of the absorption of a metallic grain in a uniform oscillating field. The second
remark concerns the amplitude of the deformation. Linear response requires that the matrix
elements of the perturbation between neighbouring levels be small compared with their spacing
$\Delta$. For $X=\int\dd^2x\,T_{xy}$ the Lehmann representation gives
$\overline{|X_{mn}|^2}\simeq V\eta\,T\Delta/\pi$ for neighbouring levels, so that the strain
must satisfy
\begin{equation}
|h_{xy}|\ll\Bigl(\frac{\Delta}{V\eta\,T}\Bigr)^{1/2}\sim\Bigl(\frac{\Delta(T)}{S_0\,T}\Bigr)^{1/2},
\end{equation}
which is exponentially small in $S_0$. For larger amplitudes the drive itself mixes
neighbouring levels, and dissipation proceeds through Landau--Zener transitions
\cite{Wilkinson1988}. The vanishing of $\eta_{\rm DC}$ thus refers to the limit of vanishing
amplitude, taken before $\omega\to0$. The third remark is that in $2+1$ dimensions hydrodynamic
fluctuations give the viscosity a contribution proportional to $\log(1/\omega)$
\cite{KovtunMooreRomatschke2011,Kovtun2012}. This is a one-loop effect in the bulk, of relative
order $Tw/\eta^2\propto T/(\mu\mathcal N)$ up to the logarithm. It is parametrically larger than
the non-perturbative effects studied here, but it is negligible for $T\lesssim E_{\rm gap}$,
where $T/\mu\lesssim1/S_0$, and on the torus the logarithm is cut off at
$\omega\sim E_{\rm Thou}$.

\subsection{\texorpdfstring{Outlook}{Outlook}}
\label{sec:outlook}

Our analysis can be extended in several directions, and most of them can
be pursued with the tools developed here.

The first concerns the dimension of the probe operator. We have studied the shear mode, which
couples to an operator with $\Delta_\cO=1$ in the throat. For a general $\Delta_\cO$ only the
vertex changes, $\Gamma(1\pm ik_1\pm ik_2)\to\Gamma(\Delta_\cO\pm ik_1\pm ik_2)/\Gamma(2\Delta_\cO)$,
and the arguments of Sections~\ref{sec:repulsion}--\ref{sec:airy} go through for the
throat correlator $\cG^>$. What does not carry over automatically is the relation of this
correlator to the boundary observable: the matching \eqref{eq:UV-IR} between the throat and the
exterior region, the contact terms and, for operators that overlap with conserved densities, the
mixing with the corresponding hydrodynamic modes depend on the operator and have to be worked
out case by case. In the bulk double-scaling limit the level-repulsion factor $R_r$ is the same for
all $\Delta_\cO$, while the thermal weight that multiplies it depends on $\Delta_\cO$ through the
vertex. In the edge regime the momenta are of order $e^{-S_0/3}$, the vertex becomes the constant
$\Gamma(\Delta_\cO)^4/\Gamma(2\Delta_\cO)$, and the edge function $\cE$ is therefore universal up
to normalization. What does depend on $\Delta_\cO$ is the temperature dependence of the
classical answer. In the classical AdS$_2$ geometry $\operatorname{Im}\cG^R\propto\omega^{2\Delta_\cO-1}$
at $T=0$, and $\operatorname{Im}\cG^R\propto\omega\,T^{2\Delta_\cO-2}$ for $\omega\ll T$. Only for
$\Delta_\cO=1$ does the corresponding DC coefficient stay finite and non-zero as $T\to0$.

This observation leads directly to other transport coefficients. The charge conductivity of the
extremal Reissner--Nordstr\"om brane has a delta function at $\omega=0$, due to momentum
conservation, and a regular part that is controlled by the throat. At zero temperature the
regular part scales as $\omega^2$, because the relevant operator in the throat has dimension
$\Delta_\cO=2$ \cite{EdalatiJottarLeigh2010c}. The Schwarzian two-point function is known for any
$\Delta_\cO$ \cite{MertensTuriaciVerlinde2017}, and the quantum corrections to the conductivity
below $E_{\rm gap}$ have been computed in
\cite{LiuNianPandoZayas2024,MaulikPandoZayasZhang2026,DasMaulikPandoZayasZhang2026}. Their fate
at the level-spacing scale can be determined in the same way as here. Thermal and thermoelectric transport, which involve the same mixed sector of metric
and gauge-field fluctuations, could be analysed along the same lines.

A third direction is to compute the viscosity at finite $S_0$ directly, without matching two
scaling limits. In this paper we have obtained it in two limits: the bulk double-scaling limit,
with $r$ and $CT$ fixed, and the edge limit, with $\kappa T$ and $\kappa\omega$ fixed. The two
overlap for $e^{-2S_0/3}\ll CT\ll1$. To draw a single curve at finite $S_0$
(Figure~\ref{fig:composite}) we glued them with the composite \eqref{eq:composite} of matched
asymptotic expansions, bulk plus edge minus their common limit. This composite is uniformly
valid only up to corrections that vanish as $S_0\to\infty$, and it is not an exact result at any
finite $S_0$. The gluing can be avoided. The string equation of Section~\ref{sec:string}
determines the exact one- and two-level statistics of JT gravity at finite $S_0$, for all
energies, and its kernel obeys a differential equation of its own
\cite{Johnson2020,Johnson2020c,Johnson2025kernel}. The sine and Airy kernels are two limits of
this kernel. Evaluating the quenched Kubo formula \eqref{eq:P-quenched-fredholm} with it, and
with the full Schwarzian vertex, would give $\eta(\omega,T)$ at all temperatures in a single
computation. It would also give the corrections of relative order
$e^{-2S_0/3}$ to the edge functions, and extend the comparison of Section~\ref{sec:string}
beyond the edge limit.

A fourth direction is a numerical test in a microscopic model. The SYK model has a Schwarzian
regime at low temperature and random-matrix level statistics, with a Dyson index
($\beta_{\rm D}=1,2$ or $4$) that depends on the number $N$ of fermions
\cite{MaldacenaStanford2016,GarciaGarciaVerbaarschot2016,CotlerEtAl2017}. Unlike a black brane,
whose $e^{S_0}$ is astronomically large, an SYK model that can be diagonalized exactly has a
moderate $e^{S_0}$. The frequencies and temperatures at which level repulsion becomes important,
including the edge regime $CT\sim e^{-2S_0/3}$, are then within numerical reach. The
low-temperature Kubo response of a simple operator at finite $N$ should be described by the
functions $\eta_{\rm DS}$ and $\cE$, with $e^{S_0}$ and $C$ read off from the density of states,
and with the sine and Airy kernels of the appropriate symmetry class.

Finally, our results raise a question about the interpretation of the low-temperature growth of
$\eta/s$. In \cite{KanargiasKiritsisMurthyPapadoulakiPorfyriadis2025} the large value of $\eta/s$
at low temperature was taken as a hint that the dynamics may become glassy. In the edge regime we
find that the growth of the quenched $4\pi\eta/s$ is driven mainly by the vanishing of the
entropy of a system with a non-degenerate ground state, $S_{\rm q}\propto T^3$, while $\eta$
itself decreases (linearly in $T$ at fixed $\omega/T$). The growth is moreover a property
of the ratio of quenched averages: the average of the single-sample ratio,
\eqref{eq:ratio-of-sample}, does not grow at fixed $\omega/T$. It would be interesting to find observables that distinguish this mechanism
from genuinely glassy behaviour, for instance the full frequency dependence of $\eta(\omega)$ at
$\omega\lesssim T$, or higher-point functions of the stress tensor. A related question is how the
coupling of the brane to an environment, for instance to Hawking radiation in the case of an
evaporating black hole, broadens the levels. A linewidth larger than $\Delta(T)$ restores the
disk result through the smeared viscosity \eqref{eq:smeared}, so the competition between this
width and the level spacing decides which of the regimes of Tables~\ref{tab:summary} and~\ref{tab:summary-edge} is realized
in a given physical situation.

More broadly, our analysis shows that non-perturbative quantum gravity does not simply
``correct'' the transport coefficients of near-extremal branes. It changes the question that
is being asked.
Below the level-spacing scale transport becomes a statement about the spectral statistics
of the black hole. The universal content of the infrared physics is then no longer a single
number, such as $1/4\pi$, but a set of scaling functions of $T$, $\omega$ and $e^{S_0}$, which
encode considerably richer physics.

\section*{Acknowledgements}

We wish to thank Roberto Emparan, Blaise Gouteraux, Elias Kiritsis, Thomas Mertens and Sameer Murthy for discussions on topics related to this work.
P.~B.\ acknowledges financial support from the European Research Council (grant BHHQG-101040024), funded by the European Union. This research was supported in part by the Research Foundation -- Flanders (FWO) through a Grant for a Long Stay Abroad (grant number V431626N). Views and opinions expressed are those of the authors only and do not necessarily reflect those of the European Union or the European Research Council. Neither the European Union nor the granting authority can be held responsible for them. Finally, P.~B.\ thanks the Institute for Advanced Study for its hospitality during the completion of part of this work.
We acknowledge the use of Claude for proofreading and for assistance with coding and numerics.

\clearpage
\appendix

\section{Moments}
\label{app:moments}

All moments needed in the text are elementary. The thermal weight $g_T(k)$ is a Gaussian
centred at $\bar k=2\pi CT$ multiplied by elementary factors. We define
\begin{equation}
\bar k=2\pi CT,\qquad \lambda=\sqrt{\frac{\pi CT}{2}},\qquad u_\pm=\frac{k_0\pm\bar k}{\sqrt{2CT}} .
\end{equation}
The truncated moments
\begin{equation}
\cI_n(CT;k_0)=\int_{k_0}^\infty\!\dd k\;k^n\,e^{-k^2/2CT}\sinh(2\pi k)
\end{equation}
are then
\begin{align}
\cI_1(CT;k_0)&=\frac{e^{2\pi^2CT}}{2}\Bigl\{CT\bigl(e^{-u_-^2}-e^{-u_+^2}\bigr)+\bar k\,\lambda\bigl[\erfc u_-+\erfc u_+\bigr]\Bigr\},\\
\cI_2(CT;k_0)&=\frac{e^{2\pi^2CT}}{2}\Bigl\{CT\bigl[(k_0+\bar k)e^{-u_-^2}-(k_0-\bar k)e^{-u_+^2}\bigr]\nonumber\\
&\qquad\qquad+(CT+\bar k^2)\,\lambda\bigl[\erfc u_--\erfc u_+\bigr]\Bigr\},
\end{align}
which reduce to \eqref{eq:I12} at $k_0=0$. The moments of $\sinh^3$ follow from
$\sinh^3x=\frac14(\sinh3x-3\sinh x)$. With
\begin{equation}
\begin{aligned}
A_1(a,CT)&=\int_0^\infty\!\dd k\;k\,e^{-k^2/2CT}\sinh(ak)=a\,CT\sqrt{\frac{\pi CT}{2}}\;e^{a^2CT/2},\\
A_2(a,CT)&=\int_0^\infty\!\dd k\;k^2e^{-k^2/2CT}\sinh(ak)\\
&=a\,(CT)^2+\sqrt{\frac\pi2}\,(CT)^{3/2}\bigl(1+a^2CT\bigr)e^{a^2CT/2}\erf\Bigl(a\sqrt{\frac{CT}{2}}\Bigr),
\end{aligned}
\end{equation}
one has
\begin{equation}
K_n(CT)=\tfrac14\bigl[A_n(6\pi,CT)-3A_n(2\pi,CT)\bigr],\qquad n=1,2 .
\end{equation}
These expressions agree with direct quadrature to relative accuracy $10^{-13}$.

\section{Weil--Petersson data}
\label{app:WP}

We collect here the higher-genus data used in Section~\ref{sec:normalization}, and first
define the symbols. $V_{g,n}(b_1,\dots,b_n)$ is the Weil--Petersson volume of the moduli space
of hyperbolic Riemann surfaces of genus $g$ with $n$ geodesic boundaries of lengths
$b_1,\dots,b_n$ \cite{Mirzakhani2007}, in the normalization of \cite{SaadShenkerStanford2019}.
$Z_{g,n}$ is the contribution of the surfaces of genus $g$ with $n$ asymptotic boundaries to the
JT path integral, with its weight $e^{(2-2g-n)S_0}$ stripped off. Only $n=1$ is needed here. The
genus expansion of the partition function is
\begin{equation}
\vev{Z(T)}\simeq\sum_{g\ge0}e^{(1-2g)S_0}Z_{g,1}(CT),\qquad Z_{0,1}=e^{-S_0}Z_{\rm disk} ,
\end{equation}
and for $g\ge1$ the term $Z_{g,1}$ is obtained by gluing a trumpet to a surface with one
geodesic boundary of length $b$,
\begin{equation}
Z_{g,1}(CT)=\int_0^\infty\!b\,\dd b\;Z_{\rm trumpet}(CT,b)\,V_{g,1}(b),\qquad
Z_{\rm trumpet}(CT,b)=\sqrt{\frac{CT}{2\pi}}\;e^{-CT\,b^2/2} .
\end{equation}
The coefficients in \eqref{eq:Z-ratio} are the ratios $a_g(CT)=Z_{g,1}(CT)/Z_{0,1}(CT)$. The
symbols $Z^{\Ai}_{g,1}$, defined in \eqref{eq:Z-airy-genus}, denote the same quantities for the
Airy density.

With $V_{1,1}(b)=(b^2+4\pi^2)/48$ and
$V_{2,1}(b)=(4\pi^2+b^2)(12\pi^2+b^2)(6960\pi^4+384\pi^2b^2+5b^4)/2211840$
\cite{Mirzakhani2007,SaadShenkerStanford2019},
\begin{align}
\sqrt{2\pi}\,Z_{1,1}(CT)&=\frac{\pi^2}{12\sqrt{CT}}+\frac{1}{24(CT)^{3/2}},\\
\sqrt{2\pi}\,Z_{2,1}(CT)&=\frac{29\pi^8}{192\sqrt{CT}}+\frac{169\pi^6}{1440(CT)^{3/2}}+\frac{139\pi^4}{2880(CT)^{5/2}}\nonumber\\
&\quad+\frac{29\pi^2}{2880(CT)^{7/2}}+\frac{1}{1152(CT)^{9/2}},
\end{align}
so that
\begin{equation}
\begin{aligned}
a_2(CT)=e^{-2\pi^2CT}\Bigl[&\frac{29\pi^8}{192(CT)^2}+\frac{169\pi^6}{1440(CT)^3}+\frac{139\pi^4}{2880(CT)^4}\\
&+\frac{29\pi^2}{2880(CT)^5}+\frac{1}{1152(CT)^6}\Bigr].
\end{aligned}
\label{eq:a2}
\end{equation}
The most singular terms, $1/24$ and $1/1152=1/(24^2\,2!)$, are those of $Z_{\Ai}$. The next
ones, $\pi^2/12$ and $29\pi^2/2880$, are reproduced by $e^{2\pi^2CT}e^{\mathcal F(CT)}$ with
$\mathcal F=\frac{e^{-2S_0}}{24(CT)^3}+\frac{\pi^2e^{-4S_0}}{120(CT)^5}$ from Section~\ref{sec:normalization}.
For example, $\frac{2\pi^2}{1152}+\frac{\pi^2}{120}=\frac{29\pi^2}{2880}$.

\section{The oscillatory term of the one-level density}
\label{app:osc}

In this appendix we derive the amplitude $1/4\pi E$ of the oscillatory term in
\eqref{eq:sss-density}, which was obtained in \cite{SaadShenkerStanford2019}, in the
Schr\"odinger formulation of Section~\ref{sec:string-eq}. The derivation shows that the
amplitude depends only on the distance from the edge of the spectrum, for any leading density
$\rho_0$.

In that formulation $\vev{\rho(E)}=\int_{-\infty}^0\psi(E,x)^2\,\dd x$, eq.~\eqref{eq:K-string}.
Let $p(x)=\sqrt{E-u_0(x)}$. In the classically allowed region the WKB eigenfunction, normalized
to a delta function in the energy, gives
\begin{equation}
\psi(E,x)^2=\frac{1+\cos2\varphi(x)}{2\pi\hbar\,p(x)},\qquad \varphi'(x)=\frac{p(x)}{\hbar}.
\end{equation}
The non-oscillating term integrates to $\rho_0(E)$, eq.~\eqref{eq:rho0-WKB}. For the oscillating
term we use $\cos2\varphi\,\dd x=\hbar\,\dd(\sin2\varphi)/2p$ and integrate by parts. Up to terms
that are suppressed by a further power of $\hbar$, the result is the contribution of the
endpoint $x=0$,
\begin{equation}
\int_{-\infty}^0\!\dd x\,\frac{\cos2\varphi(x)}{2\pi\hbar\,p(x)}\simeq\frac{\sin2\varphi(0)}{4\pi\,p(0)^2} .
\end{equation}
Since $u_0(0)=0$ is the edge of the spectrum, $p(0)^2=E$. The phase is fixed by the connection
formula at the turning point and by $\int p\,\dd x=\pi\hbar N(E)$, which give
$2\varphi(0)=2\pi N(E)-\pi/2$. The endpoint contribution is therefore
$-\cos\bigl(2\pi N(E)\bigr)/4\pi E$, the second term of \eqref{eq:sss-density}, for any leading
density $\rho_0$.

A formula with the denominator $2\pi\rho_0(E)\int^E\dd E'/\rho_0(E')$ in place of $4\pi E$ was
obtained for general spectral curves in \cite{DiUbaldoPolicastro2023}, from a hydrodynamic
description of the eigenvalues. The two expressions coincide for $\rho_0\propto\sqrt E$, where
both reduce to the Airy result \eqref{eq:rho-airy} of \cite{SaadShenkerStanford2019}, and
therefore near the edge of any spectrum; they differ away from the edge when $\rho_0$ is not
proportional to $\sqrt E$. As a check, we have computed $\int_{-\infty}^0\psi^2\,\dd x$
numerically for Schr\"odinger problems whose leading density is not $\propto\sqrt E$, with
$\hbar=2\times10^{-3}$. The amplitude of the oscillation agrees with $1/4\pi E$ to better than
$0.03\%$. In the cases we considered, the alternative form differs from it by up to $37\%$.

\section{A spurious term in bulk--edge matching}
\label{app:matching}

Consider the uniform density obtained by adding to the Airy density \eqref{eq:rho-airy} the
difference between the JT and Airy oscillations, \eqref{eq:sss-density-JT} and the second term
of \eqref{eq:rho-airy}. With $\dd E=k\,\dd k/C$, its contribution to $Z$ is
\begin{equation}
\Delta Z_{\rm bulk}=-\frac1{2\pi}\int_0^\infty\!\frac{\dd k}{k}\;e^{-k^2/2CT}
\Bigl[\cos\Phi(k)-\cos\bigl(\tfrac23e^{S_0}k^3\bigr)\Bigr].
\end{equation}
The bracket vanishes rapidly as $k\to0$, so the integral converges. Near $k=0$,
$\Phi=e^{S_0}\bigl(\frac23k^3+\frac{4\pi^2}{15}k^5+\dots\bigr)$. Define $v$ by
$\frac23v^3=e^{-S_0}\Phi$, so that $v=k\bigl(1+\frac{2\pi^2}{15}k^2+\dots\bigr)$ and
$\dd k/k=\dd v/v-\frac{4\pi^2}{15}v\,\dd v+\dots$. The $\dd v/v$ term cancels against the Airy
oscillation, and the Mellin integral
$\int_0^\infty v\cos(\frac23\lambda v^3)\,\dd v=\Gamma(\frac23)\lambda^{-2/3}/[6(2/3)^{2/3}]$
gives
\begin{equation}
\Delta Z_{\rm bulk}=\frac{2\pi}{15}\,\frac{\Gamma(2/3)}{6\,(2/3)^{2/3}}\,e^{-2S_0/3}+\cO(e^{-4S_0/3})
=0.1239\,e^{-2S_0/3}.
\end{equation}
At $CT=1$ and $S_0=8$, $10$, $12$, direct quadrature gives $6.16\times10^{-4}$,
$1.589\times10^{-4}$ and $4.164\times10^{-5}$. The formula gives $5.98\times10^{-4}$,
$1.576\times10^{-4}$ and $4.156\times10^{-5}$. The dependence on $CT$ enters only at the next
order. The whole contribution comes from $k\sim e^{-S_0/3}$, where neither oscillatory form is
valid.

\section{Numerical methods for the edge regime}
\label{app:numerics}

\paragraph{Discretization.} For the quenched quantities we discretize $\zeta\in[-8,b]$ with composite Gauss--Legendre
panels of width $\min\bigl(1,\,0.8\cdot2\pi/\sqrt{\max(\zeta,1)},\,2\theta\bigr)$ and $8$ nodes
each. This resolves both the Airy oscillations and the thermal step of $w_x$. The upper
limit $b=\theta(\log x_{\max}+30)$ makes $1-w_x$ negligible at the end of the interval.
Excluding $\zeta<-8$ removes samples of probability below $10^{-14}$. With quadrature
weights $\varpi_i$, the symmetric Nystr\"om matrix is
$B_{ij}=\sqrt{\varpi_i(1-w_x(\zeta_i))}\,\hat K(\zeta_i,\zeta_j)\sqrt{\varpi_j(1-w_x(\zeta_j))}$.
We diagonalize it, which gives $D(x)=\prod_i(1-\lambda_i)$ and the resolvent
$M_x=\hat K+\hat K\sqrt{1-w_x}\,(1-B)^{-1}\sqrt{1-w_x}\,\hat K$ at arbitrary points,
including the shifted points $\zeta+\nu$.

\paragraph{The $x$ integral.} We integrate in $\log x$ with step $0.3$ over the range
\begin{equation}
\min\bigl(-\log Z_{\rm est}-18,\,-5.5/\theta-5\bigr)\;\le\;\log x\;\le\;\max\bigl(6.5/\theta,\,-\log Z_{\rm est}\bigr)+6 ,
\end{equation}
where $\log Z_{\rm est}=\max\bigl(-\vev{\zeta_0}/\theta,\,\log[\theta^{3/2}/2\sqrt\pi]\bigr)$ is a rough estimate of $\log Z$.
The lower limit must capture samples whose ground state lies as low as $\zeta_0=-5.5$
(probability $\sim10^{-7}$). A cutoff that is too high shows up as a violation of the sum
rule \eqref{eq:sum-rule-num} at the $10^{-3}$ level, through the right tail of the Tracy--Widom
distribution.

\paragraph{Checks.} The code reproduces the known values of the Tracy--Widom distribution
\cite{TracyWidom1994}, $F_2(0)=0.969373$ and $F_2(-2)=0.413224$, and the mean
$\vev{\zeta_0}=1.771087$ of the lowest level. The sum rule
\begin{equation}
\int\dd x\,D(x)\int\dd\zeta\,e^{-\zeta/\theta}w_x(\zeta)M_x(\zeta,\zeta)=1
\label{eq:sum-rule-num}
\end{equation}
holds to $10^{-12}$ for $\theta\le0.3$ and to $2\times10^{-8}$ for $\theta\ge0.4$. Changing the number
of nodes per panel ($8\to12$), the step in $\log x$ ($0.3\to0.1$) or the cutoff ($30\to45$)
changes $S_{\rm q}$ and $\cP_{\rm q}$ by less than $10^{-5}$. For $\theta\ge2$ the quenched
results approach the annealed ones as predicted by \eqref{eq:S-quenched}.

\paragraph{Annealed integrals.} The grid described above is used for the quenched
quantities only. The interval $\zeta\ge-8$ is sufficient for them, because they are averages over
typical samples. It would not be sufficient for the annealed quantities at low temperature,
which are dominated by the saddle at $\zeta_*=-1/4\theta^2$ (Section~\ref{sec:edge-entropy}): for
$\theta=0.1$ the saddle is at $\zeta_*=-25$, and an integral cut off at $\zeta=-8$ would retain
only a fraction of order $10^{-16}$ of $Z_{\Ai}$. The annealed integrals \eqref{eq:P-annealed}
are therefore evaluated separately, by Gauss--Legendre quadrature with $10$ nodes per panel on
the interval $a\le\zeta\le40\,\theta+20+\nu$, with the panel widths given above and the
temperature-dependent lower limit
\begin{equation}
a=\min\Bigl(-\frac{3}{16\,\theta^2}-12,\;-\frac{2}{5\,\theta^2}-8\Bigr)\quad(\theta<1),\qquad
a=-14\quad(\theta\ge1),
\end{equation}
which lies well below the saddle: $a=-48$ at $\theta=0.1$ and $a=-18$ at $\theta=0.2$. The
Boltzmann factor is combined with $1/Z_{\Ai}$ in the exponent, to avoid overflow. With this grid
the Laplace transform of $\hat\rho$ reproduces the closed form of $Z_{\Ai}$ in
\eqref{eq:P-annealed} to better than $10^{-11}$ for all $\theta\ge0.1$. Lowering $a$ by a further
$20$ and using $14$ nodes per panel changes $\cP_{\rm a}$ by less than $10^{-6}$ in relative
terms. The annealed entropy is the closed form \eqref{eq:S-annealed}.

\paragraph{Selected values.} Table~\ref{tab:airy} collects the entropies and the viscosity
scaling function of the soft-edge completion at selected temperatures.

\begin{table}[!ht]
\centering
\small
\begin{tabular}{ccccccc}
\toprule
$\theta$ & $S_{\rm q}$ & $S_{\rm a}$ & $S_{\rm disk}$ & $\cE_{\rm q}(0.1\theta,\theta)$ & $\cE_{\rm q}(\theta,\theta)$ & $\cE_{\rm a}(\theta,\theta)$\\
\midrule
0.10 & 0.0035 & $-169.9$ & $-3.219$ & 0.00070 & 0.0641 & $<10^{-30}$ \\
0.20 & 0.0237 & $-23.01$ & $-2.180$ & 0.00139 & 0.1269 & $2.1\times10^{-7}$ \\
0.30 & 0.0644 & $-7.744$ & $-1.572$ & 0.00211 & 0.1883 & 0.0005 \\
0.50 & 0.1908 & $-2.139$ & $-0.805$ & 0.00368 & 0.3102 & 0.0295 \\
0.80 & 0.4332 & $-0.426$ & $-0.100$ & 0.00687 & 0.4897 & 0.1928 \\
1.00 & 0.6034 & $\phantom{-}0.068$ & $\phantom{-}0.234$ & 0.00983 & 0.5863 & 0.3374 \\
1.50 & 1.0073 & $\phantom{-}0.793$ & $\phantom{-}0.843$ & 0.02122 & 0.6566 & 0.5563 \\
2.00 & 1.3577 & $\phantom{-}1.253$ & $\phantom{-}1.274$ & 0.03927 & 0.5977 & 0.5632 \\
3.00 & 1.9109 & $\phantom{-}1.876$ & $\phantom{-}1.882$ & 0.09539 & 0.4957 & 0.4876 \\
4.00 & 2.3265 & $\phantom{-}2.311$ & $\phantom{-}2.314$ & 0.1685 & 0.4307 & 0.4277 \\
\bottomrule
\end{tabular}
\caption{Edge-regime entropies and viscosity scaling function \eqref{eq:E-def} of the
soft-edge (Airy) completion. $S_{\rm disk}=\frac32\log\theta+\frac32-\log2\sqrt\pi$ is the one-loop
entropy in edge variables.}
\label{tab:airy}
\end{table}

\paragraph{Walled completions.} For the quenched quantities of Section~\ref{sec:string} we
use the same discretization with a panel boundary at $\sigma$. We set $1-w_x=1$ below $\sigma$,
which removes the samples with a level below the wall, and divide $D(x)$ by
$F_2(-\sigma)=\det(1-\chi_\sigma\hat K)$, computed on the same nodes. For $\sigma<0$ the lower
limit of $\log x$ is lowered to $\sigma/\theta-14$, since samples with a level at the wall have
$Z\sim e^{-\sigma/\theta}$. The conditioned kernel
\eqref{eq:K-cond}, used for the annealed quantities, is evaluated with $12$ Gauss--Legendre
nodes per panel of width $0.5$ on $[-14,\sigma]$. Without a wall the code reproduces the values
of Table~\ref{tab:airy} to all digits shown. To solve \eqref{eq:P34} we write $\cR=e^g$ and
solve $g''=2(e^g-x-\sigma)-\frac12g'^2$ as a boundary-value problem on $x\in[-14,24]$. On the
left we impose the Airy behaviour $g'=-2\Ai'(z)/\Ai(z)$ with $z=-(x+\sigma)$. On the right we
impose $\cR=x+\sigma-1/[4(x+\sigma)^2]$. These are the two asymptotic regimes of the
Hastings--McLeod solution in \eqref{eq:u-HM}. The
eigenfunctions are integrated from the left with a high-order Runge--Kutta method. They are
normalized in the asymptotic region by the WKB invariant
$\pi\bigl(p\psi^2+\psi'^2/p\bigr)$, $p=\sqrt{\zeta-u}$, which equals one for
energy-normalized states. The scripts that produce all figures and tables are available from the
authors upon request.

\end{document}